\documentclass[11pt]{article} 

\RequirePackage[margin=1in]{geometry}
\usepackage{amsfonts}
\usepackage{graphicx}
\usepackage{amsmath}
\usepackage{amsthm}
\usepackage{hhline}
\usepackage{multirow}

\usepackage{amssymb}
\usepackage{amsthm}
\usepackage{float}
\usepackage{bbm}
\usepackage{algorithm}
\usepackage{algpseudocode}
\usepackage{comment}
\usepackage{caption}
\usepackage{pgf}
\usepackage{tikz}
\usepackage{enumitem}
\usepackage{thmtools}
\usepackage{thm-restate}
\usepackage{cleveref}

\def\dist{\mathcal{D}}

\def\qp{A}

\def\Z{{\mathbb{Z}}}

\def\R{{\mathbb{R}}}

\def\E{{\mathbb{E}}}

\def\V{{\mathcal{V}}}
\def\setV{{\mathbb{V}}}

\def\A{{\mathcal{A}}}
\def\F{{\mathcal{F}}}
\newcommand{\argmax}{\textnormal{argmax}}

\renewcommand \vec [1]{\bm{#1}}

\def\sample{{\mathcal{S}}}

\renewcommand \vec {\mathbf}

\newtheorem{theorem}{Theorem}
\newtheorem{lemma}{Lemma}

\newtheorem{cor}{Corollary}
\newtheorem{fact}{Fact}
\newtheorem{definition}[theorem]{Definition}
\newtheorem{note}{Note}[section]

\newcommand{\cost}{{\textnormal{\texttt{cost}}}}
\newcommand{\clus}{{\Phi}}
\newcommand{\prune}{{\Psi}}
\newcommand{\merge}{{\xi}}
\newcommand{\owr}{{\textnormal{\texttt{owr}}}}

\newcommand{\slin}{{\textnormal{\texttt{slin}}}}

\DeclareMathOperator{\sign}{sgn}

\setlist{nosep}

\allowdisplaybreaks

\begin{document}

\title{Learning-Theoretic Foundations of Algorithm Configuration for Combinatorial Partitioning Problems\footnote{Authors' addresses: \texttt{\{ninamf,vaishnavh,vitercik,crwhite\}@cs.cmu.edu}.}}
 \author{Maria-Florina Balcan \and
 Vaishnavh Nagarajan \and
Ellen Vitercik \and
Colin White}
\date{\today}

\maketitle

\begin{abstract}
Max-cut, clustering, and many other partitioning problems that are of significant importance to machine learning and other scientific fields are NP-hard, a reality that has motivated researchers to develop a wealth of approximation algorithms and heuristics. Although the best algorithm to use typically depends on the specific application domain, a worst-case analysis is often used to compare algorithms. This may be misleading if worst-case instances occur infrequently, and thus there is a demand for optimization methods which return the algorithm configuration best suited for the given application's typical inputs. We address this problem for clustering, max-cut, and other partitioning problems, such as integer quadratic programming, by designing computationally efficient and sample efficient learning algorithms which receive samples from an application-specific distribution over problem instances and learn a partitioning algorithm with high expected performance. Our algorithms learn over common integer quadratic programming and clustering algorithm families: SDP rounding algorithms and agglomerative clustering algorithms with dynamic programming. For our sample complexity analysis, we provide tight bounds on the pseudodimension of these algorithm classes,  and show that surprisingly, even for classes of algorithms parameterized by a single parameter, the pseudo-dimension is superconstant. In this way, our work both contributes to the foundations of algorithm configuration and pushes the boundaries of learning theory, since the algorithm classes we analyze consist of multi-stage optimization procedures and are significantly more complex than classes typically studied in learning theory.
\end{abstract}

\setcounter{page}{0}
\thispagestyle{empty}

\newpage
\section{Introduction}\label{sec:intro}
NP-hard problems arise in a variety of diverse and oftentimes unrelated application domains. For example, clustering is a widely-studied NP-hard problem in
unsupervised machine learning, used to group
protein sequences by function, organize documents in databases by
subject, and choose the best locations for fire stations
in a city.
Although the underlying objective is the same, a ``typical problem instance''
in one setting may be significantly different from that in another, causing approximation algorithms to have 
inconsistent performance across the different application domains.

We study how to characterize which algorithms are best for which contexts, a task often referred to in the AI literature as \emph{algorithm
configuration}. This line of work allows researchers to compare algorithms according to an application-specific metric, such as expected performance over their problem domain, rather than a worst-case analysis. If worst-case instances occur infrequently in the application domain, then a worst-case algorithm comparison could be uninformative and misleading. We approach application-specific algorithm configuration via a learning-theoretic framework wherein an application domain is modeled as a distribution over problem instances. We then fix an infinite class of approximation algorithms for that problem and design computationally
efficient and sample efficient algorithms which learn the approximation algorithm with the best performance over the distribution, and therefore an algorithm with high performance in the specific application domain. Gupta and Roughgarden \cite{gupta2016pac} introduced this learning framework to the theory community, but it has been the primary model for algorithm configuration and portfolio selection in the artificial intelligence community for decades \cite{rice1976algorithm} and has led to breakthroughs in diverse fields including combinatorial auctions \cite{Leyton09:Empirical}, scientific computing \cite{demmel2005self}, vehicle routing \cite{caseau1999meta}, and SAT \cite{satzilla}.

In this framework, we study two important, infinite algorithm classes. First, we analyze approximation algorithms based on semidefinite programming (SDP) relaxations and randomized rounding procedures, which are used to approximate integer quadratic programs (IQPs). These algorithms can be used to find a nearly optimal solution to a variety of combinatorial partitioning problems, including the seminal max-cut and max 2-SAT problems. Second, we study agglomerative clustering
algorithms followed by a dynamic programming step to extract a good clustering. These techniques are widely used in machine learning and across many scientific disciplines for data analysis. We begin with a concrete problem description.

\medskip

\noindent \textbf{Problem description.} In this learning framework, we fix a computational problem, such as max-cut or $k$-means clustering, and assume that there exists an unknown, application-specific distribution $\mathcal{D}$ over a set of problem instances $\Pi$. We denote an upper bound on the size of the problem instances in the support of $\mathcal{D}$ by $n$. For example, the support of $\mathcal{D}$ might be a set of social networks over $n$ individuals, and the researcher's goal is to choose an algorithm with which to perform a series of clustering analyses.
Next, we fix a class of algorithms $\mathcal{A}$.
Given a cost function $\cost: \mathcal{A} \times \Pi \to [0, H]$, the learner's goal is to find an algorithm $h \in \mathcal{A}$ that approximately optimizes the expected cost with respect to the distribution $\mathcal{D}$, as formalized below.
\begin{definition} [\cite{gupta2016pac}]\label{def:PAC}
A learning algorithm $L$ $(\epsilon,\delta)$\emph{-learns the algorithm class $\mathcal{A}$ with respect to the cost function $\cost$} if, for every distribution $\mathcal{D}$ over $\Pi$, with probability at least $1-\delta$ over the choice of a sample $\sample \sim \mathcal{D}^m$, $L$ outputs an algorithm $\hat{h} \in \mathcal{A}$ such that $\E_{x \sim \mathcal{D}}\left[\cost\left(\hat{h}, x\right) \right] - \min_{h \in \mathcal{A}} \left \{\E_{x \sim \mathcal{D}}[\cost(h,x)]\right\} < \epsilon$.
We require that the number of samples be polynomial in $n$, $\frac{1}{\epsilon}$, and $\frac{1}{\delta}$, where $n$ is an upper bound on the size of the problem instances in the support of $\mathcal{D}$. Further, we say that $L$ is \emph{computationally efficient} if its running time is also polynomial in $n$, $\frac{1}{\epsilon}$, and $\frac{1}{\delta}$.
\end{definition}
We derive our guarantees by analyzing the pseudo-dimension of the algorithm classes we study (see Section~\ref{sec:learning_theory}, \cite{pollard1984convergence, pollard1990empirical, anthony2009neural}). We then use the structure of the problem to provide efficient algorithms for most of the classes we study.

\medskip

\noindent \textbf{SDP-based methods for integer quadratic programming.} Many NP-hard problems, such as max-cut, max-2SAT, and correlation clustering, can be represented as an integer quadratic program (IQP) of the following form. The input is an $ n\times n$ matrix $A$ with nonnegative diagonal entries and the output is a binary assignment to each variable in the set $X = \{x_1, \dots, x_n\}$ which maximizes $\sum_{i,j \in [n]} a_{ij}x_ix_j$. In this formulation, $x_i \in \{-1,1\}$ for all $i \in [n]$. (When the diagonal entries are allowed to be negative, the ratio between the semidefinite relaxation and the integral optimum can become arbitrarily large, so we restrict the domain to matrices with nonnegative diagonal entries.)

IQPs appear frequently in machine learning applications, such as MAP inference \cite{huang2014scalable,zhong2014signal,frostig2014simple} and image segmentation and correspondence problems in computer vision \cite{cour2006balanced,brendel2010segmentation}.
Max-cut is an important IQP problem, and its applications in machine learning include community detection \cite{community}, variational methods for graphical models \cite{variational}, and
graph-based semi-supervised learning \cite{wang2013semi}. The seminal Goemans-Williamson max-cut algorithm is now a textbook example of semidefinite programming \cite{goemans1995improved,williamson2011design,vazirani2013approximation}. Max-cut also arises in many other scientific domains, such as circuit design \cite{yoshimura2015uncertain} and computational biology \cite{snir2006using}.

The best approximation algorithms for IQPs relax the problem to an SDP, where the input is the same matrix $A$, but the output is a set of unit vectors maximizing $\sum_{i,j}a_{ij}\langle \vec{u}_i, \vec{u}_j\rangle$. The final step is to transform, or ``round,'' the set of vectors into an assignment of the binary variables in $X$. This assignment corresponds to a feasible solution to the original IQP. There are infinitely many rounding techniques to choose from, many of which are randomized. These algorithms make up the class of \emph{Random Projection, Randomized Rounding} algorithms (RPR$^2$), a general framework introduced by \cite{feige2006rpr}.
RPR$^2$ algorithms are known to perform well in theory and practice. When the integer quadratic program is a formulation of the max-cut problem, the class of RPR$^2$ algorithms contain the groundbreaking Goemans-Williamson algorithm, which achieves a 0.878 approximation ratio \cite{goemans1995improved}. Assuming the unique games conjecture and $P\not=NP$, this approximation is optimal to within any additive constant \cite{khot2007optimal}. More generally, if $A$ is any real-valued $n \times n$ matrix with nonnegative diagonal entries, then there exists an RPR$^2$ algorithm that achieves an approximation ratio of $\Omega(1/\log n)$ \cite{charikar2004maximizing}, and in the worst case, this ratio is tight \cite{alon2006quadratic}. Finally, if $A$ is positive semi-definite, then there exists an RPR$^2$ algorithm that achieves a $2/\pi$ approximation ratio \cite{ben2001lectures}.

We analyze several classes of RPR$^2$ rounding function classes, including $s$-linear \cite{feige2006rpr}, outward rotation \cite{zwick1999outward}, and $\tilde{\epsilon}$-discretized rounding functions \cite{o2008optimal}. 
For each class, we derive bounds on the number of samples needed to learn an approximately optimal rounding function with respect to an underlying distribution over problem instances using pseudo-dimension.
We also provide a computationally efficient and sample efficient learning algorithm for learning an approximately optimal $s$-linear or outward rotation rounding function in expectation.
We note that our results also apply to any class of RPR$^2$ algorithms where the first step is to find some set of vectors on the unit sphere, not necessarily the SDP embedding,  and then round those vectors to a binary solution. This generalization has led to faster approximation algorithms with strong empirical performance \cite{johansson2015weighted}.

\medskip

\noindent \textbf{Clustering by agglomerative algorithms with dynamic programming.} 
Given a set of $n$ datapoints and the pairwise distances between them, 
at a high level, the goal of clustering is to partition the points into groups such that distances within
each group are minimized and distances between each group are maximized.
A classic way to accomplish this task is to use an objective function.
Common clustering objective functions include $k$-means, $k$-median, and $k$-center, which we define
later on.
We focus on a very general problem where the learner's main goal is to minimize an abstract cost function such as the 
cluster purity or the clustering objective function, 
which is the case in many clustering applications such as clustering biological data \cite{filippova2012coral,meilua2007comparing}.
We study infinite classes of two-step clustering algorithms consisting of a linkage-based step and a dynamic programming step.
First, the algorithm runs one of an infinite number of linkage-based routines to construct a hierarchical tree of clusters.
Next, the algorithm runs a dynamic programming procedure to find the pruning of this tree that minimizes one of an infinite number
of clustering objectives. For example, if the clustering objective is the $k$-means objective, then the dynamic programming step will return the optimal $k$-means pruning of the cluster tree.

For the linkage-based procedure, we consider several parameterized agglomerative procedures which 
induce a spectrum of algorithms interpolating between the popular single-, average-, and complete-linkage procedures,
which are prevalent in practice \cite{awasthi2014local,saeed2003software,white2010alignment}
and known to perform nearly optimally in many settings \cite{awasthi2012center, symmetric, balcan2012clustering, grosswendt}.
For the dynamic programming step, we study an infinite class of objectives which include the standard 
$k$-means, $k$-median, and $k$-center objectives, common in applications such as information retrieval
\cite{can1993incremental,moses}.
We show how to learn the best agglomerative algorithm and pruning objective function pair, thus extending our work to multiparameter algorithms. We provide tight pseudo-dimension bounds, ranging from $\Theta(\log n)$ for simpler algorithm classes to $\Theta(n)$ for more complex algorithm classes, so our learning algorithms are sample efficient.

\medskip

\noindent \textbf{Key challenges.}
One of the key challenges in analyzing the pseudo-dimension of the algorithm classes we study is that we must develop deep insights into how changes to an algorithm's parameters affect the solution the algorithm returns on an arbitrary input. 
For example, in our clustering analysis, the cost function could be the $k$-means or $k$-median objective function, or even the distance to some ground-truth clustering. 
As we range over algorithm parameters, we alter the merge step by tuning an intricate measurement of the overall similarity of two point sets and we alter the pruning step by adjusting the way in which the combinatorially complex cluster tree is pruned. The cost of the returned clustering
may vary unpredictably. Similarly, in integer quadratic programming, if a variable flips from positive to negative, a large number of the summands in the IQP objective will also flip signs. Nevertheless, we show that in both scenarios, we can take advantage of the structure of the problems to develop our learning algorithms and bound the pseudo-dimension.

In this way, our algorithm analyses require more care than standard complexity derivations commonly found in machine learning contexts. Typically, for well-understood function classes used in machine learning, such as linear separators or other smooth curves in Euclidean spaces, there is a simple
mapping from the parameters of a specific hypothesis to its prediction on a given
example and a close connection between the distance in the parameter space
between two parameter vectors and the distance in function space between
their associated hypotheses. Roughly speaking, it is necessary to understand this connection in order to determine
how many significantly different hypotheses there are over the full range of parameters. Due to the inherent complexity of the classes we consider, connecting the parameter space to the space of approximation algorithms and their associated costs requires a
much more delicate analysis. Indeed, the key technical part of our work
involves understanding this connection from a learning-theoretic perspective. In fact, the structure we discover in our pseudo-dimension analyses allows us to develop many computationally efficient meta-algorithms for algorithm configuration due to the related concept of \emph{shattering}. A constrained pseudo-dimension of $O(\log n)$ often implies a small search space of $2^{O(\log n)} = O(n)$ in which the meta-algorithm will uncover a nearly optimal configuration.

We bolster the theory of algorithm configuration by studying algorithms for problems that are ubiquitous in machine learning and optimization: integer quadratic programming and clustering. In this paper, we develop techniques for analyzing randomized algorithms, whereas the algorithms analyzed in the previous work were deterministic. We also provide the first pseudo-dimension lower bounds in this line of work, which require an involved analysis of each algorithm family's performance on carefully constructed instances. Our lower bounds are somewhat counterintuitive, since for several of the classes we study, they are of the order $\Omega(\log n)$, even if the corresponding classes of algorithms are defined by a single real-valued parameter.
\subsection{Preliminaries and definitions}\label{sec:learning_theory}
In this section, we provide the definition of pseudo-dimension in the context of algorithm classes. Consider a class of algorithms $\mathcal{A}$ and a class of problem instances $\mathcal{X}$. Let the cost function $\cost(h,x)$ denote the abstract cost of running an algorithm $h \in \mathcal{A}$ on a problem instance $x \in \mathcal{X}$. Similarly, define the function class $\mathcal{H}_{\mathcal{A}} = \{\cost(h,\cdot): \mathcal{X} \to [0, H] \ | \ h \in \mathcal{A}\}$. Recall that a finite subset of problem instances  $S =\{ x_1, x_2, \hdots x_m \}$ is {shattered} by the function class $\mathcal{H}$,  if there exist real-valued {witnesses} $r_1,\dots,r_m$ such that for all subsets $T\subseteq S$, there exists a function $\cost\left (h_T, \cdot \right) \in \mathcal{H}$, or in other words, an algorithm $h_T \in\mathcal{A}$ such that $\cost\left(h_T, x_i\right) \leq r_i$ if and only if $i\in T$.
Then, we can define the pseudo-dimension of the algorithm class $\mathcal{A}$ to be the {pseudo-dimension} $Pdim(\mathcal{H})$ of $\mathcal{H}$ i.e., the cardinality of the largest subset of $\mathcal{X}$ shattered by $\mathcal{H}$. 

By bounding $Pdim(\mathcal{H})$, clearly we can derive sample complexity guarantees in the context of algorithm classes \cite{dudley}: for every distribution $\mathcal{D}$ over $\mathcal{X}$, every $\epsilon>0$, and
every $\delta\in(0,1]$, $m\geq c \left(\frac{H}{\epsilon} \right)^2 \left(Pdim(\mathcal{H}) +
\log \frac{1}{\delta}\right)$
for a suitable constant $c$ (independent of all other parameters),
then with probability at least $1-\delta$ over $m$ samples
$x_1,\dots,x_m \sim \mathcal{D}$, \[\left|\frac{1}{m}\sum_{i=1}^m \cost(h, x_i)-
\mathbb{E}_{x\sim\mathcal{D}}[\cost(h, x)]\right|<\epsilon\]
for every algorithm $h \in \mathcal{A}$. Therefore, if a learning algorithm receives as input a sufficiently large set of samples and returns the algorithm which performs best on that sample, we can be guaranteed that this algorithm is close to optimal with respect to the underlying distribution.
\section{SDP-based methods for integer quadratic programming}\label{sec:maxcut}
In this section, we study several IQP approximation algorithms. These classes consist of SDP rounding algorithms and are a generalization of the seminal Goemans-Williamson (GW) max-cut algorithm \cite{goemans1995improved}. We prove that it is possible to learn the optimal algorithm from a fixed class over any given application domain, and for many of the classes we study, this learning procedure is computationally efficient and sample efficient.

We focus on integer quadratic programs of the form $\sum_{i,j \in [n]}a_{ij}x_ix_j$, where the goal is to find an assignment of the binary variables $X = \{x_1, \dots, x_n\}$ maximizing this sum for a given matrix $A = (a_{ij})_{i,j \in [n]}$. Specifically, each variable in $X$ is set to either $-1$ or $1$. This problem is also known as MaxQP \cite{charikar2004maximizing}. Most algorithms with the best approximation guarantees use an SDP relaxation. The SDP relaxation has the form
\begin{equation}\label{eq:SDP}\text{maximize } \sum_{i,j \in [n]} a_{ij}\langle \vec{u}_i , \vec{u}_j\rangle \qquad \text{subject to }\vec{u}_i \in S^{n-1}.\end{equation} 
Given the set of vectors $\{\vec{u}_1, \dots, \vec{u}_n\}$, we must decide how they represent an assignment of the binary variables in $X$. In the GW algorithm, the vectors are projected onto a random vector $\vec{Z}$ drawn from the $n$-dimensional Gaussian distribution $\vec{Z}$. If the directed distance of the resulting projection is greater than 0, then the corresponding binary variable is set to 1, and otherwise it is set to $-1$.

In some cases, the GW algorithm can be improved upon by probabilistically assigning each binary variable to 1 or $-1$. In the final rounding step, any rounding function $r: \R \to [-1,1]$ can be used to specify that a variable $x_i$ is set to 1 with probability $\frac{1}{2} + \frac{1}{2}\cdot r\left(\langle \vec{Z}, \vec{u}_i \rangle\right)$ and $-1$ with probability $\frac{1}{2} - \frac{1}{2}\cdot r\left(\langle \vec{Z}, \vec{u}_i \rangle\right)$. See Algorithm~\ref{alg:GW} for the pseudocode.
\begin{algorithm}[t]
\caption{SDP rounding algorithm with rounding function $r$}\label{alg:GW}
\begin{algorithmic}[1]
\Require Matrix $A \in \R^{n \times n}$.
\State Draw a random vector $\vec{Z}$ from $\mathcal{Z}$, the $n$-dimensional Gaussian distribution.\label{step:draw}
\State Solve the SDP (\ref{eq:SDP}) for the optimal embedding $U = \left\{\vec{u}_1, \dots, \vec{u}_n\right\}$.
\State Compute set of fractional assignments $r(\langle \vec{Z}, \vec{u}_1\rangle), \dots, r(\langle \vec{Z}, \vec{u}_n\rangle)$.\label{step:fractional}
\State For all $i \in [n]$, set $x_i$ to 1 with probability $\frac{1}{2} + \frac{1}{2}\cdot r\left(\langle \vec{Z}, \vec{u}_i \rangle\right)$ and $-1$ with probability $\frac{1}{2} - \frac{1}{2}\cdot r\left(\langle \vec{Z}, \vec{u}_i \rangle\right)$.\label{step:round}
\Ensure $x_1, \dots, x_n$.
\end{algorithmic}
\end{algorithm}
Algorithm~\ref{alg:GW} is known as a \emph{Random Projection, Randomized Rounding} (RPR$^2$) algorithm, so named by the seminal work of Feige and Langberg~\cite{feige2006rpr}.

We focus on the class of $s$-linear rounding functions in this section. For the max-cut problem, Feige and Langberg~\cite{feige2006rpr} prove that when the maximium cut in the graph is not very large, a worst-case approximation ratio above the GW ratio is possible using an $s$-linear rounding function. An $s$-linear rounding function $\phi_s:\R \to [-1,1]$ is parameterized by a real-value $s > 0$. The function $\phi_s$ is defined as follows: \\
\begin{minipage}{0.4\textwidth}
\[\phi_s(y) = \begin{cases} -1 &\text{if } y < -s\\
y/s &\text{if } -s \leq y \leq s\\
1 &\text{if } y > s.\end{cases}\]
\end{minipage}
\begin{minipage}{0.5\textwidth}
\begin{center}
  \includegraphics[width=0.7\textwidth] {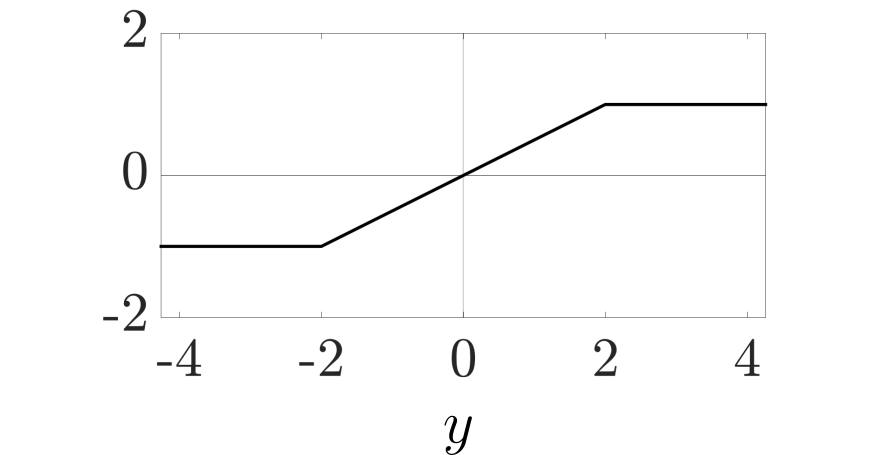}
  \captionof{figure}{A graph of the 2-linear function $\phi_2$.}\label{fig:slin2}  
  \end{center}
\end{minipage} \hfill
 \vspace{\belowdisplayskip}

Our goal is to design an algorithm $L_{slin}$ that learns a nearly-optimal $s$-linear rounding function.
 In other words, we want to find a parameter $s$ such that
the expected objective value $\sum_{i, j \in [n]} a_{ij} x_i x_j$ is maximized, where the expectation is over three sources of randomness: the matrix $A$, the vector $\vec{Z}$, and the final assignment of the variables $x_1, \dots, x_n$, which depends on $A$, $\vec{Z}$, and the choice of a parameter $s$. This expected value is thus over distributions that are both external and internal to Algorithm~\ref{alg:GW}: the unknown distribution over matrices is external and defines the algorithm's input, whereas the distribution over vectors and the distribution defining the final assignment of the variables $x_1, \dots, x_n$ are internal to Algorithm~\ref{alg:GW}.
We call this expected value the \emph{true quality\footnote{In this section, we refer to a parameter's ``quality'' rather than ``cost'' because we want to find a parameter that maximizes this value.} of the parameter $s$}.

Since the distribution $\dist$ over matrices is unknown, we cannot evaluate the true quality of any parameter, so we use samples to find a nearly optimal parameter. We draw samples from the first two sources of randomness:  the distribution over matrices and the distribution over vectors. Thus, our set of samples has the form $\sample = \left\{\left(A^{(1)}, \vec{Z}^{(1)}\right), \dots, \left(A^{(m)}, \vec{Z}^{(m)}\right)\right\} \sim \left(\dist \times \mathcal{Z}\right)^m.$ In this way, to ease our analysis, we sample the distribution over Gaussians --- an internal source of randomness --- rather than analyzing its expected value directly. Given these samples, we define the \emph{empirical quality} of a parameter $s$ to be the expected value of the solution returned by Algorithm~\ref{alg:GW} given $\qp$ as input when it uses the hyperplane $\vec{Z}$ and the $s$-linear rounding function $\phi_{s}$ in Step~\ref{step:fractional}, averaged over all $(A, \vec{Z}) \in \sample$. At a high level, upon sampling from the first two sources of randomness, we have isolated the third source of randomness, whose expectation is simple to analyze. In the following analysis, we show that every parameter's empirical quality converges to its true quality as the sample size increases, and thus the parameter with the highest empirical quality has a nearly optimal true quality.

Since the distribution over vectors is known to be Gaussian, an alternative route would be to only sample the external source of randomness $\dist$ over the matrices. We would then define the empirical quality of a parameter $s$ to be the expected value of the solution returned by Algorithm~\ref{alg:GW} given $\qp$ as input when it uses the $s$-linear rounding function $\phi_{s}$ in Step~\ref{step:fractional}, averaged over all $A \in \sample$. This would require us to incorporate the density function of a multi-dimensional Gaussian in our analysis. We abstract out this complication by sampling the Gaussian vectors and including them as a part of the learning algorithm's training set, thus simplifying the analysis significantly.

We now define the true and empirical quality of a parameter more formally. Let $p_{(i, \vec{Z}, A, s)}$ be the distribution from which the value of $x_i$ is drawn when Algorithm~\ref{alg:GW}, given $\qp$ as input, uses the hyperplane $\vec{Z}$ and the rounding function $r = \phi_s$ in Step~\ref{step:fractional}. The true quality of the parameter $s$ is $\E_{A, \vec{Z} \sim \dist \times \mathcal{Z}}\left[\E_{x_i \sim p_{(i, \vec{Z}, A, s)}}\left[\sum_{i,j} a_{ij} x_i x_j\right]\right].$\footnote{We use the abbreviated notation \[\E_{A, \vec{Z} \sim \dist \times \mathcal{Z}}\left[\E_{x_i \sim p_{(i, \vec{Z}, A, s)}}\left[\sum_{i,j} a_{ij} x_i x_j\right]\right] = \E_{A, \vec{Z} \sim \dist \times \mathcal{Z}}\left[\E_{x_1 \sim p_{(1, \vec{Z}, A, s)}, \dots, x_n \sim p_{(n, \vec{Z}, A, s)}}\left[\sum_{i,j} a_{ij} x_i x_j\right]\right].\]} Our goal is to find a parameter whose true quality is (nearly) optimal. Said another way, we want to find the value of $s$ leading to the highest expected objective value over all sources of randomness.

We do not know the distribution $\dist$ over matrices, so we also need to define the \emph{empirical quality} of the parameter $s$ given a set of samples. We will then show that this empirical quality approaches the true quality as the number of samples grows. Thus, a parameter which is nearly optimal on average over the samples will be nearly optimal in expectation as well. The definition of a parameter's empirical quality depends on a function $\slin_s$ which we now define. Let $\slin_s(A, \vec{Z})$ denote the expected value of the solution returned by Algorithm~\ref{alg:GW} given $\qp$ as input when it uses the hyperplane $\vec{Z}$ and the rounding function $r = \phi_s$ in Step~\ref{step:fractional}. The expectation is over the randomness in the assignment of each variable $x_i$ to either 1 or -1. Explicitly, $\slin_s(A, \vec{Z}) = \E_{x_i \sim p_{(i, \vec{Z}, A, s)}}\left[\sum_{i,j} a_{ij} x_i x_j\right]$. By definition, the true quality of the parameter $s$ equals $\E_{A, \vec{Z} \sim \dist \times \mathcal{Z}}\left[\slin_s(A, \vec{Z})\right]$.

We now define the empirical quality of a parameter $s$ as follows. Given a set of samples $\left(A^{(1)}, \vec{Z}^{(1)}\right), \dots, \left(A^{(m)}, \vec{Z}^{(m)}\right) \sim \dist \times \mathcal{Z}$, we define the empirical quality of the parameter $s$ to be $\frac{1}{m} \sum_{i = 1}^m \slin_s\left(A^{(i)}, \vec{Z}^{(i)}\right)$. Bounding the pseudo-dimension\footnote{Since pseudo-dimension bounds imply uniform convergence guarantees for worst-case distributions, the distribution $\mathcal{Z}$ over vectors need not be Gaussian, although this is the classic distribution of choice in the works by Goemans and Williamson~\cite{goemans1995improved} and Feige and Langberg~\cite{feige2006rpr}. Indeed, our results hold when $\mathcal{Z}$ is any arbitrary distribution over $\R^n$.} of the class of functions $\mathcal{H}_{slin} = \left\{\slin_s : s > 0\right\}$, we bound the number of samples sufficient to ensure that with high probability, for all parameters $s$, the true quality of $s$ nearly matches its expected quality. In other words, $\frac{1}{m}\sum_{i = 1}^m \slin_s\left(A^{(i)}, \vec{Z}^{(i)}\right)$ nearly matches $\E_{A, \vec{Z} \sim \dist \times \mathcal{Z}}\left[\slin_s(A, \vec{Z})\right]$. Thus, if we find the parameter $\hat{s}$ that maximizes $\frac{1}{m}\sum_{i = 1}^m \slin_s\left(A^{(i)}, \vec{Z}^{(i)}\right)$, then the true quality of $\hat{s}$ is nearly optimal, i.e., $\max_{s > 0}\E_{A, \vec{Z} \sim \dist \times \mathcal{Z}} \left[\slin_{s}(A, \vec{Z})\right]$ is close to $\E_{A, \vec{Z} \sim \dist \times \mathcal{Z}} \left[\slin_{\hat{s}}(A, \vec{Z})\right]$. In Theorem~\ref{thm:slin_ERM}, we provide a sample efficient and computationally efficient algorithm for finding $\hat{s}$.

We begin by characterizing the analytic form of $\slin_s$, which allows us to bound the pseudo-dimension of $\mathcal{H}_{slin}$.

\begin{lemma}\label{lem:exp}
Given a matrix $\qp$ and a vector $\vec{Z}$, let $\slin_s(A, \vec{Z})$ denote the expected value of the solution returned by Algorithm~\ref{alg:GW} given $\qp$ as input when it uses the hyperplane $\vec{Z}$ and the rounding function $r = \phi_s$ in Step~\ref{step:fractional}. The expectation is over the randomness in the assignment of each variable $x_i$ to either 1 or -1. Then \[\slin_s(\qp, \vec{Z}) = \sum_{i = 1}^n a_{ii}^2 + \sum_{i \not = j} a_{ij} \phi_s(\langle\vec{Z}, \vec{u}_i \rangle) \cdot \phi_s(\langle \vec{Z}, \vec{u}_j \rangle).\]
\end{lemma}
\begin{proof}
Let $p_{(i, \vec{Z}, A, s)}$ be the distribution from which the value of $x_i$ is drawn when Algorithm~\ref{alg:GW}, given $\qp$ as input, uses the hyperplane $\vec{Z}$ and the rounding function $r = \phi_s$ in Step~\ref{step:fractional}. We know that \[\E_{x_i \sim p_{(i, \vec{Z}, A, s)}}[x_i] = \frac{1}{2} + \frac{1}{2}\cdot \phi_s\left(\langle \vec{Z}, \vec{u}_i \rangle\right) - \left(\frac{1}{2} - \frac{1}{2}\cdot \phi_s\left(\langle \vec{Z}, \vec{u}_i \rangle\right)\right) = \phi_s\left(\langle \vec{Z}, \vec{u}_i \rangle\right).\]

The expected value of the solution returned by Algorithm~\ref{alg:GW} given $\qp$ as input when it uses the hyperplane $\vec{Z}$ and the rounding function $r = \phi_s$ in Step~\ref{step:fractional} is \begin{align*}\E_{x_i \sim p_{(i, \vec{Z}, A, s)}}\left[ \sum_{i,j \in [n]} a_{ij}x_ix_j\right]&=\sum_{i,j \in [n]}\E_{x_i,x_j}\left[ a_{ij}x_ix_j\right]\\
&= \sum_{i = 1}^n a_{ii}^2 \E_{x_i}[x_i^2] + \sum_{i \not= j}a_{ij}\E_{x_i,x_j}\left[x_ix_j\right].\end{align*} Since the support of $p_{(i, \vec{Z}, A, s)}$ is $\{-1, 1\}$, we know that \[\sum_{i = 1}^n a_{ii}^2 \E_{x_i}[x_i^2] + \sum_{i \not= j}a_{ij}\E_{x_i,x_j}\left[x_ix_j\right] =
\sum_{i = 1}^n a_{ii}^2 + \sum_{i \not= j}a_{ij}\E_{x_i,x_j}\left[x_ix_j\right].\] The draw $x_i \sim p_{(i, \vec{Z}, A, s)}$ is independent from the draw $x_j \sim p_{(j, \vec{Z}, A, s)}$, so \[\sum_{i = 1}^n a_{ii}^2 + \sum_{i \not= j}a_{ij}\E_{x_i,x_j}\left[x_ix_j\right] =  \sum_{i = 1}^n a_{ii}^2 + \sum_{i \not= j}a_{ij}\E_{x_i\sim p_{(i, \vec{Z}, A, s)}}\left[x_i\right] \E_{x_j\sim p_{(j, \vec{Z}, A, s)}} \left[x_j\right].\] Since $\E_{x_i \sim p_{(i, \vec{Z}, A, s)}}[x_i] = \phi_s\left(\langle \vec{Z}, \vec{u}_i \rangle\right)$, this means that 
\[\slin_s(A, \vec{Z})=\sum_{i = 1}^n a_{ii}^2 + \sum_{i \not= j}a_{ij}\E_{x_i}\left[x_i\right] \E_{x_j} \left[x_j\right] = \sum_{i = 1}^n a_{ii}^2 + \sum_{i \not= j}a_{ij}\phi_s(\langle \vec{Z}, \vec{u}_i\rangle) \cdot \phi_s(\langle \vec{Z}, \vec{u}_j\rangle).\] Putting all of these equalities together, the lemma statement holds.
\end{proof}

Lemma~\ref{lem:exp} allows us to prove that the functions in $\mathcal{H}_{slin}$ have a particularly simple form, which facilitates our pseudo-dimension analysis. Roughly speaking, for a fixed matrix $A$ and vector $\vec{Z}$, each function in $\mathcal{H}_{slin}$ is a piecewise, inverse-quadratic function of the parameter $s$. To present this lemma, we use the following notation: given a tuple $\left(\qp, \vec{Z}\right)$, let $\slin_{\qp, \vec{Z}} : \R \to \R$ be defined such that $\slin_{\qp, \vec{Z}}(s) = \slin_s\left(\qp, \vec{Z}\right)$.

\begin{lemma}\label{lem:piecewise_quad}
For any matrix $A$ and vector $\vec{Z}$, the function $\slin_{\qp, \vec{Z}}:\R_{>0} \to \R$ is made up of $n+1$ piecewise components of the form $\frac{a}{s^2} + \frac{b}{s} + c$ for some $a,b,c \in \R$. Moreover, if the border between two components falls at some $s \in \R_{>0}$, then it must be that $s = \left|\langle \vec{u}_i, \vec{Z}\rangle \right|$ for some $\vec{u}_i$ in the optimal 
SDP embedding of $\qp$. 
\end{lemma}
\begin{proof}
Let $X = \left\{\vec{u}_1, \dots, \vec{u}_n\right\} \subset S^{n-1}$ be the optimal embedding 
of $\qp$.
We may write $\slin_{\qp, \vec{Z}}(s) = \sum_{i= 1}^n a_{ii}^2 + \sum_{i\not=j}a_{ij}\phi_s\left(v_i\right) \cdot \phi_s\left( v_j \right),$ where $v_i = \langle \vec{u}_i , \vec{Z}\rangle$ and $v_j = \langle \vec{u}_j , \vec{Z}\rangle$.
For any $i \in [n]$, the specific form of $\phi_s\left(v_i\right)$ depends solely on whether $|v_i| \leq s$ or $|v_i| > s$ (recall that $s > 0$, by definition). So long as $\left|v_i\right| > s$, we know $\phi_s\left(v_i\right) = \pm 1$, where the sign depends on the sign of $v_i$. Otherwise, when $|v_i| < s$, $\phi_s(v_i) = v_i / s$. Therefore, if we order the set of real values $\left\{\left|v_i \right|, \dots, \left|v_n\right|\right\}$, then so long as $s$ falls between two consecutive elements of this ordering, the form of $\slin_{\qp, \vec{Z}}(s)$ is fixed. In particular, each summand is either a constant, a constant multiplied by $\frac{1}{s}$, or a constant multiplied by $\frac{1}{s^2}$. This means that we may partition the positive real line into $n+1$ intervals where the form of $\slin_{\qp, \vec{Z}}(s)$ is a fixed quadratic function, as claimed. \end{proof}

Lemma~\ref{lem:piecewise_quad} allows us to prove the following bound on Pdim$(\mathcal{H}_{slin})$.

\begin{lemma}\label{lem:slin_theta}
Pdim$(\mathcal{H}_{slin}) = \Theta(\log n)$.
\end{lemma}
Lemma~\ref{lem:slin_theta} follows from Lemmas~\ref{lem:slin_upper} and \ref{lem:slin_lower}, where we prove Pdim$(\mathcal{H}_{slin}) = O(\log n)$ and Pdim$(\mathcal{H}_{slin}) = \Omega(\log n)$.

\begin{lemma}\label{lem:slin_upper}
Pdim$(\mathcal{H}_{slin}) = O(\log n)$.
\end{lemma}
\begin{proof}
We prove this upper bound by showing that if a set $\sample$ of size $m$ is shatterable, then $m = O(\log n)$. This means that the largest shatterable set must be of size $O(\log n)$, so the pseudo-dimension of $\mathcal{H}_{slin}$ is $O(\log n)$. We arrive at this bound by fixing a tuple $\left(\qp^{(i)}, \vec{Z}^{(i)}\right) \in \sample$ and analyzing $\slin_{\qp, \vec{Z}}(s)$. In particular, we make use of Lemma~\ref{lem:piecewise_quad}, from which we know that $\slin_{\qp, \vec{Z}}(s)$ is composed of $n+1$ piecewise quadratic components. Therefore, if $r_i$ is the witness corresponding to the element $\left(\qp^{(i)}, \vec{Z}^{(i)}\right)$, we can partition the positive real line into at most $3(n+1)$ intervals where $\slin_{\qp, \vec{Z}}(s)$ is always either less than its witness $r_i$ or greater than $r_i$ as $s$ varies over one fixed interval. The constant 3 term comes from the fact that for a single, continuous quadratic component of $\slin_{\qp, \vec{Z}}(s)$, the function may equal $r_i$ at most twice, so there are at most three subintervals where the function is less than or greater than $r_i$.

\begin{figure}
  \centering
  \includegraphics[width=.75\textwidth] {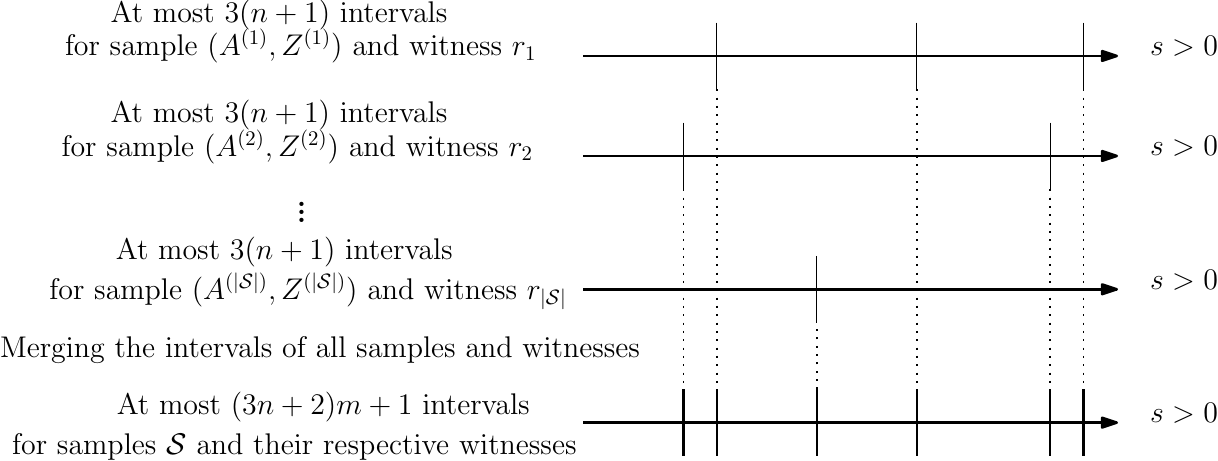}
  \caption{Partitioning $s > 0$ into intervals given a set $\mathcal{S}$ of $m$ tuples $\left(A^{(i)}, \vec{Z}^{(i)}\right)$ and witnesses $r_i$ such that within each interval for each $i$, $\slin_{A^{(i)},Z^{(i)}}(s)$ is always greater than $r_i$ or lesser than $r_i$.}\label{fig:s-intervals}
\end{figure}
Now, $\sample$ consists of $m$ tuples $\left(\qp^{(i)}, \vec{Z}^{(i)}\right)$, each of which corresponds to its own partition of the positive real line. If we merge these partitions (as shown in Figure~\ref{fig:s-intervals}), simple algebra shows that we are left with at most $(3n+2)m+1$ intervals such that for all $i \in [m]$, $\slin_{\qp^{(i)}, \vec{Z}^{(i)}}(s)$ is always either less than its witness $r_i$ or greater than $r_i$ as $s$ varies over one fixed interval. In other words, in one interval, the binary labeling of $\sample$, defined by whether each sample is less than or greater than its witness, is fixed. This means that if $\sample$ is shatterable, the $2^m$ values of $s$ which induce all $2^m$ binary labelings of $\sample$ must come from distinct intervals.
Therefore $2^m \leq (3n+2)m+1$, so $m = O(\log n)$.\end{proof}

\begin{restatable}{relem}{slinLower}\label{lem:slin_lower}
Pdim$(\mathcal{H}_{slin}) = \Omega(\log n)$.
\end{restatable}
\begin{proof}[Proof sketch]
In order to prove that the pseudo dimension of $\mathcal{H}_{slin}$ is at least $c \log n$ for some $c$, we present a set $\mathcal{S} = \left\{ (A^{(1)}, \vec{Z}^{(1)}), \dots, (\qp^{(m)}, \vec{Z}^{(m)})\right\}$ of $m = c\log n$ matrices and projection vectors that can be shattered by $\mathcal{H}_{slin}$. In other words, there exist $m$ witnesses $r_1, \dots, r_m$ and $2^m = n^c$ values $s_1, \dots, s_{n^c}$ such that for all $T \subseteq [m]$, there exists $s_T$ such that if $j \in T$, then $\slin_{s_T}(\qp^{(j)}, \vec{Z}^{(j)}) > r_j$ and if $j \not\in T$, then $\slin_{s_T}(\qp^{(j)}, \vec{Z}^{(j)}) \leq r_j$.

To build $\sample$, we use the same matrix $\qp$ for all $\qp^{(j)}$ and we vary $\vec{Z}^{(j)}$. We set $\qp$ to be a max-cut instance based on a graph composed of $\lfloor n/4 \rfloor$ disjoint copies of $K_4$. Via a careful choice of the vectors $\vec{Z}^{(j)}$ and witnesses $r_j$, we pick out $2^m$ critical values of $s$, which we call $C$, such that $\slin_{\qp, \vec{Z}^{(1)}}(s)$ switches from above to below its witness for every other element of the critical values in $C$. Meanwhile,  $\slin_{\qp, \vec{Z}^{(2)}}(s)$ switches from above to below its witness half as often as $\slin_{\qp, \vec{Z}^{(1)}}(s)$. Similarly, $\slin_{\qp, \vec{Z}^{(3)}}(s)$ switches from above to below its witness half as often as $\slin_{\qp, \vec{Z}^{(2)}}(s)$, and so on. 
Therefore, we achieve every binary labeling of $\sample$ using the functions $\left\{\slin_s \ | \ s \in C\right\}$, so $\sample$ is shattered.
\end{proof}

Our lower bound is particularly strong because it holds for a family of positive semidefinite matrices, rather than a more general family of real-valued matrices. We now prove that our learning algorithm, Algorithm~\ref{alg:ERM} is correct, computationally efficient, and sample efficient.

\begin{algorithm}
\caption{An algorithm for finding an empirical value maximizing $s$-linear rounding function}\label{alg:ERM}
\begin{algorithmic}[1]
\Require Sample $\sample = \{(\qp^{(1)}, \vec{Z}^{(1)}), \dots, (\qp^{(m)}, \vec{Z}^{(m)})\}$
\State For all $i$, solve for the SDP embedding $U^{(i)}$ of $\qp^{(i)}$, where $U^{(i)} = \left\{\vec{u}_1^{(i)}, \dots, \vec{u}_n^{(i)}\right\}$.
\State Let $T = \left\{s_1, \dots, s_{|T|}\right\}$ be the set of all values $s>0$ such that there exists a pair of indices $j \in [n], i \in [m]$ with $\left|\left\langle \vec{Z}^{(i)},\vec{u}_j^{(i)}\right\rangle\right| = s.$\label{step:threshold}
\State For $i \in [|T|-1]$, let $\hat{s}_{i}$ be the value in $[s_i, s_{i+1}]$ which maximizes $\frac{1}{m} \sum_{i = 1}^m \slin_{\qp^{(i)},\vec{Z}^{(i)}}(s)$.\label{step:consec}
\State Let $\hat{s}$ be the value in $\{\hat{s}_1, \dots, \hat{s}_{|T|-1}\}$ that maximizes $\frac{1}{m} \sum_{i = 1}^m \slin_{\qp^{(i)},\vec{Z}^{(i)}}(s)$.\label{step:global}
\Ensure $\hat{s}$
\end{algorithmic}
\end{algorithm}

\begin{theorem}\label{thm:slin_ERM}
Let $H = \sup_{A \in \textnormal{supp}(\dist)} ||A||_c$, where $||\cdot||_c$ is the cut norm and supp$(\dist)$ denotes the support of $\dist$.\footnote{$H$ is thus an upper bound on the value of $\slin_s(\qp, \vec{Z})$ for any $s > 0$ and any $(\qp, \vec{Z})$ in the support of $\mathcal{D} \times \mathcal{Z}$.} Given a sample of size $m = O\left(\left(\frac{H}{\epsilon}\right)^2 \left(\log \left({n}\right) + \log\frac{1}{\delta}\right)\right)$ drawn from $\left(\mathcal{D} \times \mathcal{Z}\right)^m$, let $\hat{s}$ be the output of Algorithm~\ref{alg:ERM}. With probability at least $1 - \delta$, the true quality of $\hat{s}$ is $\epsilon$-close optimal: \[\max_{s > 0} \E_{A \sim \dist, \vec{Z} \sim \mathcal{Z}}\left[\slin_s(A, \vec{Z})\right] - \E_{A \sim \dist, \vec{Z} \sim \mathcal{Z}}\left[\slin_{\hat{s}}(A, \vec{Z})\right] \leq \epsilon.\]
\end{theorem}

\begin{proof}
Let $\sample = \left\{\left(\qp^{(1)}, \vec{Z}^{(1)}\right), \dots, \left(\qp^{(m)}, \vec{Z}^{(m)}\right)\right\}$ be a sample of size $m$. First, we prove that Algorithm~\ref{alg:ERM} on input $\sample$ returns the value $\hat{s}$ which maximizes $\frac{1}{m} \sum_{i = 1}^m \slin_s\left(\qp^{(i)},\vec{Z}^{(i)}\right)$ in polynomial time. In Lemma~\ref{lem:piecewise_quad}, we prove that each function $\slin_{\qp^{(i)},\vec{Z}^{(i)}}(s)$ is made up of at most $n+1$ piecewise components of the form $\frac{a}{s^2} + \frac{b}{s} + c$ for some $a,b,c \in \R$. Therefore, $\frac{1}{m} \sum_{i = 1}^m \slin_s\left(\qp^{(i)},\vec{Z}^{(i)}\right)$ is made up of at most $mn+1$ piecewise components of the form $\frac{a}{s^2} + \frac{b}{s} + c$ as well. Moreover, by Lemma~\ref{lem:piecewise_quad}, if the border between two components falls at some $s \in \R_{>0}$, 
then it must be that $\left|\left\langle \vec{Z}^{(i)},\vec{u}_j^{(i)}\right\rangle\right| = s$ for some $\vec{u}_j^{(i)}$ in the optimal max-cut SDP embedding of $\qp^{(i)}$. These are the thresholds which are computed in Step~\ref{step:threshold} of Algorithm~\ref{alg:ERM}.
Therefore, as we increase $s$ starting at 0, $s$ will be a fixed inverse-quadratic function between the thresholds, so it is simple to find the optimal value of $s$ between any pair of consecutive thresholds (Step~\ref{step:consec}), and then the value maximizing $\frac{1}{m} \sum_{i = 1}^m \slin_s\left(\qp^{(i)},\vec{Z}^{(i)}\right)$ (Step~\ref{step:global}), which is the global optimum. 

Next, from Lemma~\ref{lem:slin_upper}
we have that with $m = O \left(\left(\frac{H}{\epsilon}\right)^2 \left( \log n + \log \frac{1}{\delta}\right) \right)$ samples, with probability at least $1-\delta$, for all $ s> 0$, \[\left|\frac{1}{m} \sum_{i = 1}^m \slin_s\left(A^{(i)},\vec{Z}^{(i)}\right) - \underset{\left(A,\vec{Z}\right) \sim \mathcal{D} \times \mathcal{Z}}{\E}\left[\slin_s\left(A,\vec{Z}\right)\right]\right| < \frac{\epsilon}{2}.\]
Since this is true for the parameter $\hat{s}$ returned by Algorithm~\ref{alg:ERM} and for the optimal parameter $s^* = \argmax_{s > 0} \E_{A \sim \dist, \vec{Z} \sim \mathcal{Z}}\left[\slin_s(A, \vec{Z})\right]$, we know that with probability at least $1-\delta$, \[\E_{A \sim \dist, \vec{Z} \sim \mathcal{Z}}\left[\slin_{s^*}(A, \vec{Z})\right] - \E_{A \sim \dist, \vec{Z} \sim \mathcal{Z}}\left[\slin_{\hat{s}}(A, \vec{Z})\right] \leq \epsilon.\] \end{proof}

In Appendix~\ref{app:maxcut-more}, we consider other rounding functions, including $\tilde{\epsilon}$-discretized rounding functions \cite{o2008optimal} and outward rotation algorithms \cite{zwick1999outward}.
\section{Agglomerative algorithms with dynamic programming}\label{sec:clustering}
We begin with an overview of agglomerative algorithms with dynamic programming,
which include many widely-studied clustering algorithms, 
and then we define several parameterized classes of such algorithms.
As in the previous section, we prove it is possible to learn the optimal
algorithm from a fixed class for a specific application, and for many of
the classes we analyze, this procedure is computationally efficient and sample efficient.
We focus on agglomerative algorithms with dynamic programming for
\emph{clustering} problems.
A clustering instance $\V = (V,d)$ consists of a set $V$ of $n$ points and a distance metric $d : V \times V \to \mathbb{R}_{\geq 0}$ specifying all pairwise distances between these points. The overall goal of clustering is to partition the points into groups such that distances within
each group are minimized and distances between each group are maximized.
Clustering is typically performed using an objective function $\Phi$, such as $k$-means, $k$-median, $k$-center, or the distance to the ground truth clustering (a scenario we discuss in more detail in Section~\ref{sec:pruning}).
Formally, an objective function $\Phi$ takes as input a set of points
$\textbf{c}=\{c_1,\dots,c_k\}\subseteq V$ which we call centers, as well as a partition $\mathcal{C}=\{C_1,\dots,C_k\}$ of $V$ which we call a clustering.
We define the rich class of clustering objectives $\Phi^{(p)}(\mathcal{C},\textbf{c})=\sum_{i=1}^k(\sum_{q\in C_i}d(q,c_i)^p)^{1/p}$
for $p\in [1,\infty)\cup\{\infty\}$.
The $k$-means, $k$-median, and $k$-center objective functions are $\Phi^{(2)}$, $\Phi^{(1)}$, and $\Phi^{(\infty)}$, respectively.\footnote{There have been several papers that provide theoretical guarantees for clustering under this family of objective functions for other values of p. For instance, see Gupta and Tangwongsan's work \cite{gupta2008simpler} which provides an $O(p)$ approximation algorithm when $p<\log n$ and Bateni et al.'s work \cite{bateni2014distributed} which studies distributed clustering algorithms.}

Next, we define agglomerative clustering algorithms with dynamic programming, which are prevalent in practice
\cite{awasthi2014local,saeed2003software,white2010alignment}
and enjoy strong theoretical guarantees in a variety of settings
\cite{awasthi2012center, symmetric, balcan2012clustering, grosswendt}.
Examples of these algorithms include the popular \emph{single-}, \emph{complete-},
and \emph{average-linkage} algorithms with dynamic programming.

An agglomerative clustering algorithm with dynamic programming is defined by two functions: a merge function and a pruning function. A merge function $\merge(A,B) \to \R_{\geq 0}$ defines a distance between two sets of points $A,B \subseteq V$. 
The algorithm builds a \emph{cluster tree} $\mathcal{T}$ 
by starting with $n$ singleton leaf nodes, 
and iteratively merging the two sets with minimum distance until
there is a single node remaining, consisting of the set $V$.
The children of any node $T$ in this tree correspond to the two sets of points that were merged to form $T$ during the sequence of merges.  Common choices for the merge function $\merge$ include
$\min_{a\in A,b\in B} d(a,b)$ (single linkage),  $\frac{1}{|A|\cdot |B|}\sum_{a\in A,b\in B}d(a,b)$ (average linkage) and $\max_{a\in A,b\in B} d(a,b)$ (complete linkage).

A pruning function $\prune$ takes as input a $k'$-\emph{pruning} of any subtree of $\mathcal{T}$ and returns a score $\R_{\geq 0}$ for that pruning. 
A $k'$-pruning for a subtree $T$ is a partition of the points contained in $T$'s root into $k'$ clusters such that each cluster is an internal node of $T$.    
Pruning functions may be similar to objective functions, 
though the input is a subtree. The $k$-means, -median, and -center objectives are standard pruning functions.
The algorithm returns the $k$-pruning of the tree $\mathcal{T}$ that is optimal according to $\prune$,
which can be found in polynomial time using dynamic programming.
Algorithm~\ref{alg:linkage} details how the merge function and pruning function work together to form an agglomerative clustering algorithm with dynamic programming. 
In the dynamic programming step, to find the  $1$-pruning of any node $T$, we only need to find the best center $c \in T$.  When $k' > 1$, we recursively find the best $k'$-pruning of $T$ by considering different combinations of the best $i'$-pruning of the left child $T_L$ and the best $(k'-i')$-pruning of the right child $T_R$ for $i' \in \{1, \dots, k-1\}$ and choosing the best combination. 

\begin{algorithm}
\caption{Agglomerative algorithm with dynamic programming}\label{alg:linkage}
\begin{algorithmic}[1]
\Require Clustering instance $\mathcal{V} = (V,d)$, merge function $\merge$, pruning function $\prune$.
\State  \textbf{Agglomerative merge step to build a cluster tree $\mathcal{T}$ according to $\merge$}: \label{step:merge}
\begin{itemize}
\item Start with $n$ singleton sets $\{v\}$ for each $v\in V$.
\item Iteratively merge the two sets $A$ and $B$ which minimize $\merge(A,B)$ until a single set remains.
\item Let $\mathcal{T}$ denote the cluster tree corresponding to the sequence of merges.
\end{itemize}
\State \textbf{Dynamic programming to find the $k$-pruning of $\mathcal{T}$ minimizing $\prune$ }:  \label{step:dp}
\begin{itemize}
\item For each node $T$, find the best $k'$-pruning of the subtree rooted at $T$ in $\mathcal{T}$, denoted by $\left(\mathcal{C}_{T,k'}, \mathbf{c}_{T,k'}\right)$ according to
following dynamic programming recursion:
\[
\prune\left(\mathcal{C}_{T,k'}, \mathbf{c}_{T,k'}\right) = 
\begin{cases}
\min_{c\in T} \prune\left(\{T\},c\right) & \text{if }k' = 1, \\
\min_{i'\in[k'-1]}\prune\left(\mathcal{C}_{T_L,i'} \cup \mathcal{C}_{T_R,k'-i'}, \mathbf{c}_{T_L,i'} \cup \mathbf{c}_{T_R,k'-i'}\right) & 
\text{otherwise.}\\
\end{cases}
\]
where $T_L$ and $T_R$ denote the left and right children of $T$, respectively.
\end{itemize}
\Ensure The best $k$-pruning of the root node $T_{\text{root}}$ of $\mathcal{T}$. 
\end{algorithmic}
\end{algorithm}

\begin{figure}
  \centering
  \includegraphics[width=.62\textwidth] {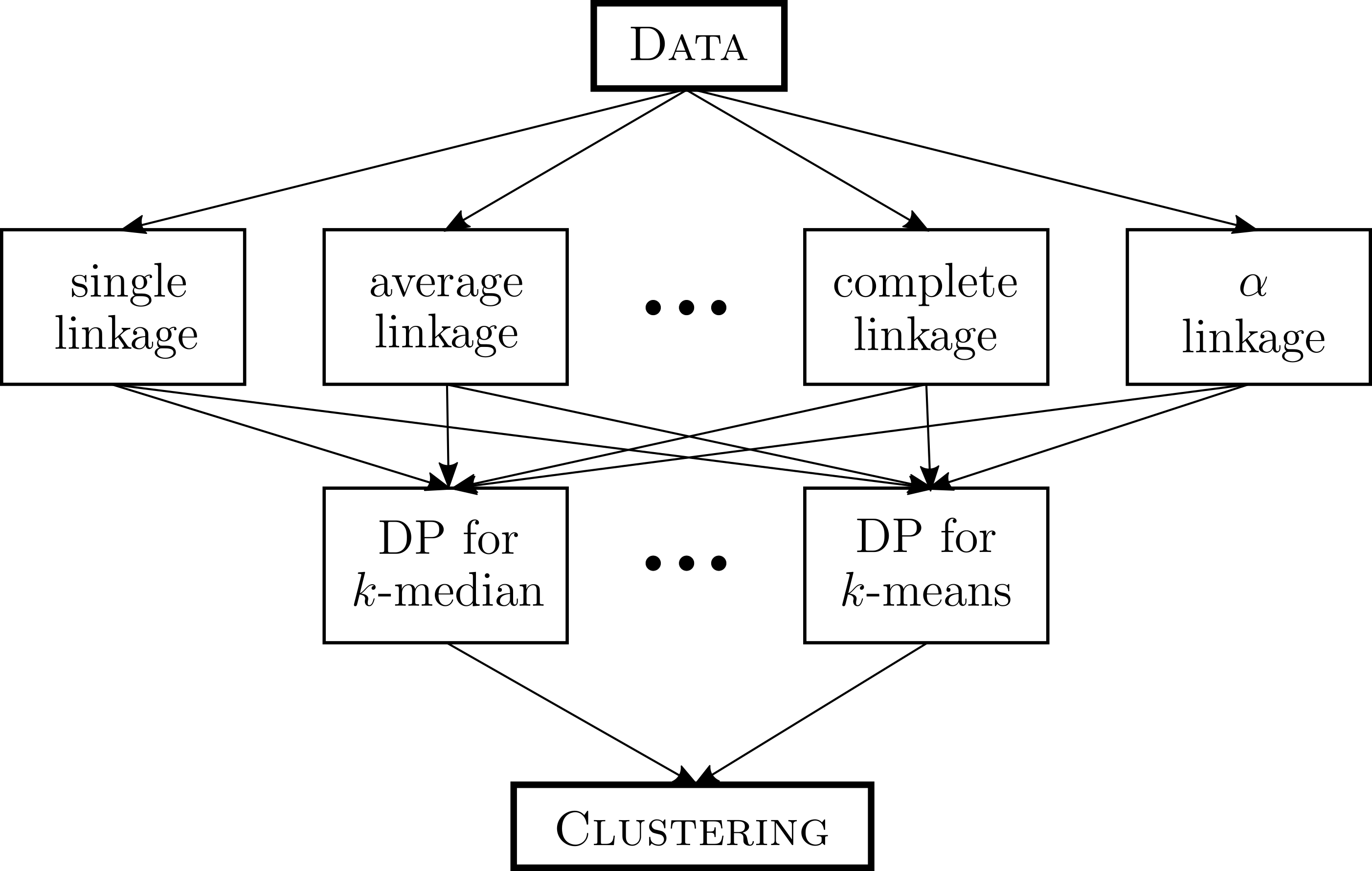}
  \caption{A schematic for a class of agglomerative clustering algorithms with dynamic programming.}\label{fig:cluster_flow}
\end{figure}

Pictorially, Figure~\ref{fig:cluster_flow} depicts an array of available choices when designing an agglomerative clustering algorithm with dynamic programming. Each path in the chart corresponds to an alternative choice of a merging function $\merge$ and pruning function $\prune$. The algorithm designer's goal is to determine the path that is optimal for her specific application domain. 

In Section~\ref{sec:simple}, we analyze several classes of algorithms where the merge function comes from an infinite family of functions while the pruning function is an arbitrary, fixed function. 
In Section~\ref{sec:pruning}, we expand our analysis to include algorithms defined over an infinite family of pruning functions in conjunction with any family of merge functions. 
Our results hold even when there is a fixed preprocessing step that precedes the
agglomerative merge step (as long as it is independent of $\merge$ and $\prune$),
therefore our analysis carries over to algorithms such as in \cite{balcan2012clustering}.

\subsection{Linkage-based merge functions}\label{sec:simple}

We now define three infinite families of merge functions
and provide sample complexity bounds for these families with
any fixed but arbitrary pruning function.
The families $\mathcal{A}_1$ and $\mathcal{A}_3$
consist of merge functions $\merge(A,B)$ that depend on
the minimum and maximum of all pairwise distances between $A$ and $B$.
The second family, denoted by $\mathcal{A}_2$, depends on all pairwise distances between $A$ and $B$.
All classes are parameterized by a single value $\alpha$.
\begin{align*}
\mathcal{A}_1&=\left\lbrace\left. \left(
\min_{u \in A, v \in B}(d(u,v))^{\alpha} + \max_{u \in A, v \in B}(d(u,v))^
\alpha \right)^{1/\alpha}\, \right| \, \alpha\in\mathbb{R}\cup\{\infty, -\infty\}\right\rbrace,\\
\mathcal{A}_2&=\left\lbrace\left(
\left. \frac{1}{|A||B|}\sum_{u \in A, v \in B} \left(d(u, v)\right)^{\alpha}\right)^{1/\alpha} \, \right| \, \alpha \in \mathbb{R} \cup \{\infty, -\infty \}\right\rbrace,\\
\mathcal{A}_3&=\left\lbrace \left. \alpha\min_{u\in A,v\in B}d(u,v)+(1-\alpha)\max_{u\in A,v\in B}d(u,v)\, \right| \,
\alpha\in[0,1]\right\rbrace.
\end{align*}

For $b\in\{1,2,3\}$, we define $\mathcal{A}_b(\alpha)$ as the merge
function in $\mathcal{A}_b$ defined by $\alpha$.
$\mathcal{A}_1$ and $\mathcal{A}_3$ define spectra of merge functions ranging
from single-linkage ($\mathcal{A}_1(-\infty)$ and $\mathcal{A}_3(1)$) to complete-linkage ($\mathcal{A}_1(\infty)$ and $\mathcal{A}_3(0)$). The class
$\mathcal{A}_2$ defines a spectrum which includes average-linkage in addition to single- and complete-linkage.
Given a pruning function
$\prune$, we denote $\left(\mathcal{A}_b(\alpha),\prune\right)$ as the algorithm which
builds a cluster tree using $\mathcal{A}_b(\alpha)$, and then prunes the tree according
to $\prune$. We use the notation $\A_b \times \{\prune\}$ to denote the set of all such algorithms.
To reduce notation,  when $\prune$ is clear from context, we often refer to the algorithm
$(\mathcal{A}_b(\alpha),\prune)$
as $\mathcal{A}_b(\alpha)$
and the set of algorithms 
$\{(\mathcal{A}_b(\alpha),\prune)\mid\alpha\in\mathbb{R}\cup\{-\infty,\infty\}\}$ as $\mathcal{A}_b$.
For example, when the cost function is $\Phi^{(p)}$, then we always set $\prune$ to minimize
the $\Phi^{(p)}$ objective, so the pruning function is clear from context.

Recall that for a given class of merge functions and a $\cost$ function (a generic clustering objective $\clus$), our goal is to learn a near-optimal value of $\alpha$  in expectation over an unknown distribution of clustering instances. One might wonder if there is some $\alpha$ that is optimal across all instances, which would preclude the need for a learning algorithm. 
In Theorem~\ref{thm:alpha-justification}, 
we prove that this is not the case; for each $p\in[1,\infty)\cup\{\infty\}$ and 
$b \in \{1,2,3\}$, given any $\alpha$, 
there exists a distribution over clustering instances for which 
$\mathcal{A}_b(\alpha)$ is the best algorithm in $\mathcal{A}_b$ 
with respect to $\Phi^{(p)}$. Crucially, this means that even if the algorithm designer sets $p$ to be 1, 2, or $\infty$ as is typical in practice, the optimal choice of the tunable parameter $\alpha$ could be any real value. The optimal value of $\alpha$ depends on the underlying, unknown distribution, and must be learned, no matter the value of $p$.

To formally describe this result, we set up notation similar to Section \ref{sec:maxcut}.  Let $\setV$ denote the set of all clustering instances over at most $n$ points.  With a slight abuse of notation, we will use $\clus_{\mathcal{A}_b(\alpha),\prune}(\V)$ to denote the abstract cost of the clustering produced by $(\mathcal{A}_b(\alpha),\prune)$ on the instance $\V$.

\begin{restatable}{rethm}{alphaJustification}\label{thm:alpha-justification}
For $b\in\{1,2,3\}$ and a permissible value of $\alpha$ for $\mathcal{A}_b$,
there exists a distribution $\mathcal{D}$ over clustering instances $\setV$ such that $\E_{\V \sim \mathcal{D}} \left[\clus^{(p)}_{\mathcal{A}_b(\alpha)}(\V) \right] < \E_{\V \sim \mathcal{D}} \left[\clus^{(p)}_{\mathcal{A}_b(\alpha')}(\V) \right] $ for all permissible values of $\alpha' \neq \alpha$ for $\mathcal{A}_b$.
\end{restatable}

For all omitted proofs in this section, see Appendix~\ref{app:clustering}.
Another natural question to ask is whether a discretized set of the parameter space will always contain some parameter that is approximately optimal
(for instance, an $\epsilon$-net of the parameter space).
In Corollary \ref{cor:discretization}, we show this is not possible: for any data-independent discretization $D=\{d_1,\dots,d_m\}$ of the parameter space,
there exists an infinite family of clustering instances such that all $\alpha\in D$ will output a clustering that
is an $\Omega(n)$ factor worse than the optimal value of $\alpha$.
First we prove the main structural idea behind this result.

\begin{restatable}{rethm}{discretization}\label{thm:discretization}
For $b\in \{1,2,3\}$, for all $\frac{1}{3}<x<y<\frac{2}{3}$, $n>10$, and $p\in O(1)$, there exists a clustering instance
$\V$ such that for all $\alpha\in [x,y]$, $\clus_{\mathcal{A}_b(\alpha)}^{(p)}(\V)\in O(1)$, and for all $\alpha\notin [x,y]$,
$\clus_{\mathcal{A}_b(\alpha)}^{(p)}(\V)\in \Omega(n)$. 
\end{restatable}

\begin{proof}[Proof sketch]
Given $\frac{1}{3}<x<y<\frac{2}{3}$ and $n>10$, we will construct an instance $\V$ with the
desired properties. We set $k=2$.
Here is a high-level description of our construction $\V = (V,d)$. 
There will be two gadgets. 
Gadget 1 contains points $x_1$, $y_1$, $x_1'$, $y_1'$, and $z_1$.
Gadget 2 contains points $x_2$, $y_2$, $x_2'$, $y_2'$, and $z_2$.
We will define the distances so the following merges take place.
Initially, $x_1$ merges to $y_1$, $x_1'$ merges to $y_1'$,
$x_2$ merges to $y_2$, and $x_2'$ merges to $y_2'$.
Then the sets are $\{x_1,y_1\}$, $\{x_1',y_1'\}$, $\{z_1\}$,
$\{x_2,y_2\}$, $\{x_2',y_2'\}$, and $\{z_2\}$.
Next, $z_1$ will merge to $\{x_1,y_1\}$ if $\alpha<x$, and otherwise it will merge to $\{x_1',y_1'\}$.
Similarly, $z_2$ will merge to $\{x_2,y_2\}$ if $\alpha<y$, and
otherwise it will merge to $\{x_2',y_2'\}$.
Finally, the sets containing $\{x_1,y_1\}$ and $\{x_2,y_2\}$ will merge, 
and the sets containing $\{x_1',y_1'\}$ and $\{x_2',y_2'\}$ will merge.

Therefore, the situation is as follows. If $\alpha\in [x,y]$, then the last two sets in the
merge tree will each contain exactly one of the points $\{z_1,z_2\}$. If $\alpha\notin [x,y]$,
then if we again look at the last two sets in the merge tree, one of the sets will contain
both points $\{z_1,z_2\}$.
Since these are the last two sets in the merge tree, the pruning step is not able to output
a clustering with $z_1$ and $z_2$ in different clusters.
To finish the proof, we give a high weight to points $z_1$ and $z_2$ by placing $\frac{n-8}{2}$ points in the same
location as $z_1$, and $\frac{n-8}{2}$ points in the same location as $z_2$.
Note this does not affect the merge equations.
When $z_1$ and $z_2$ are in different clusters, the optimal centers for $k=2$ are at $z_1$ and
$z_2$, and the cost is just the cost of the remaining points,
$\{x_1,x_1',y_1,y_1',x_2,x_2',y_2,y_2'\}$, and all distances will be between 1 and 6, 
so the total cost is at most $8\cdot 6^p$.
When $z_1$ and $z_2$ are in the same cluster, the center will be distance at least 2 from
either $z_1$ or $z_2$ (or both), so the cost is at least $ \frac{n-8}{2}\cdot 2^p\in \Omega(n)$.

When setting the distances, the main idea is to set the distances to $z_1$ and $z_2$ so that
the merge decisions switch exactly at $\alpha=x$ and $\alpha=y$.
For example, for the case of $\mathcal{A}_3$, we set
$d(x_1,z_1)=2.4$, $d(x_1',z_1)=2.6$, and $d(y_1,z_1)=d(y_1',z_1)=2.6-0.2x$.
Therefore, the corresponding merge equation is 
\begin{align*}
\alpha\cdot 2.4+(1-\alpha)\cdot 2.6&\lessgtr\alpha\cdot(2.6-.2x)+(1-\alpha)\cdot (2.6-0.2x)\\
\alpha&\lessgtr x
\end{align*}
For the case of $\mathcal{A}_1$ and $\mathcal{A}_2$,
we set $d(y_1,z_1)=d(y_1',z_1)=\left(\frac{1}{2}\left(2.4^x+2.6^x\right)\right)^\frac{1}{x}$
to achieve the same effect.
\end{proof}

Now we can prove Corollary \ref{cor:discretization}.

\begin{cor} \label{cor:discretization}
For $b\in \{1,2,3\}$ and $p\in O(1)$, given a finite discretization $D=\{d_1,\dots,d_m\}$ of the parameter space,
there exists a constant $c$ such that
for all $n>10$, there exists a clustering instance $\V$ of size $n$ such that
$c\cdot n\cdot\min_{\alpha\in [0,1]}\clus_{\mathcal{A}_b(\alpha)}^{(p)}(\V)<\min_{\alpha\in D}\clus_{\mathcal{A}_b(\alpha)}^{(p)}(\V)$.
\end{cor}

\begin{proof}
Given a discretization $D=\{d_1,\dots,d_m\}$, note that $[0=d_0,d_1],[d_1,d_2],\dots,[d_m,d_{m+1}=1]$ is a partition of
the parameter space $[0,1]$. 
Choose an interval $[d_i,d_{i+1}]$ which has nonempty intersection with $(\frac{1}{3},\frac{2}{3})$.
Now define a new interval $[d_i',d_{i+1}']$ such that $d_i'=\max(d_i,\frac{1}{3})$ and $d'_{i+1}=\min(d_{i+1},\frac{2}{3})$.
We set $x=d_i'+\frac{d_{i+1}'-d_i'}{3}$ and $y=d_{i+1}'-\frac{d_{i+1}'-d_i'}{3}$.
By construction, we have $[x,y]\subseteq (d_i,d_{i+1})$ and $[x,y]\subseteq (\frac{1}{3},\frac{2}{3})$,
and it follows that $D\cap [x,y]=\emptyset$.
Now for each $n>10$, we use Theorem \ref{thm:discretization} with $x$ and $y$ as defined above
to obtain $\V$ such that for all $\alpha\in [x,y]$, $\clus_{\mathcal{A}_b(\alpha)}^{(p)}(\V)\in O(1)$, 
and for all $\alpha\notin [x,y]$ (including all of $D$), $\clus_{\mathcal{A}_b(\alpha)}^{(p)}(\V)\in \Omega(n)$. 
This completes the proof.
\end{proof}

 Now for an arbitrary objective function $\Phi$ and arbitrary pruning function $\prune$,
we analyze the complexity of the classes 
\begin{align*}
\mathcal{H}_{\mathcal{A}_1, \prune, \Phi} &= \left\{\left.\clus_{\mathcal{A}_1(\alpha),\prune} : \setV \to \R_{\geq 0} \, \right| \, \alpha \in \mathbb{R}\cup\{\infty, -\infty\}\right\},\\
\mathcal{H}_{\mathcal{A}_2, \prune, \Phi} &= \left\{\left.\clus_{\mathcal{A}_2(\alpha),\prune} : \setV \to \R_{\geq 0} \, \right| \, \alpha \in \mathbb{R}\cup\{\infty, -\infty\}\right\}\text{, and}\\
\mathcal{H}_{\mathcal{A}_3, \prune, \Phi} &= \left\{\left. \clus_{\mathcal{A}_3(\alpha),\prune} : \setV \to \R_{\geq 0} \, \right| \, \alpha \in [0,1] \right\}. 
\end{align*} 
In our analysis we will often fix a tuple $\V = (V,d)$ and use the notation $\clus_{\mathcal{A}_b, \Psi,\V}(\alpha)$ to analyze how $\clus_{\mathcal{A}_b(\alpha), \Psi}(\V)$ changes as a function of $\alpha$. 
We start with $\mathcal{A}_1$ and $\mathcal{A}_3$.

\begin{theorem} \label{thm:vc_13}
For all objective functions\footnote{Recall that when the cost function is $\Phi^{(p)}$, we always set the pruning function $\Psi$ to minimize the $\Phi^{(p)}$ objective, so we drop $\Psi$ from the subscript of $\mathcal{H}$.} $\Phi^{(p)}$, 
Pdim$\left(\mathcal{H}_{\mathcal{A}_1,\Phi^{(p)}}\right)=\Theta(\log n)$ and 
Pdim$\left(\mathcal{H}_{\mathcal{A}_3,\Phi^{(p)}}\right)=\Theta(\log n)$. 
For all other objective functions\footnote{Recall that although $k$-means, $k$-median, and $k$-center are the most popular choices,
the algorithm designer can use other objective functions such as the distance to the ground truth clustering (which we discuss further in Section~\ref{sec:pruning}).
} $\Phi$
and all pruning functions $\prune$, 
Pdim$\left(\mathcal{H}_{\mathcal{A}_1, \prune,\Phi}\right)=O(\log n)$ and 
Pdim$\left(\mathcal{H}_{\mathcal{A}_3, \prune,\Phi}\right)=O(\log n)$.
\end{theorem}

This theorem follows from Lemma~\ref{lem:a13upper} and Lemma~\ref{lem:a13lower}.
We begin with the following structural lemma, 
which will help us prove Lemma~\ref{lem:a13upper}.

\begin{restatable}{relem}{aOneStruct}\label{lem:a1_struct}
For any pruning function $\prune$, the function $\clus_{\mathcal{A}_1, \Psi,\V}: \R\cup\{-\infty,\infty\} \to \R_{>0}$ is made up of $O(n^8)$ piecewise constant components.
\end{restatable}

\begin{proof}[Proof sketch]
Note that for $\alpha \neq \alpha'$, the clustering returned by $\mathcal{A}_1(\alpha)$ and the associated cost are both identical to that of $\mathcal{A}_1(\alpha')$ if both the algorithms construct the same merge tree. As we range $\alpha$ across $\mathbb{R}$ and observe the run of the algorithm for each $\alpha$, at what values of $\alpha$ do we expect $\A_1(\alpha)$ to produce different merge trees?  To answer this, suppose that at some point in the run of algorithm $\A_1(\alpha)$, there are two pairs of subsets of $V$, $(A,B)$ and $(X,Y)$, that could potentially merge.  There exist eight points $p,p'\in A$, $q,q'\in B$, $x,x'\in X$, and $y,y'\in Y$ such that the decision of which pair to merge 
depends on the sign of
$\left((d(p,q))^{\alpha} + d(p',q')^\alpha \right)^{1/\alpha}-
\left((d(x,y))^{\alpha} + d(x',y')^\alpha \right)^{1/\alpha}$.
Using a consequence of Rolle's Theorem, which we provide in 
Appendix \ref{app:clustering}, we show that the sign of the above expression
as a function of $\alpha$ flips at most four times across $\mathbb{R}$.
Since each merge decision is defined by eight points, iterating over all
$(A,B)$ and $(X,Y)$ it follows that we can identify all $O(n^8)$ unique 8-tuples of points which correspond to a value of $\alpha$ at which some decision flips. This means we can divide 
$\mathbb{R}\cup\{-\infty,\infty\}$ into $O(n^8)$ intervals over each of which the merge tree, and therefore the output of 
$\clus_{\mathcal{A}_1, \Psi,\V}(\alpha)$, is fixed.
\end{proof}

In Appendix~\ref{app:clustering}, we show a corresponding statement for $\mathcal{A}_3$ (Lemma~\ref{lem:a3_struct}).
These lemmas allow us to upper bound the pseudo-dimension of $\mathcal{H}_{\mathcal{A}_1,\prune,\clus}$ and $\mathcal{H}_{\mathcal{A}_3,\prune,\clus}$ 
by $O(\log n)$ in a manner similar to Lemma~\ref{lem:slin_upper}, where we prove a pseudo-dimension upper bound on the class of $s$-linear SDP rounding algorithms. 
Thus we obtain the following lemma.

\begin{restatable}{relem}{aOneThreeUpper}\label{lem:a13upper}
For any objective function $\clus$ and any pruning function $\Psi$, Pdim$(\mathcal{H}_{\mathcal{A}_1, \Psi,\clus})=O(\log n)$
and Pdim$(\mathcal{H}_{\mathcal{A}_3, \Psi,\clus})=O(\log n)$.
\end{restatable}

Next, we give lower bounds for the pseudo-dimension of the two classes.

\begin{restatable}{relem}{aOneThreeLower}\label{lem:a13lower}
For any objective function $\Phi^{(p)}$, Pdim$\left(\mathcal{H}_{\mathcal{A}_1, \Phi^{(p)}}\right)=\Omega(\log n)$
and Pdim$\left(\mathcal{H}_{\mathcal{A}_3, \Phi^{(p)}}\right)=\Omega(\log n)$.
\end{restatable}

\begin{proof}[Proof sketch]
We give a general proof outline that applies to both classes. 
Let $b \in \{1, 3\}$.
We construct a set $S=\left\{\V^{(1)},\dots,\V^{(m)}\right\}$ of $m=\log n-3$ clustering instances that can be shattered by $\mathcal{A}_b$.
There are $2^m=n/8$ possible labelings for this set, so we need to show there are $n/8$ choices of $\alpha$ such that each of these labelings is achievable by some $\mathcal{A}_{b}(\alpha)$ for some $\alpha$.
The crux of the proof lies in showing that given a sequence $\alpha_0<\alpha_1<\cdots<\alpha_{n'}<\alpha_{n'+1}$ (where $n' = \Omega(n)$), it is possible to design an instance $\V=(V,d)$ over $n$ points and choose a witness $r$ such that $\clus_{\mathcal{A}_b(\alpha)}(\V)$ alternates $n'/2$ times above and below $r$ as $\alpha$ traverses the sequence of intervals $(\alpha_i,\alpha_{i+1})$.

Here is a high level description of our construction. There will be two ``main'' points, 
$a$ and $a'$ in $V$.
The rest of the points are defined in groups of 6: $(x_i,y_i,z_i,x_i',y_i',z_i')$, for 
$1\leq i\leq (n-2)/6$.
We will define the distances between all points such that initially for all $\A_b(\alpha)$, $x_i$ merges to $y_i$ to form the set $A_i$, 
and $x_i'$ merges to $y_i'$ to form the set $A_i'$. As for $(z_i,z_i')$, depending on whether $\alpha < \alpha_i$ or not, $\A_b(\alpha)$ merges the points $z_i$ and $z_i'$ with the sets $A_i$ and $A_i'$ respectively or vice versa. This means that there are $(n-2)/6$ values of $\alpha$ such that $\A_{b}(\alpha)$ has a unique behavior in the merge step. Finally, for all $\alpha$, sets $A_i$ merge to $\{a\}$,  and sets $A_i'$ merge to $\{a'\}$. Let $A=\{a\}\cup\bigcup_i A_i$ and $A'=\{a'\}\cup\bigcup_i A_i'$. There will be $(n-2)/6$ intervals $(\alpha_i,\alpha_{i+1})$ for which $\A_b(\alpha)$ returns a unique partition $\{A,A'\}$. By carefully setting the distances, we cause the cost $\Phi(\{A,A'\})$ to oscillate above and
below a specified value $r$ along these intervals.
\end{proof}

The upper bound on the pseudo-dimension implies a computationally efficient and sample efficient
learning algorithm for $\mathcal{A}_b$ for $b\in\{1,3\}$.
See Algorithm~\ref{alg:a13-erm}.
First, we know that 
$m =  \tilde O\left(\left(H/\epsilon\right)^2\right)$ samples are sufficient to $(\epsilon,\delta)$-learn the optimal algorithm in $\mathcal{A}_b$.
Next, as a consequence of Lemmas~\ref{lem:a1_struct} and~\ref{lem:a3_struct}, 
the range of feasible values of $\alpha$ can be partitioned into $O(m n^8)$ intervals, such that the output of $\mathcal{A}_b(\alpha)$ is fixed over the entire set of samples on a given interval. Moreover, these intervals are easy to compute.
Therefore, a learning algorithm can iterate over the set of intervals, and for each interval $I$, choose an arbitrary $\alpha \in I$ and compute the average cost of $\mathcal{A}_b(\alpha)$ evaluated on the samples. The algorithm then outputs the $\alpha$ that minimizes the average cost.

\begin{algorithm} [t]
\caption{An algorithm for finding an empirical cost minimizing algorithm in $\A_1$ or $\A_3$}
\label{alg:a13-erm}
\begin{algorithmic}[1]
\Require {Sample $\sample = \left\lbrace \V^{(1)}, \hdots, \V^{(m)} \right\rbrace $, $b\in\{1,3\}$, pruning function $\Psi$, objective function $\Phi$.}
\State Let $T=\emptyset$. For each sample $\V^{(i)} = \left(V^{(i)},d^{(i)}\right)\in \sample$, and for each ordered set of 8 points
$\{v_1,\dots,v_8\}\subseteq V^{(i)}$, solve for $\alpha$ (if a solution exists) in the following equation and add the solutions to $T$:
$d(v_1,v_2)^\alpha+d(v_3,v_4)^\alpha=d(v_5,v_6)^\alpha+d(v_7,v_8)^\alpha.$
\[
\begin{array}{lrcl}
\text{If }b=1: &  d(v_1,v_2)^\alpha+d(v_3,v_4)^\alpha&=&
d(v_5,v_6)^\alpha+d(v_7,v_8)^\alpha.
\\
\text{If }b=3:  & \alpha d(v_1,v_2)+(1-\alpha)d(v_3,v_4)&=&\alpha d(v_5,v_6)+(1-\alpha)d(v_7,v_8).
\end{array}
\] 
\label{step:find_alphas}
\State Order the elements of set $T \cup \{-\infty, +\infty \}$ as $\alpha_1< \hdots < \alpha_{|T|}$.
For each $0\leq i\leq |T|$, pick an arbitrary $\alpha$ in the interval $(\alpha_i,\alpha_{i+1})$ 
and run $\A_b(\alpha)$ on all clustering instances in $\sample$ to compute $\sum_{i = 1}^m \Phi_{\A_b(\alpha), \prune}\left(\V^{(i)}\right)$. Let $\hat{\alpha}$ be the value which minimizes $\sum_{i = 1}^m \Phi_{\A_b(\alpha), \prune}\left(\V^{(i)}\right)$.
\Ensure{$\hat{\alpha}$}
\end{algorithmic} 
\end{algorithm}

\begin{theorem}\label{thm:a13algo}
Let $\Phi$ be a clustering objective and let $\prune$ be a pruning function computable in
polynomial time. Given an input sample of size $m = O\left(\left(\frac{H}{\epsilon}\right)^2\left(\log n + \log \frac{1}{\delta} \right)\right)$,
and a value $b \in \{1,3\}$,  
Algorithm~\ref{alg:a13-erm} $(\epsilon,\delta)$-learns the class 
$\mathcal{A}_b \times \{\prune\}$ with respect to the cost function $\Phi$ and it is computationally efficient.
\end{theorem}

\begin{proof}
Algorithm \ref{alg:a13-erm} finds the empirically best $\alpha$
by solving for the $O(mn^8)$ discontinuities of
$\sum_{\V \in \sample} \Phi_{\A_b(\alpha)}(\V)$ and evaluating the function over
the corresponding intervals, which are guaranteed to be constant by 
Lemmas~\ref{lem:a1_struct} and \ref{lem:a3_struct}.  
Therefore, we can pick any arbitrary $\alpha$ within each interval to evaluate the empirical cost over all samples, 
and find the empirically best $\alpha$. 
This can be done in polynomial time because there are polynomially
many intervals, and the runtime of $\A_b(\alpha)$ on a given instance
is polynomial time.

Then it follows from Theorem \ref{thm:vc_13}
that $m$ samples are sufficient for Algorithm \ref{alg:a13-erm} to $(\epsilon,\delta)$-learn the optimal
algorithm in $\mathcal{A}_b$ for $b\in\{1,3\}$.
\end{proof}

Now we turn to $\mathcal{A}_2$. We obtain the following bounds on the pseudo-dimension.

\begin{theorem} \label{thm:vc_2}
For any objective function\footnote{Recall that when the cost function is $\Phi^{(p)}$, we always set the pruning function $\Psi$ to minimize the $\Phi^{(p)}$ objective, so we drop $\Psi$ from the subscript of $\mathcal{H}$.} $\Phi^{(p)}$, Pdim$(\mathcal{H}_{\mathcal{A}_2, \Phi^{(p)}})=\Theta(n)$. For all other objective functions $\Phi$ and all pruning functions
$\prune$, Pdim$\left(\mathcal{H}_{\mathcal{A}_2, \prune, \Phi}\right)=O(n)$.
\end{theorem}

This theorem follows from Lemmas~\ref{lem:overall_linkage} and~\ref{lem:general_lb}.

\begin{lemma}\label{lem:overall_linkage}
For all objective functions $\Phi$ and all pruning functions $\prune$, 
Pdim$\left(\mathcal{H}_{\mathcal{A}_2, \prune,\Phi}\right)=O(n)$.
\end{lemma}

\begin{proof}

Recall the proof of Lemma~\ref{lem:a13upper}. 
We are interested in studying how the merge trees constructed by $\A_2(\alpha)$ changes over $m$ instances as we increase $\alpha$ over $\mathbb{R}$. To do this, as in the proof of Lemma~\ref{lem:a13upper}, 
we fix an instance and consider two pairs of sets $A,B$ and $X,Y$ that could be potentially merged. Now, the decision to merge one pair before the other is determined by the sign of the expression 
$ \frac{1}{|A||B|}\sum_{p \in A, q \in B} (d(p,q))^\alpha - \frac{1}{|X||Y|}\sum_{x \in X, y \in Y} (d(x,y))^\alpha$. First note that this expression has $O(n^2)$ terms, and 
by a consequence of Rolle's Theorem which we provide in Appendix~\ref{app:clustering},
it has $O(n^2)$ roots. Therefore, as we iterate over the $O\left(\left(3^{n}\right)^2\right)$ possible pairs $(A,B)$ and $(X,Y)$, we can determine $O\left(3^{2n}\right)$ unique expressions each with $O(n^2)$ values of $\alpha$ at which the corresponding decision flips. Thus we can divide $\mathbb{R}$ into $O\left(n^2 3^{2n}\right)$ intervals over each of which the output of $\Phi_{\A_3,\V}(\alpha)$ is fixed. In fact, suppose $\sample = \left\{\V^{(1)}, \dots, \V^{(m)}\right\}$ is a shatterable set of size $m$ with witnesses $r_1, \dots, r_m$. We can divide $\mathbb{R}$ into $O\left(m n^2 3^{2n}\right)$ intervals over each of which $\Phi_{\A_2,\V^{(i)}}(\alpha)$ is fixed for all $i \in [m]$ and therefore the corresponding labeling of $\sample$ according to whether or not $\Phi_{\A_2(\alpha)}\left(\V^{(i)}\right)\leq r_i$ is fixed as well for all $i \in [m]$. This means that $\mathcal{H}_{\mathcal{A}_2}$ can achieve only  $O\left(m n^2 3^{2n}\right)$ labelings, which is at least $2^m$ for a shatterable set $\sample$, so $m = O(n)$.
\end{proof}

\begin{restatable}{relem}{generalLB}\label{lem:general_lb}
For all objective functions $\Phi^{(p)}$, Pdim$\left(\mathcal{H}_{\mathcal{A}_2, \Phi^{(p)}}\right)=\Omega(n)$.
\end{restatable}

\begin{proof}[Proof sketch] 
The crux of the proof is to show that there exists  a clustering instance $\V$ over $n$ points, a witness $r$, and a set of $\alpha$'s
$1=\alpha_0<\alpha_1<\cdots<\alpha_{2^N}<\alpha_{2^N+1}=3$, where $N= \lfloor(n-8)/4\rfloor$,
such that $\Phi_{\mathcal{A}_2,\V}(\alpha)$ oscillates above and below 
$r$ along the sequence of intervals $(\alpha_i,\alpha_{i+1})$. We finish the proof in a manner similar to Lemma~\ref{lem:a13lower} by constructing instances with fewer oscillations.

To construct $\V$, first we define two pairs of points
which merge together regardless of the value of $\alpha$.
Call these merged pairs $A$ and $B$.
Next, we define a sequence of points $p_i$ and $q_i$ for $1\leq i\leq N$ with distances set such that merges involving points in this sequence occur one after the other. In particular, each $p_i$ merges with one of $A$ or $B$ while $q_i$ merges with the other.
Therefore, there are potentially $2^N$ distinct merge trees which can be created.
Using induction to precisely set the distances,
we show there are $2^N$ distinct values of $\alpha$, each corresponding to a unique merge tree,
thus enabling $\mathcal{A}_2$ to achieve all possible merge tree behaviors.
Finally, we carefully add more points to the instance to control the oscillation of the cost function over these intervals as desired.
\end{proof}

\subsection{Dynamic programming pruning functions}
\label{sec:pruning}
In the previous section, we analyzed several classes of linkage-based merge functions assuming a fixed pruning function in the dynamic programming step of the standard linkage-based clustering algorithm, i.e. Step~\ref{step:dp} of Algorithm~\ref{alg:linkage}. In this section, we analyze an infinite class of dynamic programming pruning functions and derive comprehensive sample complexity guarantees for learning the best merge function and  pruning function in conjunction.

By allowing an application-specific choice of a pruning function, we significantly generalize the standard linkage-based clustering algorithm framework. Recall that in the algorithm selection model, we instantiated the $\cost$ function to be a generic clustering objective $\clus$.
In the standard clustering algorithm framework, where $\clus$ is defined to be any general $\clus^{(p)}$ (which include objectives like $k$-means), the best choice of the pruning function for the algorithm selector is $\clus^{(p)}$ itself as it would return the optimal pruning of the cluster tree for that instantiation of $\cost$. However, when the goal of the algorithm selector is, for example, to provide solutions that are close to a ground truth clustering for each problem instance, the best choice for the pruning function is not obvious. In this case, we assume that the learning algorithm's training data consists of clustering instances that have been labeled by an expert according to the ground truth clustering. For example, this ground truth clustering might be a partition of a set of images based on their subject, or a partition of a set of proteins by function. On a fresh input data, we no longer have access to the expert or the ground truth, so we cannot hope to prune a cluster tree based on distance to the ground truth.\footnote{If $\clus$ is the distance to ground truth clustering, then $\clus$ cannot be directly measured when the
clustering algorithm is used on new data. However, we assume that the learning algorithm has access to training data which consists of clustering instances labeled by the ground truth clustering. The learning algorithm uses this data to optimize the parameters defining the clustering algorithm family. With high probability, on a new input drawn from the same distribution as the training data, the clustering algorithm will return a clustering that is close to the unknown ground truth clustering.}

Instead, the algorithm selector must empirically evaluate how well pruning according to alternative objective functions, such as $k$-means or $k$-median, approximate the ground truth clustering on the labeled training data. In this way, we instantiate $\cost$ to be the distance of a clustering from the ground truth clustering. We guarantee that the empirically best pruning function from a class of computable objectives is near-optimal in expectation over new problem instances drawn from the same distribution as the training data. Crucially, we are able to make this guarantee even though it is not possible to compute the $\cost$ of the algorithm's output on these fresh instances because the ground truth clustering is unknown.

Along these lines, we can also handle the case where the training data consists of clustering instances, each of which has been clustered according to an objective function that is NP-hard to compute. In this scenario, our learning algorithm returns a pruning objective function that is efficiently computable and which best approximates the NP-hard objective on the training data, and therefore will best approximate the NP-hard objective on future data.
Hence, in this section, we analyze a richer class of algorithms defined by a class of merge functions and a class of pruning functions. The learner now has to learn the best combination of merge and pruning functions from this class.

To define this more general class of agglomerative clustering algorithms, let $\A$ denote a generic class of linkage-based merge functions (such as any of the classes $\A_b$ defined in Section~\ref{sec:simple}) parameterized by  $\alpha$. We also define a rich class of center-based clustering objectives for the dynamic programming step: $\mathcal{F} = \left\{\prune^{(p)} \ | \ p > 0\right\}$ where $\prune^{(p)}$ takes as input a
partition $\mathcal{C} = \{C_1, C_2, \hdots, C_{k'} \}$ of $n'$ points and a set of centers $\mathbf{c} = \{c_1, c_2, \hdots, c_{k'}\}$ such that $c_i \in C_i$. The function $\prune^{(p)}$ is defined such that \begin{equation}\prune^{(p)}(\mathcal{C}, \mathbf{c}) = \sqrt[p]{\sum_{C_i \in \mathcal{C}} \sum_{q \in C_i} (d(q,c_i))^p}.\label{eq:prun_func}\end{equation}
Note that the definition of $\prune^{(p)}$ is identical to $\clus^{(p)}$, but we use this different notation
so as not to confuse the dynamic programming function with the clustering objective function.
Let $\A(\alpha)$ denote the $\alpha$-linkage merge function from $\A$ and $\mathcal{F}(p)$ denote the pruning function $\prune^{(p)}$.
Earlier, for an abstract objective $\clus$, we bounded the pseudodimension  of $\mathcal{H}_{\A,\prune, \clus} = \left\{\clus_{\A(\alpha),\prune} : \setV \to \R_{\geq 0} \right\}$, where $\clus_{\A(\alpha),\prune}(\V)$ denoted the cost of the clustering produced by building the cluster tree on $\V$ using the merge function $\A(\alpha)$ and then pruning the tree using a fixed pruning function $\prune$. Now, we are interested in doubly-parameterized algorithms of the form $(\A(\alpha), \mathcal{F}(p))$ which uses the merge function $\A(\alpha)$ to build a cluster tree and then use the pruning function $\mathcal{F}(p)$ to prune it. To analyze the resulting  class of algorithms, which we denote by $\A \times \mathcal{F}$, we have to bound the pseudodimension of  $\mathcal{H}_{\A, \mathcal{F}, \clus}$, which consists of all functions $\clus_{\A(\alpha), \mathcal{F}(p)} : \setV \to \R_{\geq 0}$.

Recall that in order to show that pseudodimension of $\mathcal{H}_{\A, \prune,\clus}$ is upper bounded by a positive integer $d$, we proved that, given a sample of $m$ clustering instances over $n$ nodes, we can split the real line into at most $O\left(m2^{d}\right)$ intervals such that as $\alpha$ ranges over a single interval, the $m$ cluster trees returned by the $\alpha$-linkage merge function are fixed.
To extend this analysis to $\mathcal{H}_{{\A}, \mathcal{F},\clus}$, we first prove a similar fact in Lemma~\ref{lem:pruning-intervals}. Namely, given a single {\emph {cluster tree}}, we can split the real line into a fixed number of intervals such that as $p$ ranges over a single interval, the pruning returned by using the function $\prune^{(p)}$ is fixed. We then show in 
Theorem~\ref{thm:sample_comp_prun} 
how to combine this analysis of the rich class of dynamic programming algorithms with our previous analysis of the possible merge functions to obtain a comprehensive analysis of agglomerative algorithms with dynamic programming.

\begin{figure}
  \centering
  \includegraphics[width=.3\textwidth] {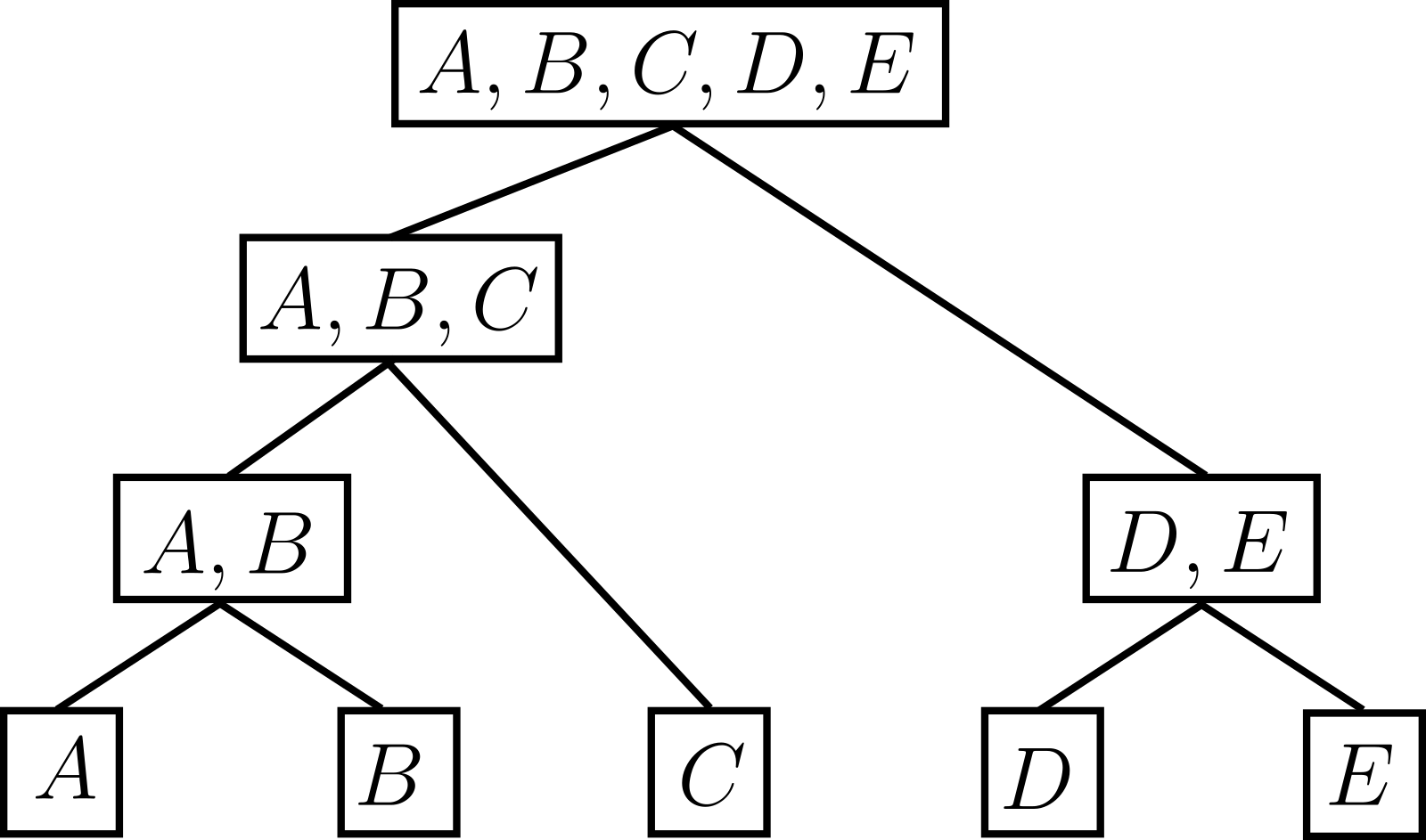}
  \caption{Cluster tree corresponding to Table~\ref{tab:dp}.}\label{fig:clustertree}
\end{figure}

\begin{table}
\centering
\footnotesize
\begin{tabular}{|l|l||l|l|l|l|l|l|l|l|l|}
\hline
   && $A$ & $B$ & $C$ & $D$ & $E$ & $A,B$          & $D,E$          & $A,B,C$             & $A,B,C,D,E$                       \\
  \hhline{|=|=||=|=|=|=|=|=|=|=|=|}
\multirow{2}{*}{1} &Clusters& $\{A\}$ & $\{B\}$ & $\{C\}$ & $\{D\}$ & $\{E\}$ & $\{A,B\}$          & $\{D,E\}$          & $\{A,B,C\}$             & $\{A,B,C,D,E\}$             \\      \hhline{~----------}

  &Centers& $\{A\}$ & $\{B\}$ & $\{C\}$ & $\{D\}$ & $\{E\}$ & $\{A\}$          & $\{E\}$          & $\{C\}$             & $\{C\}$             \\ \hline
\multirow{2}{*}{2} &Clusters&     &     &     &     &     & $\{A\}, \{B\}$ & $\{D\}, \{E\}$ & $\{A,B\}, \{C\}$    & $\{A,B,C\}, \{D,E\}$    \\\hhline{~----------}
 &Centers&     &     &     &     &     & $\{A\}, \{B\}$ & $\{D\}, \{E\}$ & $\{A,C\}$    & $\{C,E\}$    \\\hline
\multirow{2}{*}{3} &Clusters&     &     &     &     &     &                &                & $\{A\},\{B\},\{C\}$ & $\{A,B\},\{C\},\{D,E\}$\\\hhline{~----------}
 &Centers&     &     &     &     &     &                &                & $\{A,B,C\}$ & $\{A,C,E\}$\\\hline
\end{tabular}
\caption{Example dynamic programming table corresponding to the cluster tree in Figure~\ref{fig:clustertree} for $k=3$.}\label{tab:dp}
\end{table}
We visualize the dynamic programming step of Algorithm~\ref{alg:linkage} with pruning function $\prune^{(p)}$ using a table such as Table~\ref{tab:dp}, which corresponds to the cluster tree in Figure~\ref{fig:clustertree}. Each row of the table corresponds to a sub-clustering value $k' \leq k$, and each column corresponds to a node of the corresponding cluster tree.  In the column corresponding to node $T$ and the row corresponding to the value $k'$, we fill in the cell with the partition of $T$ into $k'$ clusters that corresponds to the best $k'$-pruning of the subtree rooted at $T$, $\left(\mathcal{C}_{T,k'}, \mathbf{c}_{T,k'}\right)$ as defined in Step~\ref{step:dp} of Algorithm~\ref{alg:linkage}.

\begin{lemma}\label{lem:pruning-intervals}
Given a cluster tree $\mathcal{T}$ for a clustering instance $\V = (V,d)$ of $n$ points, the positive real line can be partitioned into a set $\mathcal{I}$ of $O(n^{2(k+1)}k^{2k})$ intervals such that for any $I \in \mathcal{I}$, the cluster tree pruning according to $\prune^{(p)}$ is identical for all $p \in I$.
\end{lemma}

\begin{proof}
To prove this claim, we will examine the dynamic programming (DP) table corresponding to the given cluster tree and the pruning function $\prune^{(p)}$ as $p$ ranges over the positive real line. As the theorem implies, we will show that we can split the positive real line into a set of intervals so that on a fixed interval $I$, as $p$ ranges over $I$, the DP table under $\prune^{(p)}$ corresponding to the cluster tree is invariant. No matter which $p \in I$ we choose, the DP table under $\prune^{(p)}$ will be identical, and therefore the resulting clustering will be identical. After all, the output clustering is the bottom-right-most cell of the DP table since that corresponds to the best $k$-pruning of the node containing all points (see Table~\ref{tab:dp} for an example). We will prove that the total number of intervals is bounded by $O(n^{2(k+1)}k^{2k})$.

We will prove this lemma using induction on the row number $k'$ of the DP table. Our inductive hypothesis will be the following. \emph{The positive real line can be partitioned into a set $\mathcal{I}^{(k')}$ of $O\left(n^2\prod_{j = 1}^{k'} n^2 j\right)$ intervals such that for any $I^{(k')} \in \mathcal{I}^{(k')}$, as $p$ ranges over $I^{(k')}$, the first $k'$ rows of the DP table corresponding to $\prune^{(p)}$ are invariant.} Notice that this means that the positive real line can be partitioned into a set $\mathcal{I}$ of $O\left(n^2\prod_{j = 1}^k n^2 j^2\right) = O\left(n^{2(k+1)}k^{2k}\right)$ intervals such that for any $I \in \mathcal{I}$, as $p$ ranges over $I$, the DP table corresponding to $\prune^{(p)}$ is invariant. Therefore, the resulting output clustering is invariant as well.

\smallskip

\emph{Base case ($k'=1$).} Let $p$ be a positive real number. Consider the first row of the DP table corresponding to $\prune^{(p)}$. Recall that each column in the DP table corresponds to a node $T$ in the clustering tree where $T \subseteq V$. In the first row of the DP table and the column corresponding to node $T$, we fill in the cell with the single node $T$ and the point $c \in T$ which minimizes $\prune^{(p)}(\{T\}, \{c \}) =\sum_{q \in T}(d(q,c))^p $. The only thing that might change as we vary $p$ is the center minimizing this objective.

Let $v_1$ and $v_2$ be two points in $T$. The point $v_1$ is a better candidate for the center of $T$ than $v_2$ if and only if 
$\prune^{(p)}(\{T\}, \{v_1 \}) \leq \prune^{(p)}(\{T\}, \{v_2 \}) $
which means that $\prune^{(p)}(\{T\}, \{v_1 \}) - \prune^{(p)}(\{T\}, \{v_2 \}) \leq 0 $, or in other words, $\sum_{q \in T}(d(q,v_1))^p - \sum_{q \in T}(d(q,v_2))^p \leq 0$. 
The equation $\sum_{q \in T}(d(q,v_1))^p - \sum_{q \in T}(d(q,v_2))^p$ has at most $2|T|$ zeros, so there are at most $2|T|+1$ intervals $I_1, \dots, I_t$ which partition the positive real line such that for any $I_i$, as $p$ ranges over $I_i$, whether or not 
$\prune^{(p)}(\{T\}, \{v_1 \}) \leq \prune^{(p)}(\{T\}, \{v_2 \}) $ is fixed.
 For example, see Figure~\ref{fig:partition-base-case}.
\begin{figure}
  \centering
  \includegraphics[width=\textwidth] {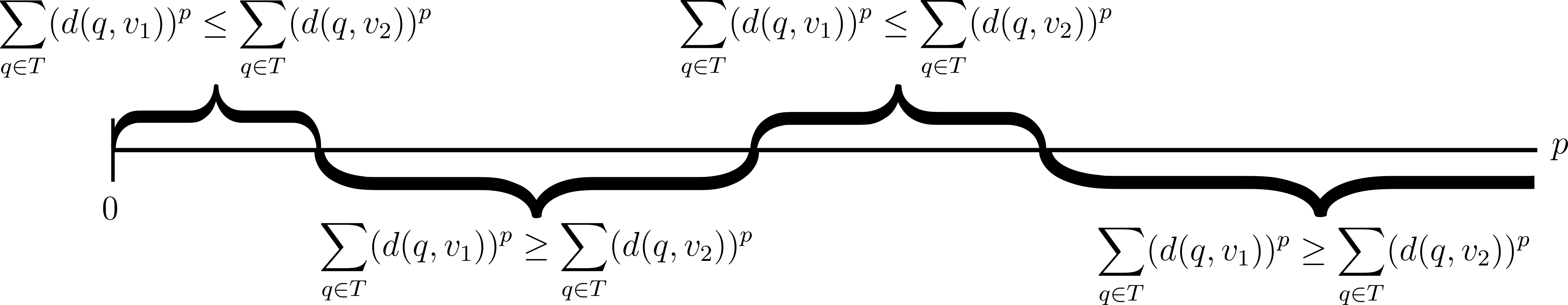}
  \caption{Partition of the positive real line based on whether or not $\sum_{q \in T}(d(q,v_1))^p \leq \sum_{q \in T}(d(q,v_2))^p$ as $p$ ranges $\R_{>0}$.}\label{fig:partition-base-case}
\end{figure}
Every pair of points in $T$ similarly partitions the positive real line into $2|T|+1$ intervals. If we merge all $|T|^2/2$ partitions --- one partition for each pair of points in $T$ --- then we are left with at most $\frac{|T|^2}{2}\cdot 2|T| + 1 = |T|^3 + 1$ intervals $I_1, \dots, I_w$ partitioning the positive real line such that for any $I_i$, as $p$ ranges over $I_i$, the point $v \in T$ which minimizes $\prune^{(p)}(\{T\}, \{v\})$ is fixed.

Since $T$ is arbitrary, we can thus partition the real line for each node $T'$ in the cluster tree. Again, this partition defines the center of the cluster $T'$ as $p$ ranges over the positive real line.
If we merge the partition for every node 
$T \in \mathcal{T}$, 
then we are left with $\left(\sum_{T \in \mathcal{T}} |T|^3\right) + 1 = O(n^4)$ intervals $I_1, \dots, I_{\ell}$ such that as $p$ ranges over any one interval $I_i$, the centers of all nodes in the cluster tree are fixed. In other words, for each $T$, the point $v_i \in T$ which minimizes
$\prune^{(p)}(\{T\}, \{v_i\})$ 
is fixed. Of course, this means that the first row of the DP table is fixed as well. Therefore, the inductive hypothesis holds for the base case.

\smallskip

\emph{Inductive step.} Consider the $k'$th row of the DP table. We know from the inductive hypothesis that the positive real line can be partitioned into a set $\mathcal{I}^{(k'-1)}$ of $O\left(n^2\prod_{j = 1}^{k'-1} n^2 j^2\right)$ intervals such that for any $I^{(k'-1)} \in \mathcal{I}^{(k'-1)}$, as $p$ ranges over $I^{(k'-1)}$, the first $k'-1$ rows of the DP table corresponding to $\prune^{(p)}$ are invariant.

Fix some interval $I^{(k'-1)} \in \mathcal{I}^{(k'-1)}$. Let $T$ be a node in the cluster tree $\mathcal{T}$
and let $T_L$ and $T_R$ be the left and right children of $T$ in $\mathcal{T}$ respectively. Notice that the pruning which belongs in the cell in the $i$th row and the 
 column corresponding to $T$ does not depend on the other cells in the $i$th row, but only on the cells in rows 1 through $i-1$. In particular, the pruning which belongs in this cell depends on the inequalities defining which $i' \in \{1, \dots, k'-1\}$ minimizes $\prune^{(p)}\left(\mathcal{C}_{T_L,i'} \cup \mathcal{C}_{T_R,k'-i'}, \mathbf{c}_{T_L,i'} \cup \mathbf{c}_{T_R,k'-i'}\right)$.
We will now examine this objective function and show that the minimizing $i'$, and therefore the optimal pruning, only changes a small number of times as $p$ ranges over $I^{(k'-1)}$.

For an arbitrary $i' \in \{1, \dots, k'-1\}$, since $i'$ and $k'-i'$ are both strictly less than $k'$, the best $i'$-pruning of  $T_L$ $(\mathcal{C}_{T_R,i'}, \mathbf{c}_{T_R,i'})$ is exactly the entry in the $i'$th row of the DP table and the column corresponding to $T_L$. Similarly, the best $k'-i'$-pruning of $T_R$, $(\mathcal{C}_{T_R,k'-i'}, \mathbf{c}_{T_R,k'-i'})$ is exactly the entry in the $k'-i'$th row of the DP table and the column corresponding to $T_R$. Crucially, these entries do not change as we vary $p \in I^{(k'-1)}$, thanks to the inductive hypothesis.

Therefore, for any $i', i'' \in \{1, \dots, k'-1\}$, we know that for all $p \in I^{(k'-1)}$, the $k'$-pruning of $T$ corresponding to the combination of the best $i'$-pruning of $T_L$ and the best $k'-i'$ pruning of $T_R$ is fixed and can be denoted as $(\mathcal{C}', \mathbf{c}')$. Similarly, the $k'$-pruning of $T$ corresponding to the combination of the best $i''$-pruning of $T_L$ and the best $k'-i''$ pruning of $T_R$ is fixed and can be denoted as $(\mathcal{C}'', \mathbf{c}'')$. Then, for any $p \in I^{(k'-1)}$, $(\mathcal{C}', \mathbf{c}')$ is a better pruning than $(\mathcal{C}'', \mathbf{c}'')$ if and only if 
$
\prune^{(p)}\left(\mathcal{C}', \mathbf{c}'\right) \leq \prune^{(p)}\left(\mathcal{C}'', \mathbf{c}''\right)
$. In order to analyze this inequality, let us consider the equivalent inequality $\left( \prune^{(p)}\left(\mathcal{C}', \mathbf{c}'\right)\right)^p \leq \left(\prune^{(p)}\left(\mathcal{C}'', \mathbf{c}''\right)\right)^p
$ i.e.,
$\left( \prune^{(p)}\left(\mathcal{C}', \mathbf{c}'\right)\right)^p - \left(\prune^{(p)}\left(\mathcal{C}'', \mathbf{c}''\right)\right)^p \leq 0
$. Now, to expand this expression let $\mathcal{C}' = \{C'_1, C'_2, \hdots, C'_{k'}\}$ and $\mathbf{c}' = \{c'_1, c'_2 \hdots, c'_{k'}\}$ and similarly
$\mathcal{C}'' = \{C''_1, C''_2, \hdots, C'_{k'}\}$ and $\mathbf{c}'' = \{c''_1, c''_2 \hdots, c''_{k'}\}$. Then, this inequality can then be written as,

\[
\sum_{i=1}^{k'} \sum_{q \in C'_i} \left( d(q, c'_i) \right)^p - 
\sum_{i=1}^{k'} \sum_{q \in C''_i} \left( d(q, c''_i) \right)^p  \leq
0.
\]

The equation $\sum_{i=1}^{k'} \sum_{q \in C'_i} \left( d(q, c'_i) \right)^p - 
\sum_{i=1}^{k'} \sum_{q \in C''_i} \left( d(q, c''_i) \right)^p $ has 
 has at most $2n$ zeros as $p$ ranges over $I^{(k'-1)}$. Therefore, there are at most $2n+1$ subintervals partitioning $I^{(k'-1)}$ such that as $p$ ranges over one subinterval, the smaller of 
$\prune^{(p)}\left(\mathcal{C}', \mathbf{c}'\right)$ and $ \prune^{(p)}\left(\mathcal{C}'', \mathbf{c}''\right) $
is fixed. In other words, as $p$ ranges over one subinterval, either the combination of the best $i'$-pruning of $T$'s left child and the best $(k'-i')$-pruning of $T$'s right child is better than the combination of the best $i''$-pruning of $T$'s left child with the best $(k-i'')$-pruning of $T$'s right child, or vice versa. For all pairs $i',i'' \in \{1, \dots, k'-1\}$, we can similarly partition $I$ into at most $2n+1$ subintervals defining the better of the two prunings. If we merge all $(k'-1)^2/2$ partitions of $I^{(k'-1)}$, we have $\frac{(k'-1)^2}{2}\cdot 2n + 1 = (k'-1)^2n+1$ total subintervals of $I^{(k'-1)}$ such that as $p$ ranges over a single subinterval, 
\[
\text{argmin}_{i' \in \{1, \dots, k'-1 \}}\prune^{(p)}\left(\mathcal{C}_{T_L,i'} \cup \mathcal{C}_{T_R,k'-i'}, \mathbf{c}_{T_L,i'} \cup \mathbf{c}_{T_R,k'-i'}\right)
\]
is fixed. Since these equations determine the entry in the $i$th row of the DP table and the column corresponding to the node $T$, we have that this entry is also fixed as $p$ ranges over a single subinterval in $I^{(k'-1)}$.

The above partition of $I^{(k'-1)}$ corresponds to only a single cell in the $k'$th row of the DP table. Considering the $k'$th row of the DP table as a whole, we must fill in at most $2n$ entries, since there are at most $2n$ columns of the DP table. For each column, there is a corresponding partition of $I^{(k'-1)}$ such that as $p$ ranges over a single subinterval in the partition, the entry in the $k'$th row and that column is fixed. If we merge all such partitions, we are left with a partition of $I^{(k'-1)}$ consisting of at most $2n^2(i-1)^2+1$ intervals such that as $p$ ranges over a single interval, the entry in every column of the $k'$th row is fixed. As these intervals are subsets of $I^{(k'-1)}$, by assumption, the first $k'-1$ rows of the DP table are also fixed. Therefore, the first $k'$ rows are fixed.

To recap, we fixed an interval $I^{(k'-1)}$ such that as $p$ ranges over $I^{(k'-1)}$, the first $k'-1$ rows of the DP table are fixed. By the inductive hypothesis, there are $O\left(n^2\prod_{j = 1}^{k'-1} n^2 j^2\right)$ such intervals. Then, we showed that $I^{(k'-1)}$ can be partitioned into $2n^2(k'-1)^2+1$ intervals such that for any one subinterval $I^{(k')}$, as $p$ ranges over $I^{(k')}$, the first $k'$ rows of the DP table are fixed. Therefore, there are $O\left(n^2\prod_{j = 1}^{k'} n^2 j^2\right)$ total intervals such that as $p$ ranges over a single interval, the first $k'$ rows of the DP table are fixed.

Aggregating this analysis over all $k$ rows of the DP table, we have that there are \[O\left(n^2\prod_{k = 1}^k n^2 {k'}^2\right) = O\left(n^{2(k+1)}k^{2k}\right)\] intervals such that the entire DP table is fixed so long as $p$ ranges over a single interval. 
\end{proof}

We are now ready to prove our main theorem in this section.

\begin{theorem}\label{thm:sample_comp_prun} Suppose there exists a positive integer $d$ such that for any clustering instance, there are at most $d$ intervals partitioning the domain of $\alpha$ such that as $\alpha$ ranges over a single interval, the cluster tree returned by the $\alpha$-linkage merge function from $\A$ is fixed. Then $Pdim\left(\mathcal{H}_{\A, \mathcal{F},\Phi}\right)=O\left(\log d + k \log n \right)$.
\end{theorem}

\begin{proof}
Let $\sample$ be a set of $m$ clustering instances. Fix a single interval of $\alpha$ (as shown along the horizontal axis in Figure~\ref{fig:p-intervals}) where the set of cluster trees returned by the $\alpha$-linkage merge function from $\A$ is fixed across all samples. We know from Lemma~\ref{lem:pruning-intervals} that we can split the real line into a fixed number of intervals such that as $p$ ranges over a single interval (as shown along the vertical axis in Figure~\ref{fig:p-intervals}), the dynamic programming (DP) table is fixed for all the samples, and therefore the resulting set of clusterings is fixed. In particular, for a fixed $\alpha$ interval, each of the $m$ samples has its own $O\left(n^{2(k+1)}k^{2k}\right)$ intervals of $p$, and when we merge them, we are left with $O\left(mn^{2(k+1)}k^{2k}\right)$ intervals such that as $p$ ranges over a single interval, each DP table for each sample is fixed, and therefore the resulting clustering for each sample is fixed. Since there are $O\left(md\right)$ such $\alpha$ intervals, each inducing $O\left(mn^{2(k+1)}k^{2k}\right)$ such $p$ intervals  in total, we have $O\left(dm^2n^{2(k+1)}k^{2k}\right)$ cells in $\R^2$ such that if $(\alpha,p)$ is in one fixed cell, the resulting clustering across all samples is fixed. If $\mathcal{H}_{\A, \mathcal{F}, \Phi}$ shatters $\sample$, then it must be that $2^m = O\left(dm^2n^{2(k+1)}k^{2k}\right)$, which means that $m = O\left(\log\left(dn^{2(k+1)}k^{2k}\right)\right) = O\left(\log d + k \log n\right)$.
\end{proof}

\begin{theorem}\label{thm:prune_merg_algo} Suppose the conditions of Theorem~\ref{thm:sample_comp_prun} hold.
Given a sample of size \[m = O\left(\left(\frac{H}{\epsilon}\right)^2\log \frac{dn}{\delta}\right)\] and a clustering objective $\clus$, it is possible to $(\epsilon,\delta)$-learn the class of algorithms $\A \times \F$ with respect to the cost function $\clus$. Moreover, this procedure is efficient if the following conditions hold:
\begin{enumerate}
\item The integer $k$ is constant, which ensures that the partition of $p$ values is polynomial in $n$.
\item The integer $d$ is polynomial in $n$, which ensures that the partition of $\alpha$ values is polynomial in $n$.
\item It is possible to efficiently compute the partition of $\alpha$ into intervals so that on a single interval $I$, for all $\alpha \in I$, the $m$ cluster trees returned by $\alpha$-linkage performed on $\sample$ are fixed.
\end{enumerate}
\end{theorem}
\begin{proof}
A technique for finding the empirically best algorithm from $\A \times \F$ follows naturally from Lemma~\ref{lem:pruning-intervals}; we partition the range of feasible values of $\alpha$ as described in Section~\ref{sec:simple}, and for each resulting interval of $\alpha$, we  find the fixed set of cluster trees on the samples. We then partition the values of $p$ as discussed in the proof for Lemma~\ref{lem:pruning-intervals}. For each interval of $p$, we use $\prune^{(p)}$ to prune the trees and determine the fixed empirical cost corresponding to that interval of $p$ and $\alpha$. This is illustrated in Figure~\ref{fig:p-intervals}.
\begin{figure}
  \centering
  \includegraphics[width=.75\textwidth] {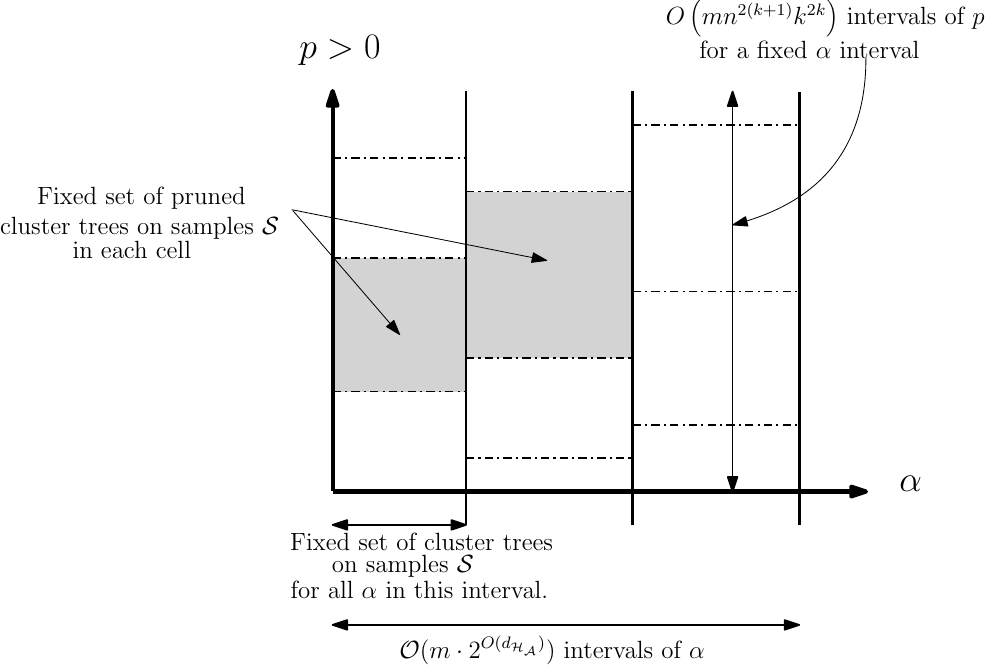}
  \caption{Illustration of the partition of the parameter space as described in the proof of Theorem~\ref{thm:prune_merg_algo}.}
  \label{fig:p-intervals}
\end{figure}
Iterating over all partitions of the parameter space, we can find parameters with the best empirical cost.
In Theorem~\ref{thm:sample_comp_prun}, we use Lemma~\ref{lem:pruning-intervals} to show that Pdim$\left(\mathcal{H}_{\A, \mathcal{F},\Phi}\right)=O\left(\log d + k \log n \right)$ and thus arrive at our sample complexity bound when $k$ is constant.
\end{proof}
\section{Discussion and open questions}
In this work, we show how
to learn near-optimal algorithms over several infinite, rich classes of SDP rounding algorithms and agglomerative clustering algorithms with dynamic programming. 
We provide computationally efficient and sample efficient learning algorithms for many of these problems and we push the boundaries of learning theory by developing techniques to
compute the pseudo-dimension of intricate, multi-stage classes
of IQP approximation algorithms and clustering algorithms. We derive tight pseudo-dimension bounds for the classes we study, which lead to strong sample complexity guarantees. We hope that our techniques will lead to theoretical guarantees in other areas where empirical methods for algorithm configuration have been developed.

There are many open avenues for future research in this area. In this work, we focused on algorithm families containing only computationally efficient algorithms. However, oftentimes in empirical AI research, the algorithm families in question contain procedures that are too slow to run to completion on many training instances. In this situation, we would not be able to determine the exact empirical cost of an algorithm on the training set. Could we still make strong, provable guarantees for application-specific algorithm configuration in this scenario? This work also leaves open the potential for data-dependent bounds over well-behaved distributions, such as those over clustering instances satisfying some form of stability, be it approximation stability, perturbation resilience, or so on.

\smallskip

\noindent \textbf{Acknowledgments.} This work was supported in part by grants NSF-CCF 1535967, NSF
CCF-1422910, NSF IIS-1618714, a Sloan Fellowship, a Microsoft Research Fellowship, a NSF Graduate Research Fellowship, a Microsoft Research Women's Fellowship, and a National
Defense Science and Engineering Graduate (NDSEG) fellowship.

We thank Sanjoy Dasgupta, Travis Dick, Anupam Gupta, and 
Ryan O'Donnell for useful discussions.

\bibliographystyle{plain}
\bibliography{bibliography}

\appendix

\section{Proofs from Section~\ref{sec:maxcut} on SDP-based methods for IQPs}\label{app:maxcut}
\slinLower*
\begin{proof}
In order to prove that the pseudo dimension of $\mathcal{H}_{slin}$ is at least $c \log n$ for some $c$, we must present a set $\mathcal{S} = \left\{ \left(\qp^{(1)}, \vec{Z}^{(1)}\right), \dots, \left(\qp^{(m)}, \vec{Z}^{(m)}\right)\right\}$ of $m = c\log n$ graphs and projection vectors that can be shattered by $\mathcal{H}_{slin}$. In other words, there exist $m$ witnesses $r_1, \dots, r_m$ and $2^m = n^c$ $s$ values $H = \left\{s_1, \dots, s_{n^c}\right\}$ such that for all $T \subseteq [m]$, there exists $s_T \in H$ such that if $j \in T$, then $\slin_{S_T}\left(\qp^{(j)}, \vec{Z}^{(j)}\right) > r_j$ and if $j \not\in T$, then $\slin_{S_T}\left(\qp^{(j)}, \vec{Z}^{(j)}\right) \leq r_j$.

To build $\sample$, we will use the same graph $\qp$ for all $\qp^{(j)}$ and we will vary $\vec{Z}^{(j)}$. We set $\qp$ to be the graph composed of $\lfloor n/4 \rfloor$ disjoint copies of $K_4$. If $n=4$, then a simple calculation confirms that an optimal max-cut SDP embedding of $\qp$ is \[\left\{\begin{pmatrix}
1\\
0\\
0\\
0\\
\end{pmatrix},\begin{pmatrix}
-1/3\\
2\sqrt{2}/3\\
0\\
0\\
\end{pmatrix}
,\begin{pmatrix}
-1/3\\
-\sqrt{2}/3\\
\sqrt{2/3}\\
0\\
\end{pmatrix}
,\begin{pmatrix}
-1/3\\
-\sqrt{2}/3\\
-\sqrt{2/3}\\
0\\
\end{pmatrix}\right\}.\]
Therefore, for $n > 4$, an optimal embedding is the set of $n$ vectors $SDP(\qp)$ such that for all $i \in \left\{0, \dots, \lfloor n/4\rfloor - 1\right\}$, \[\vec{e}_{4i + 1}, -\frac{1}{3}\vec{e}_{4i + 1} + \frac{2\sqrt{2}}{3}\vec{e}_{4i+2}, -\frac{1}{3}\vec{e}_{4i + 1} - \frac{\sqrt{2}}{3}\vec{e}_{4i+2}+ \sqrt{\frac{2}{3}}\vec{e}_{4i+3}, -\frac{1}{3}\vec{e}_{4i + 1} - \frac{\sqrt{2}}{3}\vec{e}_{4i+2}- \sqrt{\frac{2}{3}}\vec{e}_{4i+3}\] are elements $SDP(\qp)$.

We now define the set of $m$ vectors $\vec{Z}^{(j)}$. First, we set $\vec{Z}^{(1)}$ to be the vector \[\vec{Z}^{(1)} = \left(7^0,5\cdot 7^0,5\cdot 7^0,7^0,7^1,5\cdot 7^1,5\cdot 7^1,7^1,7^2,5\cdot 7^2,5\cdot 7^2,7^2,7^3,5\cdot 7^3,5\cdot 7^3,7^3,\dots\right).\] In other words, it is the concatenation the vector $7^i(1,5,5,1)$ for all $i > 0$. Next, $\vec{Z}^{(2)}$ is defined as \[\vec{Z}^{(2)} = \left(7^0,5\cdot 7^0,5\cdot 7^0,7^0,0,0,0,0,7^2,5\cdot 7^2,5\cdot 7^2,7^2,0,0,0,0,\dots\right),\] so $\vec{Z}^{(2)}$ is the same as $\vec{Z}^{(1)}$ for all even powers of 7, and otherwise its entries are 0. In a similar vein, \[\vec{Z}^{(3)} = \left(7^0,5\cdot 7^0,5\cdot 7^0,7^0,0,0,0,0,0,0,0,0,0,0,0,0,7^4,5\cdot 7^4,5\cdot 7^4,7^4, \dots\right).\] To pin down this pattern, we set $\vec{Z}^{(j)}$ to be the same as $\vec{Z}^{(1)}$ for all entries of the form $7^{i2^{j-1}}(1,5,5,1)$ for $i \geq 0$, and otherwise its entries are 0.

We set the following positive, increasing constants which will appear throughout the remaining analysis:
\[\begin{array}{lrl}
a =& (1,0,0,0) \cdot (1,5,5,1) &= 1\\
b =& (-1/3, -\sqrt{2}/3, \sqrt{2/3}, 0)\cdot (1,5,5,1) &= 5\sqrt{2/3} - \frac{5 \sqrt{2} +1}{3}\\
c =& (-1/3, 2\sqrt{2}/3, 0, 0) \cdot (1,5,5,1) &= \frac{10 \sqrt{2} - 1}{3}\\
d =& \left|(-1/3, -\sqrt{2}/3, \sqrt{2/3}, 0)\cdot (1,5,5,1)\right| &= 5\sqrt{2/3} + \frac{5 \sqrt{2} +1}{3}.\\
\end{array}\]

We also set $\tilde{c} = b+c+bc-d-bd-cd$ and we claim that the witnesses \begin{align*}
r_1 &= \frac{1}{2} - \frac{1}{3n}\left(\frac{b}{c^2}-1\right)&\\
r_j &=  \frac{1}{2} - \frac{\tilde{c}}{3n7^{2^{j-1}-2}d^2} &j > 1
\end{align*} are sufficient to prove that this set is shatterable, and we will spend the remainder of the proof showing that this is true.

Now, the domain of $\slin_{\qp, \vec{Z}^{(j)}}(s)$ can be split into intervals on which it has a simple, fixed form. These intervals begin at $1$ and have the form $\left[ 7^{i2^{j-1}}, 7^{(i+1)2^{j-1}}\right)$, for $i \geq 0$.
\begin{figure}
\includegraphics[width=.4\textwidth]{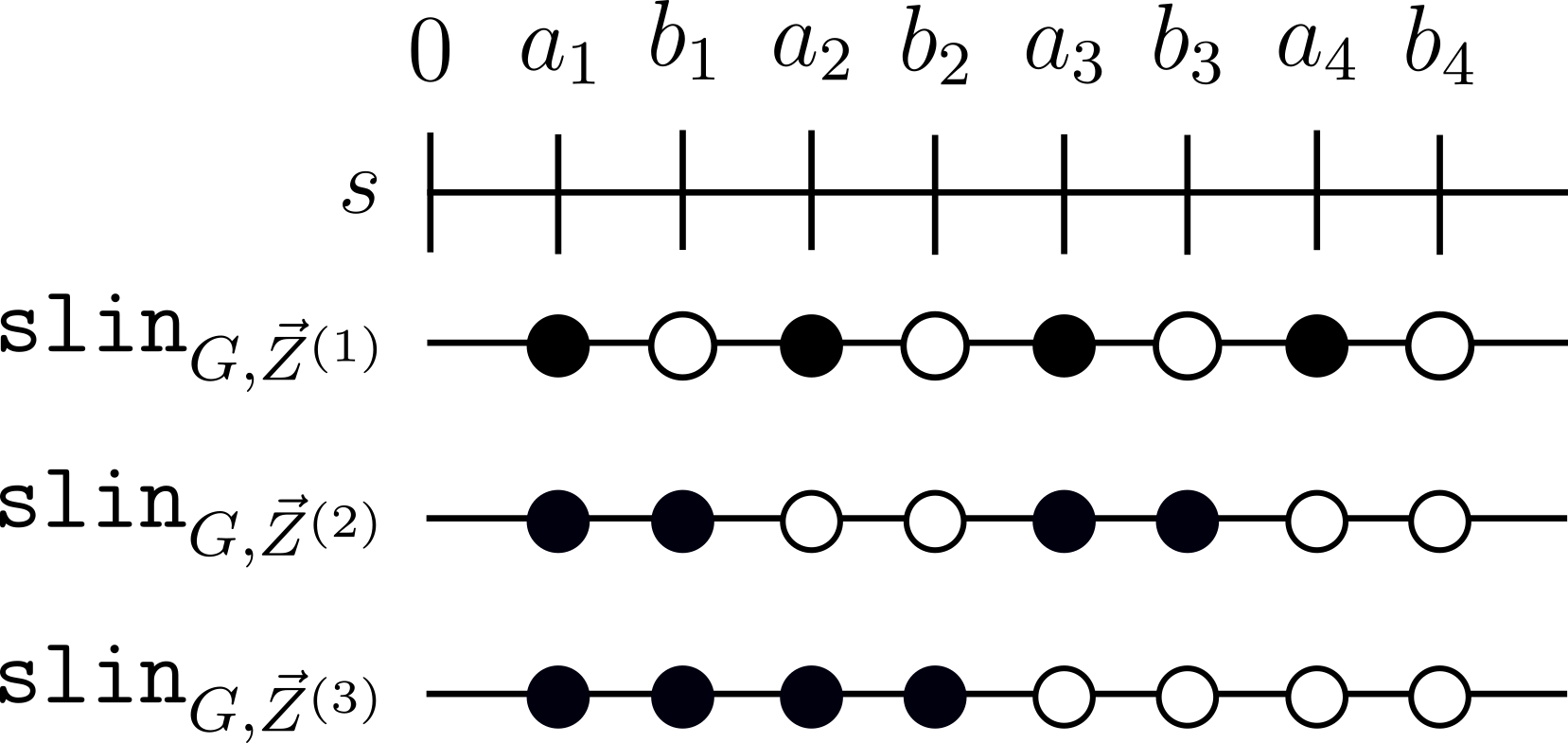}\centering
\caption{Depiction of $\slin_{\qp, \vec{Z}^{(j)}}(s)$ as $s$ increases from 0. A black dot means that $\slin_{\qp, \vec{Z}^{(j)}}(s) \geq r_j$ and a white dot means that $\slin_{\qp, \vec{Z}^{(j)}}(s) < r_j$. Here, $a_i = 7^{i-1}c$ and $b_i = 7^{i-1}d$.}\label{fig:dots}
\end{figure}
It is straightforward matter of calculations to check that for $s \in \left[ 7^{i2^{j-1}}, 7^{(i+1)2^{j-1}}\right)$, \begin{align*}&\slin_{\qp, \vec{Z}^{(j)}}(s) = \frac{1}{2}\\ &-\begin{cases}\frac{1}{3n}\left(\frac{1}{s^2}\left[\tilde{c}\sum_{k = 0}^{i-1}7^{2k2^{j-1}} \right]+ \frac{7^{2i2^{j-1}}}{s} - 1\right) &\text{if } s \in \left[ 7^{i2^{j-1}}, 7^{i2^{j-1}}b\right)\\
\frac{1}{3n}\left(\frac{1}{s^2}\left[\tilde{c}\sum_{k = 0}^{i-1}7^{2k2^{j-1}} +  7^{2i2^{j-1}}b\right] - 1\right) &\text{if } s \in \left[ 7^{i2^{j-1}}b, 7^{i2^{j-1}}c\right)\\
\frac{1}{3n}\left(\frac{1}{s^2}\left[\tilde{c}\sum_{k = 0}^{i-1}7^{2k2^{j-1}} +  7^{2i2^{j-1}}(b+c+bc)\right] - y\right)  &\text{if } s \in \left[ 7^{i2^{j-1}}c, 7^{i2^{j-1}}d\right)\\
\frac{1}{3n}\left(\frac{1}{s^2}\left[\tilde{c}\sum_{k = 0}^{i}7^{2k2^{j-1}} \right]\right) &\text{if } s \in \left[ 7^{i2^{j-1}}d, 7^{(i+1)2^{j-1}}\right),
\end{cases}\end{align*} where $y = \frac{7^{2i2^{j-1}}(1+b+c)}{s}$. (We note here that the power of 7 pattern was chosen so that these intervals are well defined, since $7^id < 7^{i+1}$.)

We call the following increasing sequence of numbers \emph{points of interest}, which we use to prove that this set is shattered: $\left\{7^0c, 7^0d, 7^1c, 7^1d, 7^2c, 7^2d, \dots, 7^ic, 7^id, \dots\right\}$

We make two claims about these points of interest: 
\begin{enumerate}
\item $\slin_{\qp, \vec{Z}^{(1)}}(s)$ is above its witness whenever $s=7^ic$ and it is below its witness whenever $s = 7^id$ for $i \geq 0$.
\item Let $j > 1$ and consider $\slin_{\qp, \vec{Z}^{(j)}}(s)$. There are $2^j$ points of interest per interval 
\[\left[ 7^{i2^{j-1}}, 7^{(i+1)2^{j-1}}\right).\] On the first half of these points of interest, $\slin_{\qp, \vec{Z}^{(j)}}(s)$ is greater than its witness and on the second half, $\slin_{\qp, \vec{Z}^{(j)}}(s)$ is less than its witness.
\end{enumerate}

These claims are illustrated by the dots in Figure~\ref{fig:dots}. Together, these claims imply that $\sample$ can be shattered because for any vector $\vec{b} \in \{0,1\}^m,$ there exists a point of interest $s$ such that $\slin_s(\sample)$ induces the binary labeling $\vec{b}$ on $\sample$.

The first claim is true because \begin{align*}\slin_{\qp, \vec{Z}^{(1)}}\left(7^ic\right) &= \frac{1}{2} - \frac{1}{3n}\left(\frac{1}{7^{2i}c^2}\left[\tilde{c}\sum_{k = 0}^{i-1}7^{2k} +  7^{2i}b\right] - 1\right)\\
&= \frac{1}{2} - \frac{1}{3n}\left(\frac{1}{7^{2i}c^2}\left[\tilde{c}\cdot\frac{7^{2i}-1}{7^{2}-1} +  7^{2i}b\right] - 1\right),\end{align*} which is an increasing function of $i$, so it is minimized when $i=0$, where \[\slin_{\qp, \vec{Z}^{(1)}}\left(7^0c\right)=\frac{1}{2} - \frac{1}{3n}\left(\frac{b}{c^2} - 1\right) = r_1\] so $\slin_{\qp, \vec{Z}^{(1)}}\left(7^ic\right)$ is always at least its witness. Further, \begin{align*}\slin_{\qp, \vec{Z}^{(1)}}\left(7^id\right) &=\frac{1}{2} - \frac{1}{3n}\left(\frac{1}{7^{2i}d^2}\left[\tilde{c}\sum_{k = 0}^{i}7^{2k} \right]\right)\\
&= \frac{1}{2} - \frac{1}{3n}\left(\frac{1}{7^{2i}d^2}\left[\tilde{c}\cdot \frac{7^{2(i+1)}-1}{48} \right]\right)),\end{align*} which is again an increasing function in $i$, with a limit of \[\frac{1}{2} - \frac{49\tilde{c}}{144nd^2} < r_1.\] Therefore, $\slin_{\qp, \vec{Z}^{(1)}}\left(7^id\right)$ is always less than its witness, and we may conclude that the first claim is always true.

For the second claim, notice that \[7^{i2^{j-1}}c < 7^{i2^{j-1}}d < 7^{i2^{j-1}+1}c < 7^{i2^{j-1}+1}d < 7^{i2^{j-1}+2}c \cdots < 7^{i2^{j-1} + 2^{j-1}}c = 7^{(i+1)2^{j-1}}c,\] so there are $2^j$ points of interest per interval $\left[ 7^{i2^{j-1}}c, 7^{(i+1)2^{j-1}}c\right)$, as claimed. The first two points of interest, $ 7^{i2^{j-1}}c$ and $7^{i2^{j-1}}d$, fall in an interval where $\slin_{\qp, \vec{Z}^{(j)}}$ is decreasing in $s$. Therefore, it is minimized when $s = 7^{i2^{j-1}}d$, where
\begin{align*} \slin_{\qp, \vec{Z}^{(j)}}\left(7^{i2^{j-1}}d\right) &= \frac{1}{2} - \frac{1}{3n}\left(\frac{1}{7^{2i2^{j-1}}d^2}\left[\tilde{c}\sum_{k = 0}^{i}7^{2k2^{j-1}} \right]\right)\\
&= \frac{1}{2} - \frac{1}{3n}\left(\frac{1}{7^{2i2^{j-1}}d^2}\left[\tilde{c}\cdot \frac{7^{(i+1)2^j}-1}{7^{2^j}-1}\right]\right).\end{align*} Simple calculations show that $\slin_{\qp, \vec{Z}^{(j)}}\left(7^{i2^{j-1}}d\right)$ is an increasing function in $i$, so it is minimized when $i=0$, where $\slin_{\qp, \vec{Z}^{(j)}}\left(d\right)=\frac{1}{2} - \frac{\tilde{c}}{3nd^2} > r_j$, as desired.

The remaining points of interest fall in the interval $\left[ 7^{i2^{j-1}}d, 7^{(i+1)2^{j-1}}\right)$, so $\slin_{\qp, \vec{Z}^{(j)}}(s)$ has the form $\frac{1}{2} - \frac{1}{3n}\left(\frac{1}{s^2}\left[\tilde{c}\sum_{k = 0}^{i}7^{2k2^{j-1}} \right]\right)$. This segment of the function has a negative derivative, so it is decreasing.

If $j=2$, then the points of interest we already considered, $7^{i2^{j-1}}c$ and $7^{i2^{j-1}}d$, make up half of the $2^j$ points of interest in the interval $\left[ 7^{i2^{j-1}}c, 7^{(i+1)2^{j-1}}c\right)$. Therefore, we only need to show that when $s$ equals $7^{i2^{j-1} + 1}c$ and $7^{i2^{j-1}+1}d$, then $\slin_{\qp, \vec{Z}^{(j)}}(s)$ is less than its witness. As we saw, $\slin_{\qp, \vec{Z}^{(j)}}$ is decreasing on this segment, so it is enough to show that $\slin_{\qp, \vec{Z}^{(j)}}\left(7^{i2^{j-1} + 1}c\right)$ is less than its witness. To this end, \begin{align*}\slin_{\qp, \vec{Z}^{(j)}}\left(7^{i2^{j-1} + 1}c\right) &= \frac{1}{2} - \frac{1}{3n}\left(\frac{1}{7^{2i2^{j-1} + 2}c^2}\left[\tilde{c}\sum_{k = 0}^{i}7^{2k2^{j-1}} \right]\right)\\
&= \frac{1}{2} - \frac{1}{3n}\left(\frac{1}{7^{2i2^{j-1} + 2}c^2}\left[\tilde{c}\cdot \frac{7^{(i+1)2^j}-1}{7^{2^j}-1} \right]\right).
\end{align*}

This is an increasing function of $i$ with a limit of $\frac{1}{2} - \frac{1}{3n}\left(\frac{1}{7^{ 2}c^2}\left[\tilde{c}\cdot \frac{7^{2^j}}{7^{2^j}-1}\right]\right) < r_j$ when $j=2$. Therefore, when $s$ equals $7^{i2^{j-1} + 1}b$ and $7^{i2^{j-1}+1}c$, then $\slin_{\qp, \vec{Z}^{(j)}}(s)$ is less than its witness.

Finally, if $j>2$, since $\slin_{\qp, \vec{Z}^{(j)}}(s)$ is decreasing on the interval $\left[ 7^{i2^{j-1}}c, 7^{(i+1)2^{j-1}}\right)$, we must only check that at the $\left(2^{j-1}-1\right)^{th}$ point of interest $\left(7^{i2^{j-1}+2^{j-2}-1}d\right)$, $\slin_{\qp, \vec{Z}^{(j)}}(s)$ is greater than its witness and at the $\left(2^{j-1}+1\right)^{th}$ point of interest $\left(7^{i2^{j-1}+2^{j-2}}c\right)$, $\slin_{\qp, \vec{Z}^{(j)}}(s)$ is less than its witness. To this end, \begin{align*}
\slin_{\qp, \vec{Z}^{(j)}}\left(7^{i2^{j-1}+2^{j-2}-1}d\right) &= \frac{1}{2} - \frac{1}{3n}\left(\frac{1}{7^{i2^{j}+2^{j-1}-2}d^2}\left[\tilde{c}\sum_{k = 0}^{i}7^{2k2^{j-1}} \right]\right)\\
&= \frac{1}{2} - \frac{1}{3n}\left(\frac{1}{7^{i2^{j}+2^{j-1}-2}d^2}\left[\tilde{c}\cdot \frac{7^{(i+1)2^j}-1}{7^{2^j}-1} \right]\right).
\end{align*} This function is increasing in $i$, so it is minimized when $i = 0$, where \begin{align*}
\slin_{\qp, \vec{Z}^{(j)}}\left(7^{2^{j-2}-1}d\right) &= \frac{1}{2} - \frac{1}{3n}\left(\frac{1}{7^{2^{j-1}-2}d^2}\left[\tilde{c}\cdot \frac{7^{2^j}-1}{7^{2^j}-1} \right]\right)\\
&= \frac{1}{2} - \frac{1}{3n}\left(\frac{\tilde{c}}{7^{2^{j-1}-2}d^2}\right) = r_j.\\
\end{align*} Therefore, $\slin_{\qp, \vec{Z}^{(j)}}\left(7^{i2^{j-1}+2^{j-2}-1}d\right) \geq r_j$ for all $i$.

Next, \begin{align*}
\slin_{\qp, \vec{Z}^{(j)}}\left(7^{i2^{j-1}+2^{j-2}}c\right) &= \frac{1}{2} - \frac{1}{3n}\left(\frac{1}{7^{i2^{j}+2^{j-1}}c^2}\left[\tilde{c}\sum_{k = 0}^{i}7^{2k2^{j-1}} \right]\right)\\
&= \frac{1}{2} - \frac{1}{3n}\left(\frac{1}{7^{i2^{j}+2^{j-1}}c^2}\left[\tilde{c}\cdot \frac{7^{(i+1)2^j}-1}{7^{2^j}-1} \right]\right).
\end{align*} which is an increasing function in $i$, with a limit of \[\frac{1}{2} - \frac{\tilde{c}7^{2^{j-1}}}{3nc^2\left(7^{2^j}-1\right)}\] as $i$ tends toward infinity. Therefore, \[\slin_{\qp, \vec{Z}^{(j)}}\left(7^{i2^{j-1}+2^{j-2}}\right) \leq \frac{1}{2} - \frac{\tilde{c}7^{2^{j-1}}}{3nc^2\left(7^{2^j}-1\right)} < r_j\] for all $i$, so the second claim holds.
\end{proof}
\section{More algorithm classes for {\sc MaxQP}} \label{app:maxcut-more}
\subsection{$\tilde{\epsilon}$-discretized functions for max-cut}\label{sec:discretized}

The class of $\tilde{\epsilon}$-discretized rounding functions are a finite yet rich class of functions for the RPR$^2$ paradigm. They were introduced by O'Donnell and Wu~\cite{o2008optimal} as a tool for characterizing the \emph{SDP gap curve} for the max-cut problem, which we now define. Let $G$ be a graph with $n$ nodes and a weight matrix $A \in \R^{n \times n}$, where $a_{ij}$ is the weight between node $i$ and $j$ (with $a_{ij} = 0$ if no edge exists).
Recall that the binary quadratic programming formulation of the max-cut problem is \begin{equation*}\text{maximize } \sum_{i,j \in [n]} a_{ij}\left(\frac{1}{2} - \frac{1}{2}x_ix_j\right) \qquad \text{subject to }x_i \in \{-1, 1\} \forall i \in [n].\end{equation*} After all, if $x_i$ and $x_j$ are on the same side of the cut, then $x_i = x_j$, so  $\frac{1}{2} - \frac{1}{2}x_ix_j = 0$. Meanwhile, if $x_i$ and $x_j$ are on opposite sides of the cut, then $x_i \not= x_j$, so  $\frac{1}{2} - \frac{1}{2}x_ix_j = 1$. We assume, without loss of generality, that $\sum_{i,j \in [n]} a_{ij} \leq 1$. Let $Sdp(G)$ be the objective value of the SDP relaxation of $G$. It is the objective value of the solution to: \begin{equation*}\text{maximize } \sum_{i,j \in [n]} a_{ij}\left(\frac{1}{2} - \frac{1}{2}\vec{u}_i\cdot \vec{u}_j\right) \qquad \text{subject to }\vec{u}_i \in S^{n-1}.\end{equation*}

For $c \in [0,1]$, the SDP gap curve $\text{Gap}_{SDP}(c)$ is a function that measures the smallest optimal max-cut value among all graphs such that Sdp($G) \geq c$. In other words, given that Sdp($G) \geq c$, we are guaranteed that the optimal max-cut value of $G$ is at least $\text{Gap}_{SDP}(c)$. Formally,

\begin{definition}
For $\frac{1}{2} \leq s \leq c \leq 1$, we call the pair $(c,s)$ an \emph{SDP gap} if there exists a graph $G$ with Sdp($G) \geq c$ and Opt$(G) \leq s$. We define the \emph{SDP gap curve} by \[\text{Gap}_{SDP}(c) = \inf\{s \ | \ (c,s) \text{ is an SDP gap}\}.\]
\end{definition}

O'Donnell and Wu~\cite{o2008optimal} prove that if $G$ is a graph such that Sdp($G) \geq c$, if one runs Algorithm~\ref{alg:GW} iteratively with all $\tilde{\epsilon}$-discretized rounding functions (defined below), then with high probability, at least one will result in a cut with value $\text{Gap}_{SDP}(c) - \tilde{\epsilon}$.

\begin{definition}[$\tilde{\epsilon}$-discretized rounding function~\cite{o2008optimal}]
Given $\tilde{\epsilon} >0$, let $\mathcal{I}_{\tilde{\epsilon}}$ denote the (finite) partition of $\R\setminus \{0\}$ into intervals, \[\mathcal{I}_{\tilde{\epsilon}} = \left\{\pm(-\infty, -B], \pm (-B, -B + \tilde{\epsilon}^2], \pm (-B + \tilde{\epsilon}^2, -B + 2\tilde{\epsilon}^2], \dots, \pm (-2\tilde{\epsilon}^2, -\tilde{\epsilon}^2], \pm (-\tilde{\epsilon}^2, \tilde{\epsilon}^2)\right\},\] where $B = B(\tilde{\epsilon})$ is the smallest integer multiple of $\tilde{\epsilon}^2$ exceeding $\sqrt{2\ln(1/\tilde{\epsilon})}$. We say that a function $r:\R \to [-1,1]$ is \emph{$\tilde{\epsilon}$-discretized} if the following hold:
\begin{enumerate}
\item $r$ is identically $-1$ on $(-\infty, -B]$, 0 at 0, and identically 1 on $[B, \infty)$.
\item $r$'s values on the intervals in $\mathcal{I}_{\tilde{\epsilon}}$ are from the set $\tilde{\epsilon} \Z \cap (-1,1)$.
\end{enumerate}
\end{definition}

Note that there are $2^{O(1/\tilde{\epsilon}^2)}$ $\tilde{\epsilon}$-discretized functions. O'Donnell and Wu~\cite{o2008optimal} prove the following guarantee.

\begin{theorem}[Corollary 5.4 in \cite{o2008optimal}]\label{thm:disc_alg}
There is an algorithm which, given any graph $G$ with Sdp($G) \geq c$ and any $\tilde{\epsilon} >0$, runs in time poly$(|V|)2^{O(1/\tilde{\epsilon}^2)}$ and with high probability outputs a proper cut in $G$ with value at least $\text{Gap}_{SDP}(c) - \tilde{\epsilon}$.
\end{theorem}

Namely, the algorithm takes as input a graph, runs Algorithm~\ref{alg:GW} using all $\tilde{\epsilon}$-discretized rounding functions, and returns the cut with the maximum value. We define $\cost_{\tilde{\epsilon}}(G)$ to be the value of the resulting cut.

It is well-known that the pseudo-dimension of a finite function class $\mathcal{F}$ has pseudo-dimension $\log |\mathcal{F}|$. This immediately implies the following theorem.

\begin{theorem}
Given an input sample of size $m = O\left(\frac{1}{\epsilon^2}\left(\frac{1}{\tilde{\epsilon}^2} +\log \frac{1}{\delta}\right)\right)$ there exists an algorithm that $(\epsilon,\delta)$-learns the class of $\tilde{\epsilon}$-discretized rounding functions with respect to the cost function $\cost_{\tilde{\epsilon}}$.
\end{theorem}

\subsection{Outward rotations}\label{sec:outward}
Next we study a class of ``outward rotation" based algorithms proposed  by Zwick \cite{zwick1999outward}.  For the max-cut problem, outward rotations are proven to work better than the random hyperplane technique of Goemans and Williamson \cite{goemans1995improved} on graphs with ``light'' max-cuts where the max-cut does not constitute a large proportion of the edges.

The class of outward rotation algorithms is characterized by an angle $\gamma \in [0, \pi/2]$. Varying $\gamma$ results in a range of algorithms that interpolate between the random hyperplane technique of Goemans and Williamson and the na\"ive approach of outputting a random binary assignment \cite{goemans1995improved}. In essence, an outward rotation algorithm extends the optimal SDP embedding $\vec{u}_1, \dots, \vec{u}_n$ from $\mathbb{R}^n$ to $\mathbb{R}^{2n}$: the original embedding is first carried over to the first $n$ co-ordinates of a $2n$-dimensional space  while the remaining co-oordinates are set to zero. 
Suppose $\vec{e}_{n+1}, \vec{e}_{n+2}, \dots, \vec{e}_{2n} \in \R^{2n}$ are the orthonormal vectors along each of the last $n$ co-ordinates (i.e., the $(n+i)^{th}$ co-ordinate of $\vec{e}_{n+i}$ is 1 every other co-ordinate is 0). Each embedding $\vec{u}_i$ is rotated ``out'' of the original $n$-dimensional space towards $\vec{e}_{n+i}$ by an angle of $\gamma$. After performing these outward rotations, the new embedding is projected onto a random hyperplane in $\mathbb{R}^{2n}$.   The binary assignment is then defined deterministically based on the sign of the projections like in the GW algorithm \cite{goemans1995improved}.  Intuitively, the parameter $\gamma$  determines to what extent the SDP embedding is used to determine the final assignment of each variable $x_i$. See Algorithm~\ref{alg:owr} for the pseudo-code.

\begin{algorithm}
\caption{SDP rounding algorithm using $\gamma$-outward rotation}\label{alg:owr}
\begin{algorithmic}[1]
\Require Matrix $\qp \in \mathbb{R}^{n \times n}$
\State Solve the SDP (\ref{eq:SDP}) for the optimal embedding $U = \left\{\vec{u}_1, \dots, \vec{u}_n\right\}$ of $\qp$.
\State Define a new embedding $\vec{u}_i'$ in $\mathbb{R}^{2n}$ such that the first $n$ co-ordinates correspond to $\vec{u}_i \cos \gamma$ and the following $n$ co-ordinates are set to $0$ except the $(n+i)^{th}$ co-ordinate which is set to $\sin \gamma$.
\State Choose a random vector $\vec{Z} \in \R^{2n}$ according to the $2n$-dimensional Gaussian distribution.\label{step:OWR_round}
\State For each decision variable $x_i$, assign $x_i=\sign \left(\langle \vec{u}_i', \vec{Z} \rangle\right).$\label{step:OWR_assignment}
\Ensure $x_1, \hdots, x_n$.
\end{algorithmic}
\end{algorithm}

Our goal is to design an algorithm that learns a nearly-optimal outward rotation parameter $\gamma$. As in Section~\ref{sec:maxcut}, let $\dist$ be an unknown distribution over matrices $A$ and let $\mathcal{Z}$ be the $2n$-dimensional Gaussian distribution. We want to find a parameter $\gamma$ such that in expectation over $A \sim \dist$ and in expectation over $\vec{Z} \sim \mathcal{Z}$, the objective value $\sum_{i, j \in [n]} a_{ij} x_i x_j$ is maximized\footnote{Note that unlike Section~\ref{sec:maxcut}, once the vector $\vec{Z}$ is fixed, the final assignment of each variable $x_i$ in Step~\ref{step:OWR_assignment} of Algorithm~\ref{alg:owr} is deterministic.}. We call this the \emph{true quality of the parameter $\gamma$}. In other words, the true quality of the parameter $\gamma$ is $\E_{A, \vec{Z} \sim \dist \times \mathcal{Z}}\left[\sum_{i,j}a_{ij}\sign\left(\langle \vec{u}_i'  , \vec{Z}\rangle\right)\sign\left(\langle \vec{u}_j', \vec{Z}\rangle\right)\right].$ Our goal is to find a parameter whose true quality is (nearly) optimal.

We do not know the distribution $\dist$ over matrices, so we also need to define the \emph{empirical quality} of the parameter $\gamma$ given a set of samples. As in Section~\ref{sec:maxcut}, we will then show that this empirical quality approaches the true quality as the number of samples grows. Thus, a parameter which is nearly optimal on average over the samples will be nearly optimal in expectation as well. The definition of a parameter's empirical quality depends on a function $\owr_{\gamma}$ which we now define. Let $\owr_{\gamma}(A, \vec{Z})$ denote the objective value of the solution returned by Algorithm~\ref{alg:owr} given $\qp$ as input when it uses the hyperplane $\vec{Z}$ in Step~\ref{step:OWR_round}. Explicitly, $\owr_{\gamma} \left(\qp, \vec{Z} \right) = \sum_{i,j}a_{ij}\sign\left(\langle \vec{u}_i'  , \vec{Z}\rangle\right)\sign\left(\langle \vec{u}_j', \vec{Z}\rangle\right)$. By definition, the true quality of the parameter $\gamma$ equals \[\E_{A, \vec{Z} \sim \dist \times \mathcal{Z}}\left[\owr_\gamma(A, \vec{Z})\right].\]

We now define the empirical quality of a parameter $\gamma$ as follows. Given a set of samples $\left(A^{(1)}, \vec{Z}^{(1)}\right), \dots, \left(A^{(m)}, \vec{Z}^{(m)}\right) \sim \dist \times \mathcal{Z}$, we define the empirical quality of the parameter $\gamma$ to be $\frac{1}{m} \sum_{i = 1}^m \owr_{\gamma}\left(A^{(i)}, \vec{Z}^{(i)}\right)$. Bounding the pseudo-dimension of the class of functions $\mathcal{H}_{\owr} =  \left\{\owr_\gamma:  \gamma \in [0, \pi/2] \right\}$, we bound the number of samples $m$ sufficient to ensure that with high probability, for all parameters $\gamma$, the true quality of $\gamma$ nearly matches its expected quality. In other words, $\frac{1}{m}\sum_{i = 1}^m \owr_{\gamma}\left(A^{(i)}, \vec{Z}^{(i)}\right)$ nearly matches $\E_{A, \vec{Z} \sim \dist \times \mathcal{Z}}\left[\owr_{\gamma}(A, \vec{Z})\right]$. Thus, if we find the parameter $\hat{\gamma}$ that maximizes $\frac{1}{m}\sum_{i = 1}^m \owr_{\gamma}\left(A^{(i)}, \vec{Z}^{(i)}\right)$, then the true quality of $\hat{\gamma}$ is nearly optimal. In other words, $\max_{\gamma \in [0, \pi/2]}\E_{A, \vec{Z} \sim \dist \times \mathcal{Z}} \left[\owr_{\gamma}(A, \vec{Z})\right]$ is close to $\E_{A, \vec{Z} \sim \dist \times \mathcal{Z}} \left[\owr_{\hat{\gamma}}(A, \vec{Z})\right]$.

We first prove in Section \ref{sec:outward-unif-conv} that the pseudo-dimension of $\mathcal{H}_{\owr}$ is $O(\log n)$. Next, in Section~\ref{sec:outward-ERM} we present an efficient learning algorithm.

\subsubsection{The pseudo-dimension of the class of outward rotation based algorithms}
\label{sec:outward-unif-conv}

We show an upper bound on the pseudo-dimension of the class of outward rotation based algorithms. 
As in Section~\ref{sec:maxcut}, we use the following notation: given a tuple $\left(\qp, \vec{Z}\right)$, let $\owr_{\qp, \vec{Z}} : [0, \pi/2] \to \R$ be defined such that $\owr_{\qp, \vec{Z}}(\gamma) = \owr_\gamma\left(\qp, \vec{Z}\right)$. 

\begin{theorem}
\label{thm:outward-pdim}
 Pdim$(\mathcal{H}_{\textnormal{\owr}}) = O(\log n)$.
\end{theorem}

\begin{proof}
First, we claim that for any matrix $A$ and any vector $\vec{Z}$, $\owr_{A, \vec{Z}}$ is piecewise constant with at most $n$ discontinuities. Observe that as $\gamma$ increases, $\owr_{\qp, \vec{Z}}(\gamma)$ will change only when $\sign\left(\langle \vec{u}_i'  , \vec{Z}\rangle\right)$ changes for some $i \in [n]$. Now, note that $\langle \vec{u}_i'  , \vec{Z}\rangle = \langle \vec{u}_i  , \vec{Z}_{[1,\hdots,n]} \rangle \cos \gamma + z_{n+i} \sin \gamma$ where  $ \vec{Z}_{[1,\hdots,n]}$ is the projection of $\vec{Z}$ over the first $n$ co-ordindates.   Clearly, $\langle \vec{u}_i' , \vec{Z}\rangle$ is a monotone function in $\gamma \in [0,\pi/2]$ and attains zero at \[ \gamma =
\tan^{-1} \left( - \frac{\langle \vec{u}_i , \vec{Z}_{[1,\hdots,n]}\rangle}{z_{n+i} }\right).
\]
This implies that  for each $i \in [n]$, $\sign\left(\langle \vec{u}_i' , \vec{Z}\rangle\right)$  changes at most once within $[0,\pi/2]$. Therefore, $\owr_{\qp, \vec{Z}}(\gamma)$ is a piecewise constant function with at most $n$ discontinuities.

Next, suppose  Pdim$(\mathcal{H}_{\textnormal{\owr}}) = m$ and $\sample = \left\{\left(\qp^{(1)}, \vec{Z}^{(1)}\right), \dots, \left(\qp^{(m)}, \vec{Z}^{(m)}\right)\right\}$ is shatterable. This means that there exist $m$ thresholds $\{r_1, \dots, r_m\} \subset \R$ such that for each $T \subseteq [m]$, there exists a parameter $\gamma_T$ such that
$\owr_{\qp^{(i)}, \vec{Z}^{(i)}}\left(\gamma_T\right) > r_i$
if and only if $i \in T$.
Since $\owr_{\qp^{(i)}, \vec{Z}^{(i)}}$ is piecewise constant with $n$ discontinuities, there are $n$ intervals partitioning $[0, \pi/2]$ such that within a given interval, $\owr_{\qp^{(i)}, \vec{Z}^{(i)}}(\gamma)$ is invariant, so it is either greater than or less than $r_i$ (but not both). Therefore, there are at most $mn$ values of $\gamma$ defining $mn+1$ intervals such that the labels given by the witnesses for the set of $m$ samples is identical within each interval. In other words, at most $mn+1$ distinct labelings of $\sample$ are achievable for any choice of the witnesses.  However, since $\sample$ is shatterable, we need $2^m < mn + 1$. Therefore, Pdim$(\mathcal{H}_{\textnormal{\owr}}) = O(\log n)$.
\end{proof}

\subsubsection{A learning algorithm}
\label{sec:outward-ERM}

We now present
Algorithm \ref{alg:outward-ERM} that efficiently learns the best value of $\gamma$ for outward rotation with respect to samples drawn from $\mathcal{D} \times \mathcal{Z}$.

\begin{algorithm}
\caption{An algorithm for finding the empirical value maximizing $\gamma$}\label{alg:outward-ERM}
\begin{algorithmic}[1]
\Require Sample $\sample = \left\{\left(\qp^{(1)}, \vec{Z}^{(1)}\right), \dots, \left(\qp^{(m)}, \vec{Z}^{(m)}\right)\right\}$
\State Solve for $\left\{U^{(1)}, \dots, U^{(m)}\right\}$ the optimal SDP embeddings for $\qp^{(1)}, \dots, \qp^{(m)}$, where $U^{(i)} = \left\{\vec{u}_1^{(i)}, \dots, \vec{u}_n^{(i)}\right\}$.

\State Let $T = \left\{\gamma_1, \dots, \gamma_{|T|}\right\}$ be the set of all values $\gamma \in [0,\pi/2]$ such that there exists a pair of indices $i \in [n], j \in [m]$ with $\tan^{-1} \left(- \frac{\langle \vec{u}_i^{(j)}  , \vec{Z}^{(j)}_{[1,\hdots,n]}\rangle}{Z^{(j)}_{n+i} }\right) = \gamma.$\label{step:outward-threshold}
\State Let $\hat{\gamma} = \underset{\gamma \in T \cup \{\pi/2\}}{\mbox{argmax}}\left\{\frac{1}{m} \sum_{i = 1}^m \owr_\gamma\left(\qp^{(i)},\vec{Z}^{(i)}\right)\right\}$.\label{step:outward-global}
\Ensure $\hat{\gamma}$
\end{algorithmic}
\end{algorithm}

\begin{lemma}
\label{lem:owr_erm}
Algorithm~\ref{alg:outward-ERM} produces the value $\hat{\gamma}$ which maximizes $\frac{1}{m} \sum_{i = 1}^m \owr_{\qp^{(i)},\vec{Z}^{(i)}}(\gamma)$ given the sample $\sample = \left\{\left(\qp^{(1)}, \vec{Z}^{(1)}\right), \dots, \left(\qp^{(m)}, \vec{Z}^{(m)}\right)\right\}$. Algorithm~\ref{alg:outward-ERM} has running time polynomial in $m$ and $n$.
\end{lemma}

\begin{proof}
Recall from the proof of Theorem~\ref{thm:outward-pdim} that the set $T$ in Step~\ref{step:outward-threshold} of Algorithm~\ref{alg:outward-ERM} defines $T+1$ intervals over $[0,\pi/2]$ within each of which the behavior of any $\gamma$ is constant across all samples in $\sample$. Therefore, we only need to examine the performance of  a single value of $\gamma$ within each interval to exhaustively evaluate all possibilities, and single out the best one. 

Also observe that since there are only $O(mn)$ values in $T$ (in Step \ref{step:outward-threshold}) and since computing the binary assignment on a set of $m$ instances for a particular value of $\gamma$ takes polynomial time in $m$ and $n$, Step~\ref{step:outward-global} also takes only polynomial time in $m$ and $n$.
\end{proof}

Together with Theorem~\ref{thm:outward-pdim}, Lemma \ref{lem:owr_erm} implies the following theorem. 

\begin{theorem}
Let $H = \sup_{A \in \textnormal{supp}(\dist)} ||A||_c$, where $||\cdot||_c$ is the cut norm and supp$(\dist)$ denotes the support of $\dist$.\footnote{$H$ is thus an upper bound on the value of $\owr_{\gamma}(\qp, \vec{Z})$ for any $\gamma \in [0, \pi/2]$ and any $(\qp, \vec{Z})$ in the support of $\mathcal{D} \times \mathcal{Z}$.} Given a sample of size $m = O\left(\left(\frac{H}{\epsilon}\right)^2 \left(\log \left({n}\right) + \log\frac{1}{\delta}\right)\right)$ drawn from $\left(\mathcal{D} \times \mathcal{Z}\right)^m$, let $\hat{\gamma}$ be the output of Algorithm~\ref{alg:outward-ERM}. With probability at least $1 - \delta$, the true quality of $\hat{\gamma}$ is $\epsilon$-close optimal: \[\max_{\gamma \in [0, \pi/2]} \E_{A \sim \dist, \vec{Z} \sim \mathcal{Z}}\left[\owr_{\gamma}(A, \vec{Z})\right] - \E_{A \sim \dist, \vec{Z} \sim \mathcal{Z}}\left[\owr_{\hat{\gamma}}(A, \vec{Z})\right] \leq \epsilon.\]
\end{theorem}

\section{Proofs from Section~\ref{sec:clustering} on agglomerative algorithms with dynamic programming}\label{app:clustering}
\alphaJustification*

\begin{proof}
We give a general proof for all three classes 
$\mathcal{A}_1$, $\mathcal{A}_2$, and $\mathcal{A}_3$. We will point out a few places in the proof where the details for
$b=1,2,3$ are different, but the general structure of the argument is the same.
For each value of $b$, we construct a single clustering instance $\V=(V,d)$ that has the desired property; 
the distribution $\mathcal{D}$ is merely the single clustering instance with probability 1.

Consider some permissible value of $\alpha$, denoted $\alpha^*$.
Set $k=4$ and $n=210$.
The clustering instance consists of two well-separated `gadgets' of two clusters each. The class $\A_b$ results in different 2-clusterings of the first gadget depending on whether $\alpha \leq \alpha^*$ or not. Similarly, $\A_b$ results in different 2-clusterings of the second gadget depending on whether $\alpha \geq \alpha^*$ or not. By ensuring that for the first gadget $\alpha \leq \alpha^*$ results in the lowest cost 2-clustering, and for the second gadget $\alpha \geq \alpha^*$ results in the lowest cost 2-clustering, we ensure that $\alpha= \alpha^*$ is the optimal parameter overall.

The first gadget is as follows. We define five points $a_1,b_1,c_1,x_1$ and $y_1$.
For the sake of convenience, we will group the remaining points into four sets $A_1$, $B_1$, $X_1$, and $Y_1$ each containing $25$ points.
We set the distances as follows:
$d(a_1,b_1)=d(x_1,y_1)=1$, $d(a_1,c_1)=1.1$, and $d(b_1,c_1)=1.2$.
For $a\in A_1\cup B_1$, $d(c_1,a)=1.51$ and $d(a_1,a)=d(b_1,a)=1.6$. 
For $x\in X_1\cup Y_1$, $d(x_1,x)=d(y_1,x)=1.6$. 
For $a\in A_1$, $b\in B_1$, $x\in X_1$, and $y\in Y_1$, $d(a,b)=d(x,y)=1.6$.
We also define special points $x_1^*\in X_1$ and $y_1^*\in Y_1$, 
which have the same distances as the rest of the points in $X_1$ and $Y_1$ respectively,
except that $d(x_1,x_1^*)=1.51$ and $d(y_1,y_1^*)=1.51$.
If two points $p$ and $q$ belong to the same set ($A_1$, $B_1$, $X_1$, or $Y_1$), then $d(p,q)=1.5$.

The distances $d(x_1,c_1)$ and $d(y_1,c_1)$ are defined in terms of $b$ and $\alpha^*$, but they will always be between
1.1 and 1.2. For $b=1$, we set
$d(x_1,c_1)=d(y_1,c_1)=1.2-.1\cdot\alpha^*$.  
For $b=2$ and $b=3$, $d(x_1,c_1)=d(y_1,c_1)=((1.1^{\alpha^*}+1.2^{\alpha^*})/2)^{\frac{1}{\alpha^*}}$.

So far, all of the distances we have defined are in $[1,2]$, therefore they trivially satisfy the triangle inequality.
We set all of the rest of the distances to be the maximum distances allowed under the triangle inequality.
Therefore, the triangle inequality holds over the entire metric.

Now, let us analyze the merges caused by $\A_b(\alpha)$ for various values of $\alpha$. 
Regardless of the values of $\alpha$ and $b$, since the distances between the first five points are the smallest, 
merges will occur over these initially. 
In particular, regardless of $\alpha$ and $b$, $a_1$ is merged with $b_1$, and $x_1$ with $y_1$.
Next, by a simple calculation, if $\alpha\leq\alpha^*$, then $c_1$ merges with $a_1\cup b_1$.
If $\alpha>\alpha^*$, then $c_1$ merges with $x_1\cup y_1$.
Denote the set containing $a_1$ and $b_1$ by $A'_1$, and denote the set containing $x_1$ and $y_1$ by $X'_1$ 
(one of these sets will also contain $c_1$).
Between $A_1'$ and $X_1'$, the minimum distance is $\geq 1.1 + 1.1 \geq 2.2$.
All other subsequent merges (except for the very last merge)
will involve all distances smaller than 2.2, so we never need to consider $A_1'$ merging to $X_1'$.

The next smallest distances are all 1.5, so all points in $A_1$ will merge together, and similarly for $B_1$, $X_1$, and $Y_1$.
At this point, the algorithm has created six sets: $A'_1$, $X'_1$, $A_1$, $B_1$, $X_1$, and $Y_1$.
We claim that if $\alpha\leq \alpha^*$, $A'_1$ will merge to $A_1$ and $B_1$, and $X'_1$ will merge to $X_1$ and $Y_1$.
This is because the maximum distance between sets in each of these merges is 1.6, whereas
the minimum distance between $\{A_1',A_1,B_1\}$ and $\{X'_1,X_1,Y_1\}$ is $\geq 2.2$.
Therefore, for all three values of $b$, the claim holds true.

Next we claim that the 2-clustering cost of gadget 1 will be lowest for clusters $A_1'\cup A_1\cup B_1\}$ and $X'_1\cup X_1\cup Y_1$ and when $c_1\in A'_1$,
i.e., when $\alpha\leq\alpha^*$.
Clearly, since the distances within $A'_1\cup A_1\cup B_1$ and $X'_1\cup X_1\cup Y_1$ are much less than the distances across these sets, the 
best 2-clustering is $A'_1\cup A_1\cup B_1$ and $X'_1\cup X_1\cup Y_1$ (with all points at distance $\leq 1.6$ to their center). 
We proved this will be a pruning of the tree when $\alpha\leq\alpha^*$.
Therefore, we must argue the cost of this 2-clustering is lowest when $c_1\in A'_1$. 
The idea is that $c_1$ can act as a very good center for $A'_1\cup A_1\cup B_1$.
But if $c_1\in X'_1$, then the best center for $A'_1\cup A_1\cup B_1$ will be an arbitrary point in $A_1\cup B_1$.
The cost in the first case is $1.51^p\cdot 50+1.1^p+1.2^p$. The cost in the second case is $1.5^p\cdot 24+1.6^p\cdot 27$.

For $X'_1\cup X_1\cup Y_1$, the center does not change depending on $\alpha$ ($x_1^*$ and $y^*_1$ tie for the best center), 
so the only difference in the cost is whether or not to include $c_1$.
If $\alpha\leq\alpha^*$, then the cost is $1.5^p\cdot 24+1.51^p+1.6^p\cdot 26$, otherwise the cost is 
$1.5^p\cdot 24+1.51^p+1.6^p\cdot 26+(1.6+1.2-0.1\alpha^*)^p$.

Putting it all together, if $\alpha\leq\alpha^*$, the cost is $1.51^p\cdot 50+1.1^p+1.2^p+1.5^p\cdot 24+1.51^p+1.6^p\cdot 26$.
Otherwise the cost is $1.5^p\cdot 48+1.51^p+1.6^p\cdot 53+(1.6+1.2-0.1\alpha^*)^p$.
Subtracting off like terms, we conclude that the first case is always smaller because
$1.51^p\cdot 49+1.1^p+1.2^p<1.5^p\cdot 24+1.6^p\cdot 26+(1.6+1.2-0.1\alpha^*)^p$ for all $p\geq 1$.

Next, we will construct the second gadget arbitrarily far away from the first gadget.
The second gadget is very similar to the first. There are points $a_2,b_2,c_2,x_2,y_2,x^*_2,y^*_2$ and sets to $A_2$, $B_2$, $X_2$, $Y_2$.
$d(a_2,b_2)=d(x_2,y_2)=1$, $d(x_2,c_2)=1.1$, $d(y_2,c_2)=1.2$, and
for $b=1$, $d(a_2,c_2)=d(b_2,c_2)=1.2-.1\cdot\alpha^*$.
For $b=2$ or $b=3$, $d(a_2,c_2)=d(b_2,c_2)=((1.1^{\alpha^*}+1.2^{\alpha^*})/2)^{\frac{1}{\alpha^*}}$.
The rest of the distances are the same as in gadget 1. 
Then $c_2$ joins $\{a_2,b_2\}$ if $\alpha\geq\alpha^*$, not $\alpha\leq\alpha^*$.
The rest of the argument is identical.
So the conclusion we reach, is that the cost for the second gadget is much lower if $\alpha\geq\alpha^*$.

Therefore, the final cost of the 4-clustering is minimized when $\alpha=\alpha^*$,
and the proof is complete.
\end{proof}

Now we give the full details behind Theorem \ref{thm:discretization}

\discretization*

\begin{proof}
Given $\frac{1}{3}<x<y<\frac{2}{3}$ and $n>10$, we will construct an instance $\V$ with the
desired properties. We set $k=2$.

Here is a high level description of our construction $\V = (V,d)$. 
There will be two gadgets. 
Gadget 1 contains points $x_1$, $y_1$, $x_1'$, $y_1'$, and $z_1$.
Gadget 2 contains points $x_2$, $y_2$, $x_2'$, $y_2'$, and $z_2$.
We will define the distances so the following merges take place.
Initially, $x_1$ merges to $y_1$, $x_1'$ merges to $y_1'$,
$x_2$ merges to $y_2$, and $x_2'$ merges to $y_2'$.
Then the sets are $\{x_1,y_1\}$, $\{x_1',y_1'\}$, $\{z_1\}$,
$\{x_2,y_2\}$, $\{x_2',y_2'\}$, and $\{z_2\}$.
Next, $z_1$ will merge to $\{x_1,y_1\}$ if $\alpha<x$, otherwise it will merge to $\{x_1',y_1'\}$.
Similarly, $z_2$ will merge to $\{x_2,y_2\}$ if $\alpha<y$, 
otherwise it will merge to $\{x_2',y_2'\}$.
Finally, the sets containing $\{x_1,y_1\}$ and $\{x_2,y_2\}$ will merge, 
and the sets containing $\{x_1',y_1'\}$ and $\{x_2',y_2'\}$ will merge.

Therefore, the situation is as follows. If $\alpha\in [x,y]$, then the last two sets in the
merge tree will each contain exactly one of the points $\{z_1,z_2\}$. If $\alpha\notin [x,y]$,
then if we again look at the last two sets in the merge tree, one of the sets will contain
both points $\{z_1,z_2\}$.
Since these are the last two sets in the merge tree, the pruning step is not able to output
a clustering with $z_1$ and $z_2$ in different clusters.
To finish the proof, we give a high weight to points $z_1$ and $z_2$ by placing $\frac{n-8}{2}$ points in the same
location as $z_1$, and $\frac{n-8}{2}$ points in the same location as $z_2$.
Note this does not affect the merge equations.
When $z_1$ and $z_2$ are in different clusters, the optimal centers for $k=2$ are at $z_1$ and
$z_2$, and the cost is just the cost of the remaining points,
$\{x_1,x_1',y_1,y_1',x_2,x_2',y_2,y_2'\}$, and all distances will be between 1 and 6, 
so the total cost is $\leq 8\cdot 6^p$.
When $z_1$ and $z_2$ are in the same cluster, the center will be distance at least 2 from
either $z_1$ or $z_2$ (or both), so the cost is $\geq \frac{n-8}{2}\cdot 2^p\in \Omega(n)$.

Now we define the distances and prove the desired merges take place in the correct ranges of
$\alpha$ (see Figure \ref{fig:discretization}). First we consider $\mathcal{A}_3$.
We set
\begin{align*}
d(x_1,y_1)=d(x_2,y_2)=d(x_1',y_1')=d(x_2',y_2')=1.5, \\
d(x_1,z_1)=d(x_2,z_2)=2.4, \\
d(x_1',z_1)=d(x_2',z_2)=2.6, \\
d(y_1,z_1)=d(y_1',z_1)=2.6-.2x, \\
d(y_2,z_2)=d(y_2',z_2)=2.6-.2y, \\
\end{align*}

\begin{figure}\centering
\includegraphics[width=.5\textwidth]{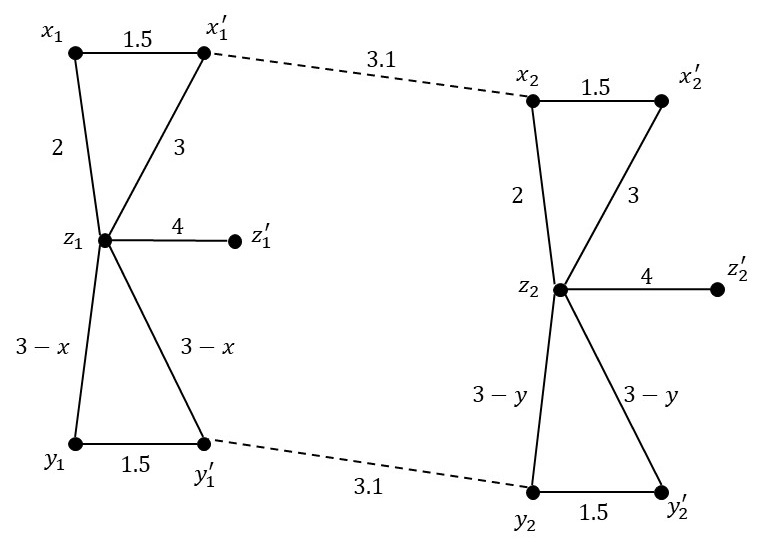}
\caption{The clustering instance used in Theorem \ref{thm:discretization}}\label{fig:discretization}
\end{figure}

We set all distances between $\{x_1,y_1\}$ and $\{x_2,y_2\}$ to 2.7.
Similarly, we set all distances between $\{x_1',y_1'\}$ and $\{x_2',y_2'\}$ to 2.7.
All other distances are the maximum allowed by the triangle inequality.

Then, the first four merges are $\{x_1,y_1\}$, $\{x_1',y_1'\}$, 
$\{x_2,y_2\}$, and $\{x_2',y_2'\}$ since these are the shortest distances.
The next-shortest distances are between 2.4 and 2.6, so $z_1$ will merge to either
$\{x_1,y_1\}$ or $\{x_1',y_1'\}$,
and $z_2$ will merge to either $\{x_2,y_2\}$ or $\{x_2',y_2'\}$.
The decision for $z_1$ corresponds to the equation
$\alpha\cdot 2.4+(1-\alpha)\cdot 2.6=\alpha\cdot (2.6-.2x)+(1-\alpha)\cdot (2.6-.2x)$,
so $z_1$ will merge to $\{x_1,y_1\}$ if $\alpha<x$, otherwise it will merge to $\{x_1',y_1'\}$.
Similarly, we conclude that $z_2$ merges to $\{x_2,y_2\}$ if $\alpha<y$,
otherwise $\{x_2',y_2'\}$.

Next, we want the set containing $\{x_1,y_1\}$ to merge to the set containing $\{x_2,y_2\}$
and the set containing $\{x_1',y_1'\}$ to merge to the set containing $\{x_2',y_2'\}$.
For both of these merges, the merge equation is $\alpha\cdot 2.7+(1-\alpha)2.7=2.7$.
However, the merge equation for $\{x_1,y_1\}$ to $\{x_1',y_1'\}$ could be as small as $\alpha\cdot 2.4+(1-\alpha)\cdot 4.8$,
which is smaller than $2.7$ for $\alpha>.875$. 
In order to ensure that this clustering instance has high cost when $\alpha>.875$,
we add a few more points close to $z_1$ and $z_2$ which will cause a cluster containing $z_1$ and $z_2$ to merge
early on, whenever $\alpha>.86$. Specifically,
we set $d(z_1,z_2)=2.4$ and add $z_1'$ and $z_2'$ such that $d(z_1,z_1')=d(z_2,z_2')=1.1$, 
and the distances to $x_1,y_1,x_1',y_1',x_2,y_2,x_2',y_2'$ are the same as for $z_1$ and $z_2$.
So $z_1$ merges to $z_1'$ and $z_2$ merges to $z_2'$,
and the merge equation for $\{z_1,z_1'\}$ and $\{z_2,z_2'\}$ is smaller than 2.7 when $\alpha>.86$.
This will ensure the clustering has high cost when $\alpha>.86$.

If $\alpha\in [x,y]$, then the last two sets in the merge tree are
$\{x_1,y_1,x_2,y_2,z_1\}$ and $\{x_1',y_1',x_2',y_2',z_2\}$,
which each contain exactly one of the points $\{z_1,z_2\}$. If $\alpha\notin [x,y]$,
then if we again look at the last two sets in the merge tree, one of the sets will contain
both points $\{z_1,z_2\}$.
Since these are the last two sets in the merge tree, the pruning step is not able to output
a clustering with $z_1$ and $z_2$ in different clusters.
When $z_1$ and $z_2$ are in different clusters, since they both have high weight,
the optimal centers for $k=2$ are at $z_1$ and $z_2$, and the cost of the remaining 8 points
is at most $8\cdot 6^p$.
When $z_1$ and $z_2$ are in the same cluster, the center is distance 2 from at least one of them,
so the cost is $\geq \frac{n-8}{2}\cdot 2^p$.
When $p$ is a constant, the difference in cost between these cases is $\Omega(n)$.

The cases for $\mathcal{A}_1$ and $\mathcal{A}_2$ are similar to the previous case. 
All distances are the same, except we set
\begin{align*}
d(y_1,z_1)=d(y_1',z_1)=\left(\frac{1}{2}\left(2.4^x+2.6^x\right)\right)^\frac{1}{x}, \\
d(y_2,z_2)=d(y_2',z_2)=\left(\frac{1}{2}\left(2.4^y+2.6^y\right)\right)^\frac{1}{y}.
\end{align*}
This ensures that $z_1$ will merge to $\{x_1,y_1\}$ if $\alpha<x$, otherwise it will merge to $\{x_1',y_1'\}$,
and $z_2$ will merge to $\{x_2,y_2\}$ if $\alpha<y$, 
otherwise it will merge to $\{x_2',y_2'\}$. 
The rest of the details of the proof are identical to the previous case.
This concludes the proof.
\end{proof}

Now we will show a structural lemma for $\mathcal{A}_3$, which
is similar to Lemma \ref{lem:a1_struct}.
Then we will provide the full details for the proof of Lemma
\ref{lem:a1_struct}.

\begin{lemma}\label{lem:a3_struct}
$\clus_{\mathcal{A}_3,\V}: [0,1] \to \R_{>0}$
is made up of $O(n^8)$ piecewise constant components.
\end{lemma}

\begin{proof}
First note that for $\alpha \neq \alpha'$, the clustering returned by $\mathcal{A}_1(\alpha)$ and the associated cost are both identical to that of $\mathcal{A}_1(\alpha')$ if both the algorithms construct the same merge tree. Now, as we increase $\alpha$ from $0$ to $1$ and observe the run of the algorithm for each $\alpha$, at what values of $\alpha$ do we expect $\A_1(\alpha)$ to produce different merge trees?  To answer this, suppose that at some point in the run of algorithm $\A_1(\alpha)$, there are two pairs of subsets of $V$, $(A,B)$ and $(X,Y)$, that could potentially merge.  There exist eight points $p,p'\in A$, $q,q'\in B$, $x,x'\in X$, and $y,y'\in Y$ such that the decision of which pair to merge 
depends on whether
$\alpha d(p,q)+(1-\alpha) d(p',q')$ or $\alpha d(x,y)+(1-\alpha)d(x',y')$
is larger. 
This is a linear equation in $\alpha$, so there is at most one value of $\alpha$ for which these expressions are equal, unless the difference of the expressions is zero for all $\alpha$. 
Assuming that
ties are broken arbitrarily but consistently, this implies that there is at most one $\alpha \in [0,1]$  such that the choice of whether to merge $(A,B)$ before $(X,Y)$ is identical for all $\alpha<\alpha'$, and similarly identical for $\alpha\geq\alpha'$. 
Since each merge decision is defined by eight points, iterating over all pairs $(A,B)$ and $(X,Y)$  it follows that we can identify all $O(n^8)$ unique 8-tuples of points which correspond to a value of $\alpha$ at which some decision flips. This means we can divide $[0,1]$ into $O(n^8)$ intervals over each of which the merge tree, and therefore the output of $\clus_{\mathcal{A}_1,\V}(\alpha)$, is fixed.
\end{proof}

Now we will provide the details of Lemma \ref{lem:a1_struct}.
In the argument for the structure of $\mathcal{H}_{\mathcal{A}_3,\clus}$, we relied on the linearity of $\mathcal{A}_3$'s merge equation to prove that for any eight points, there is exactly one value of $\alpha$ such that $\alpha d(p,q)+(1-\alpha) d(p',q') = \alpha d(x,y)+(1-\alpha)d(x',y')$. 
Now we will use Theorem \ref{thm:roots}, a consequence of Rolle's Theorem, 
to bound the values of $\alpha$ such that $\left(
(d(p,q))^{\alpha} + d(p',q')^
\alpha \right)^{1/\alpha} = \left(
(d(x,y))^{\alpha} + d(x',y')^
\alpha \right)^{1/\alpha}$.

\begin{theorem}[ex. \cite{tossavainen}]\label{thm:roots}
Let $f$ be a polynomial-exponential sum of the form $f(x) = \sum_{i = 1}^N a_i b_i^x$, where $b_i > 0$, $a_i \in \R$, and at least one $a_i$ is non-zero. The number of roots of $f$ is upper bounded by $N$.
\end{theorem}

\aOneStruct*

\begin{proof}
As was the case for $\mathcal{H}_{\mathcal{A}_3}$, 
the clustering returned by
$\mathcal{A}_1(\alpha)$ and the associated cost are identical to that of
$\mathcal{A}_1(\alpha')$ as long as both algorithms construct the same merge trees.
Our objective is to understand the behavior of $\A_1(\alpha)$ over $m$ instances. In particular, as $\alpha$ varies over $\mathbb{R}$ we want to count the number of times the algorithm outputs a different merge tree on one of these instances. 
For some instance $\V$ we will consider two pairs of sets $A, B$ and $X, Y$ that can be potentially merged. The decision to merge one pair before the other is determined by the sign of $ d^{\alpha}(p,q)+ d^{\alpha}(p',q') - d^{\alpha}(x,y)+d^{\alpha}(x',y')$. This expression, as before, is determined by  a set of $8$ points $p,p'\in A$, $q,q'\in B$, $x,x'\in X$ and $y,y' \in Y$ chosen independent of $\alpha$. 

Now, from Theorem \ref{thm:roots}, we have that the sign of the above expression as a function of $\alpha$ flips at most $4$ times across $\mathbb{R}$.  Since the expression is defined by exactly 8 points, iterating over all pairs $(A,B)$ and $(X,Y)$ we can list only $O(n^8)$ such unique expressions, each of which correspond to $O(1)$ values of $\alpha$ at which the corresponding decision flips. Thus, we can divide $\mathbb{R}$ into $O(n^8)$ intervals over each of which the output of $\clus_{\mathcal{A}_1,\V}(\alpha)$ is fixed. 
\end{proof}

Now we give the full details for Lemmas~\ref{lem:a13upper} and~\ref{lem:a13lower}.

\aOneThreeUpper*

\begin{proof}
Suppose $\sample=\left\{\V^{(1)},\dots,\V^{(m)}\right\}$ is a set of clustering instances that can be shattered by $\mathcal{H}_{\mathcal{A}_1}$ using the witnesses 
$r_1,\dots,r_m$. We must show that $m = O(\log n)$.
For each value of $\alpha \in \mathbb{R} \cup \{-\infty, \infty \}$, 
the algorithm $\mathcal{A}_1(\alpha)$
induces a binary labeling on each $\V^{(i)}$, based on whether or not 
$\clus_{\mathcal{A}_1(\alpha)}\left(\V^{(i)}\right) \leq r_i$.
From Lemma \ref{lem:a1_struct}, we know that
every sample $\V^{(i)}$ partitions $\mathbb{R} \cup \{\infty, -\infty \}$ into $O(n^8)$ intervals in this way. Merging all $m$ partitions, we can divide $\mathbb{R} \cup \{\infty, -\infty \}$ into $O(m n^8)$ intervals over each of  which $\clus_{\mathcal{A}_3,\V^{(i)}}(\alpha)$, and therefore the labeling induced by the witnesses, is fixed for all $i \in [m]$ (similar to Figure \ref{fig:dots}). This means that $\mathcal{H}_{\mathcal{A}_1}$ can achieve only $O(m n^8)$ binary labelings, which is at least $2^{m}$ since $\sample$ is shatterable, so $m = O(\log n)$.

The details for $\mathcal{H}_{\mathcal{A}_3},\clus$ are identical, by
using Lemma \ref{lem:a3_struct}.
\end{proof}

\aOneThreeLower*

We first prove this lemma for the center-based objective cost denoted by $\Phi^{(p)}$ for $p \in [1, \infty) \cup \{ \infty \}$. 
We later note how this can be extended cluster purity based cost.
We first prove the following useful statement which helps us construct general examples with desirable properties. 
In particular, the following lemma guarantees that given a sequence of values of $\alpha$ of size $O(n)$, it is possible to construct an instance $\V$ such that the cost of the output of $\A_1(\alpha)$ on $\V$ as a function of $\alpha$, that is $\Phi^{(p)}_{\A_1, \V}(\alpha)$, oscillates above and below some threshold as $\alpha$ moves along the sequence of intervals $(\alpha_i, \alpha_{i+1})$. 
Given this powerful guarantee, we can then pick appropriate sequences of $\alpha$ and generate a sample set of $\Omega(\log n)$ instances that correspond to cost functions that oscillate in a manner that helps us pick $\Omega(n)$ values of $s$ that shatters the samples.
We also show how to trade off the number of oscillations, with the difference in cost between the oscillations, using parameter $\gamma$.
However, $\gamma=1$ is sufficient to obtain a pseudo-dimension lower bound.

\begin{lemma} \label{lem:oscillate} 
Given $n\in \mathbb{N}$, $0<\gamma\leq 1$, $b\in\{1,3\}$, $\gamma\leq 1$, and given a sequence of $n'\leq\lfloor \frac{\gamma n}{7}\rfloor$ 
$\alpha$'s such that 
$.3=\alpha_0<\alpha_1<\cdots<\alpha_{n'}<\alpha_{n'+1}=.6$,
 there exists a real valued witness $r>0$ and a clustering instance $\V =(V,d)$, $|V|=n$, 
such that for $0\leq i\leq n'/2-1$, 
$\Phi^{(p)}_{\A_b(\alpha)}(\V)<\gamma \cdot r$ for $\alpha\in(\alpha_{2i},\alpha_{2i+1})$, and 
$\Phi^{(p)}_{\A_b(\alpha)}(\V)> r$ for $\alpha\in(\alpha_{2i+1},\alpha_{2i+2})$, for $k=2$.
\end{lemma}

\begin{proof}
The idea of the proof is as follows.
There will be two ``main'' points, $a$ and $a'$ in $V$.
The rest of the points are defined in groups of 6: $(x_i,y_i,z_i,x_i',y_i',z_i')$, for $1\leq i\leq (n-2)/6$.
We will define the distances between all points such that initially for all $\A_b(\alpha)$, $x_i$ merges to $y_i$ to form the set $A_i$, 
and $x_i'$ merges to $y_i'$ to form the set $A_i'$. 
As for $z_i$ and $z_i'$, depending on whether $\alpha < \alpha_i$ or not, $\A_b(\alpha)$ merges the points $z_i$ and $z_i'$ with the sets $A_i$ and $A_i'$ respectively or vice versa. This means that there are $(n-2)/6$ values of $\alpha$ such that $\A_{b}(\alpha)$ has a unique behavior in the merge step. Finally, for all $\alpha$, sets $A_i$ merge to $\{a\}$,  and sets $A_i'$ merge to $\{a'\}$. Let $A=\{a\}\cup\bigcup_i A_i$ and $A'=\{a'\}\cup\bigcup_i A_i'$. 
There will be $(n-2)/6$ intervals $(\alpha_i,\alpha_{i+1})$ for which $\A_b(\alpha)$ returns a unique partition $\{A,A'\}$. 
By carefully setting the distances, we cause the cost $\Phi(\{A,A'\})$ to oscillate above and below a specified value $r$ along these intervals.
In order to make the cost oscillate above $r$ and below $\gamma r$, we give a high weight to two points, by putting $\frac{1-\gamma}{2}$ points in the same
location as these two points. The first high-weight point is $a$, and the second high-weight point is a new point $z$.
We set the distances so that $z$ oscillates between merging to $A$ or $A'$ as we increase $\alpha$ from $.3$ to $.6$.
If $z$ merges to $A'$, then the 2-clustering cost is low because we can put centers on $a$ and $z$.
If $z$ merges to $A$, then both $a$ and $z$ are in the same cluster, incurring a large cost.

Now we will give the full details of the proof, including all distances.
First we focus on $\mathcal{A}_3$ and $\gamma=1$, and we discuss the other cases later in the proof.
First of all, in order for $d$ to be a metric, we set all distances in $[1,2]$ so that the triangle inequality is trivially satisfied.
The following are the distances of the pairs of points for $1\leq i\leq (n-2)/6$.

\begin{align*}
d(x_i,y_i)&=d(x_i',y_i')=1,\\
d(x_i,z_i)&=1.3,~d(y_i,z_i)=1.4,\\
d(x_i',z_i)&=d(y_i',z_i)=1.4-.1\cdot\alpha_i,\\
d(x_i,x_i')&=d(y_i,y_i')=2.
\end{align*}

We set the distances to $z_i'$ as follows (see Figure \ref{fig:simple-lb}).
\begin{align*}
&d(x_i,z_i')=d(y_i,z_i')=d(x_i',z_i')=d(y_i',z_i')=1.41,\\
&d(z_i,z_i')=2.
\end{align*}

\begin{figure}\centering
\includegraphics[width=.5\textwidth]{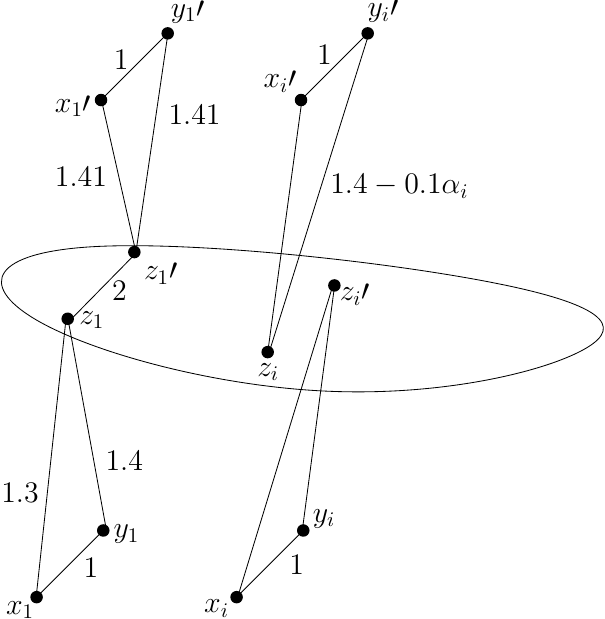}
\caption{The clustering instance used in Lemma \ref{lem:oscillate}}\label{fig:simple-lb}
\end{figure}

Then the first merges will be $x_i$ to $y_i$ and $x_i'$ to $y_i'$, no matter what $\alpha$ is set to be
(when each point is a singleton set, each pair of points with the minimum distance in the metric will merge).
Next, $z_i$ will either merge to $A_i$ or $A_i'$ based on the following equation:
\begin{align*}
&&\alpha\cdot 1.3+(1-\alpha)\cdot 1.4 &\lessgtr \alpha\cdot (1.4-.1\cdot\alpha_i)+(1-\alpha)(1.4-.1\cdot\alpha_i)\\
&\implies &1.4-.1\cdot \alpha &\lessgtr 1.4-.1\cdot\alpha_i\\
&\implies &\alpha_i &\lessgtr \alpha
\end{align*}

If $\alpha<\alpha_i$, then $z_i$ merges to $A_i'$, otherwise it will merge to $A_i$.
After one of these merges takes place, the new value for merging $A_i$ to $A_i'$ could be as small as 
$\alpha\cdot 1.3+(1-\alpha)\cdot 2=2-.6\cdot \alpha$, 
but we do not want this merge to occur.
If we ensure all subsequent merges have maximum distance less than 1.5, then $A_i$ will not merge to $A_i'$ (until $A$ and $A'$ merge in the very final step) as long as $\alpha<.6$, 
because $\alpha\cdot 1.5+(1-\alpha)\cdot 1.5=1.5<2-.6\cdot .7$.

These distances ensure $z_i'$ merges after $z_i$ regardless of the value of $\alpha$,
since $z_i$ is closer than $z_i'$ to $x_i$, $x_i'$, $y_i$, and $y_i'$.
Furthermore, $z_i'$ will merge to the opposite set of $z_i$, since we set $d(z_i,z_i')=2$.
The merge expression for $z_i'$ to merge to the opposite set is
$\alpha\cdot 1.41 + (1-\alpha)\cdot 1.41$, 
while the merge expression to the same set is
$\geq \alpha\cdot 1.41+(1-\alpha)\cdot 2$.

Now we set the distances to $a$ and $a'$ as follows.
\begin{align*}
&d(a,x_i)=d(a,y_i)=d(a',x_i')=d(a',y_i')=1.42,\\
&d(a,x_i')=d(a,y_i')=d(a',x_i)=d(a',y_i')=2.
\end{align*}

We also set all distances between $A_i$ and $A_j'$ to be 2, for all $i$ and $j$,
and all distances between $A_i$ and $A_j$ to be 1.5, for all $i\neq j$.
We will set the distances from $a$ and $a'$ to $z_i$ and $z_i'$ later, but they will all fall between $1.46$ and $1.47$.
By construction, every set $A_i$ will merge to the current superset containing $\{a\}$, because the merge
expression is $\alpha\cdot 1.42+(1-\alpha)1.5$, and \emph{any other possible merge} will have value 
$\geq \alpha\cdot 1.3+(1-\alpha)\cdot 2$, which is larger for $\alpha<.6$.
Similarly, all $A_i'$ sets will merge to $\{a'\}$.

Therefore, the final two sets in the linkage tree are $A$ and $A'$. Given $1\leq i\leq (n-2)/6$,
by construction, for $\alpha\in (\alpha_i,\alpha_{i+1})$, 
$\{z_1,\dots,z_i,z_{i+1}',\dots z_{(n-2)/6}'\}\subseteq A$ and 
$\{z_1',\dots,z_i',z_{i+1},\dots z_{(n-2)/6}\}\subseteq A'$.

Finally, we set the distances between $a$, $a'$, $z_i$, and $z_i'$ to ensure the cost function oscillates. 
\begin{align*}
\forall i, \quad d(a,z_i')&=d(a',z_i)=1.46\\
\forall 1\leq j\leq (n-2)/12,\quad d(a,z_{2j-1})&=d(a',z_{2j}')=1.47,\\
\text{and } d(a,z_{2j})&=d(a',z_{2j+1}')=(2\cdot 1.46^p-1.47^p)^{1/p}.
\end{align*}

Now we calculate the 2-clustering cost of $(A,A')$ for $\alpha$'s in different ranges.
Regardless of $\alpha$, all partitions will pay 
$\sum_i (d(a,x_i)^p+d(a,y_i)^p+d(a',x_i')^p+d(a',y_i')^p)=(n-2)/6\cdot(4\cdot 1.42^p)$,
but the distances for $z_i$ and $z_i'$ differ.
For $\alpha\in (\alpha_0,\alpha_1)$, all of the $z$'s pay $1.46^p$, 
so the cost is $(n-2)/6\cdot(4\cdot 1.42^p+2\cdot 1.46^p)$.
Denote this value by $r_{\it low}$.

When $\alpha\in (\alpha_1,\alpha_2)$, the only values that change are $z_1$ and $z_1'$,
which adds $d(a,z_1)+d(a',z_1')-d(a,z_1')-d(a',z_1)=2\cdot(1.47^p-1.46^p)>0$ to the cost
(the inequality is always true for $p\in [1,\infty]$).
Denote $r_{\it low}+2\cdot(1.47^p-1.46^p)$ by $r_{\it high}$.
When $\alpha\in (\alpha_2,\alpha_3)$, the values of $z_2$ and $z_2'$ change, 
and the cost changes by 
$d(a,z_2)+d(a',z_2')-d(a,z_2')-d(a',z_2)=2\cdot((2\cdot 1.46^p-1.47^p)-1.46^p)=-2\cdot(1.47^p-1.46^p)$, decreasing it back to $r_{\it low}$.

In general, the cost for $\alpha\in(\alpha_i,\alpha_{i+1})$ is
$r_{low}+\sum_{1\leq j\leq i} (-1)^{i+1}\cdot 2(1.47^p-1.46^p)=
r_{\it low}+(1.47^p-1.46^p)+(-1)^{i+1}\cdot (1.47^p-1.46^p)$.
If $\alpha\in (\alpha_{2j},\alpha_{2j+1})$, then the cost is $r_{\it low}$,
and if $\alpha\in (\alpha_{2j+1},\alpha_{2j+2})$, the cost is $r_{\it high}$.
We set $r=(r_{\it low}+r_{\it high})/2$, and conclude that the cost function oscillates
above and below $r$ as specified in the lemma statement.

The pruning step will clearly pick $(A,A')$ as the optimal clustering, since
the only centers with more than 3 points at distance $< 1.5$ are $a$ and $a'$, 
and $(A,A')$ are the clusters in which the most points can have $a$ and $a'$ as centers.
This argument proved the case where $n'=(n-2)/6$. If $n'<(n-2)/6$, then we set
$d(a,z_i)=d(a',z_i')=1.46$ for all $i>n'$, which ensures the cost function oscillates exactly
$n'$ times.
This completes the proof for $\mathcal{A}_3$ and $\gamma=1$.

It is straightforward to modify this proof to work for $\mathcal{A}_1$.
The only major change is to set
\begin{equation*}
d(x_i',z_i)=d(y_i',z_i)=((1.3^\alpha_i+1.4^\alpha_i)/2)^{\frac{1}{\alpha_i}}.
\end{equation*}

Now we move to the case where $\gamma<1$. 
In this case, the cost will oscillate between $>r$ and $<\gamma\cdot r$, for a value of $r$ defined later.
To accomplish this, we put large weight on $a$ and a new point $z$.
Our goal is to show that
the optimal $k=2$ pruning oscillates between putting $a$ and $z$ in the same cluster, versus different clusters,
for the intervals defined by $\alpha_0,\dots,\alpha_{n'}$.
We will use $\frac{\gamma\cdot n}{7}$ gadgets consisting of $6$ points each, to achieve $\frac{\gamma\cdot n}{7}$ intervals that oscillate.
The remaining $(1-\gamma)n$ points will be used to create a separation between the costs of the optimal 2-clustering in neighboring $\alpha$ intervals.

Now we will show how to alternate $a$ and $z$ in the same cluster versus different clusters.
Note that $a$ will always merge to the set $A$ by definition.
Next we set the distances from $z_1,\dots, z_{n'},z_1',\dots,z_{n'}'$ to $z$ so that it alternates merging to $A$ or $A'$ along $\alpha_0,\dots,\alpha_{n'}$.
We set the distances between $z$ and $z_1,\dots,z_{n'}$, $z_1',\dots,z_{n'}'$ as follows.
We nest $d(z,z_1)<\cdots<d(z,z_{n'})<d(z,z_{n'})<d(z,z_1')$.
Recall that in interval $(\alpha_i,\alpha_{i+1})$, $A$ contains $z_1,\dots,z_i,z_{i+1}',\dots,z_{n'}'$,
and $A'$ contains $z_1',\dots,z_i',z_{i+1}',\dots,z_{n'}'$.
Therefore, the merge equation in this interval is $\alpha_i d(z,z_1)+(1-\alpha_i)d(z,z_1')\lessgtr \alpha_i d(z,z_{i+1})+(1-\alpha_i) d(z,z_{i+1}')$.
We set $d(z,z_1)=1.46$, $d(z,z_1')=1.47$, and $d(z,z_i')-d(z,z_i)=\frac{1}{2^i}$.
Then we solve to find $d(z,z_i)=1.47-.01\alpha_1+\frac{\alpha_i}{2^i}$ and $d(g_2,z_i')=1.47-.01\alpha_1-\frac{1}{2^i}(1-\alpha_i)$ 
would achieve equality in the equation above. Call these values $d_i$ and $d_i'$, respectively.
If we set the distances to exactly these values, then we would have exact ties for the decision to merge $z$ to $A$ or $A'$ in all $\alpha$ intervals.
Therefore, we add small offsets of size $\epsilon=.0001$ to some of the values.
Specifically, set $d(z,z_i)=d_i$ for all $i$. For even $i$, set $d(z,z_i')=d_i'+\epsilon$, and for odd $i$, set $d(z,z_i')=d_i'-\epsilon$.
This ensures $z$ oscillates merging to $A$ or $A'$ along the $n'$ $\alpha$ intervals.

The pruning step for $k=2$ must output $A$ and $A'$, since this is the last merge that takes place in the tree.
When $z$ is in $A'$, then the optimal centers are at $a$ and $z$, and the cost of the clustering is the cost of the
$\gamma n$ points making up the gadgets, which is $O(\gamma n)$.
When $z$ is in $A$, then the center for $A$ must be distance at least 1 to either $a$ or $z$, so the cost of the clustering
is at least $\frac{1-\gamma}{2}\cdot n$. 
Therefore, the difference in cost is $\Omega\left(\frac{1-\gamma}{\gamma}\right)$.
\end{proof}

Now we can prove Lemma \ref{lem:a13lower}.

\begin{proof}[Proof of Lemma \ref{lem:a13lower}]
Given $b\in\{1,2\}$,
we prove the claim for $\mathcal{H}_{\A_b,\clus^{(p)}}$ by constructing a set of samples $\sample = \{ \V^{(1)}, \hdots, \V^{(m)}\}$ where $m=\log n - 3$ that can be shattered by $\mathcal{H}_{\A_b,\clus^{(p)}}$.
That is, we should be able to choose $2^m = n/8$ different values of $\alpha$ such that there exists some witnesses $r_1, \hdots, r_m$ with respect to which $\Phi^{(p)}_{\mathcal{A}_b(\alpha)}(\cdot)$ induces all possible labelings on $\sample$.

Choose a sequence of $2^m$ distinct $\alpha$'s arbitrarily in the range $(0,.7)$.  We will index the terms of this sequence using the notation  $\alpha_{\bf x}$ for all 
${\bf x}\in\{0,1\}^m$, such that $\alpha_{\bf x}<\alpha_{\bf y}$ iff 
${\bf x}_1{\bf x}_2\dots{\bf x}_m<{\bf y}_1{\bf y}_2{\bf y}_m$.
Then the $\alpha$'s satisfy 
\begin{equation*}
0<\alpha_{[0~\dots~0~0]}<\alpha_{[0~\dots~0~1]}<\alpha_{[0~\dots~1~0]}<\dots<\alpha_{[1~\dots~1~1]}<.7.
\end{equation*}
Given ${\bf x}$, denote by $n({\bf x})$ the vector corresponding to ${\bf x}_1{\bf x}_2\dots{\bf x}_s+1$, therefore, $\alpha_{n({\bf x})}$ is the smallest $\alpha$ greater than $\alpha_{\bf x}$.

Now, the crucial step is that  we will use Lemma \ref{lem:oscillate} to define our examples $\V^{(1)},\dots V^{(m)}$ and witnesses $r_1,\dots r_m$ so that
 when $\alpha\in(\alpha_{\bf x},\alpha_{n({\bf x})})$ the labeling induced by the witnesses on $\sample$ corresponds to the vector ${\bf x}$. This means that for $\alpha\in(\alpha_{\bf x},\alpha_{n({\bf x})})$ the cost function $\Phi^{(p)}_{\A_b(\alpha)}(\V^{(i)})$ must be greater than $r_i$  if the  $i$th term in ${\bf x}$ is $1$, and less than $r_i$ otherwise. Since there are only
$2^m=\frac{n}{8}$ ${\bf x}$'s, it implies that for any sample $\V^{(i)}$ there at most $n/8$ values of $\alpha$ at which we want its cost to flip above/below $r_i$. We can we can accomplish this using Lemma \ref{lem:oscillate} by choosing 
$\alpha_{\bf x}$'s for which  $\V^{(i)}$ is supposed to switch labels.
In this manner, we pick each $\V^{(i)}$ and $r_i$ thus creating a sample of size $\Omega(\log n)$ that is shattered by $\mathcal{H}_{\A_b,\clus^{(p)}}$.
\end{proof}

\begin{note}
Lemma \ref{lem:a13lower} assumes that the pruning step fixes a partition, and then the optimal
centers can be chosen for each cluster in the partition, but points may not switch clusters even if
they are closer to the center in another cluster.
This is desirable, for instance, in applications which much have a balanced partition.

If it is desired that the pruning step only outputs the optimal centers, 
and then the clusters are determined
by the Voronoi partition of the centers, we modify the proof as follows.
We introduce $2n'$ more points into the clustering instance: 
$c_1,\dots,c_{n'}$, and $c_1',\dots,c_{n'}'$.
Each $c_i$ will merge to cluster $A$, and each $c_i'$ will merge to cluster $A'$.
We set the distances so that $c_i$ and $c_i'$ will be the best centers for $A$ and $A'$ when
$\alpha\in (\alpha_i,\alpha_{i+1})$. 
The distances are also set up so that the cost of the Voronoi tiling induced by $c_{2i}$ and $c_{2i}'$ is
$r_{\it low}$, and the cost for $c_{2i+1}$ and $c_{2i+1}'$ is $r_{\it high}$.
This is sufficient for the argument to go through.

Furthermore, the lower bound holds even if the cost function is the symmetric distance to the ground truth clustering.
For this proof, let 
$A\cup\bigcup_i\{z_{2i},z'_{2i+1}\}$ and 
$A'\cup\bigcup_i\{z_{2i+1},z'_{2i}\}$
be the ground truth clustering. Then in each interval as $\alpha$ increases, 
the cost function switches between having $(n-2)/3$ errors and having $(n-2)/3-2$ errors.
\end{note}

Now we restate and prove Lemma~\ref{lem:general_lb}.

\generalLB*

To prove this, we start with a helper lemma.

\begin{lemma} \label{lem:general_lb_helper}
Given $n$, and setting $N=\lfloor(n-8)/2\rfloor$, then
there exists a clustering instance $\V=(V,d)$ of size
$|V|=n$ and a set of $2^N+2$ values of $\alpha$
for which $\alpha$-linkage creates a unique merge tree.
\end{lemma}

\begin{proof}
The idea of the proof is as follows.
First we define two pairs of points
which merge together regardless of the value of $\alpha$.
Call these merged pairs $A=\{p_a,q_a\}$ and $B=\{p_b,q_b\}$.
Next, we define a sequence of points $p_i$ and $q_i$ for $1\leq i\leq N$ with distances set such that merges involving points in this sequence occur one after the other. 
In particular, first $p_1$ merges to $A$ or $B$, then $q_1$ merges to the opposite set, then $p_2$ merges to $A$ or $B$ and $q_2$ merges to the opposite set, 
and so on.
Using induction to precisely set all the distances, we show that for all $1\leq i\leq N$, $p_i$ merges to $A$ or $B$ based on the value of $\alpha$,
regardless of all previous merges that took place.
Therefore, there are $2^N$ distinct merge trees which can be created.
In particular, there are $2^N$ distinct values of $\alpha$, each corresponding to a distinct merge tree,
enabling $\mathcal{A}_2$ to achieve all possible merge tree behaviors.
Finally, we carefully add more points to the instance to control the oscillation of the cost function over these intervals as desired.

Now we go into more detail on the specifics of the construction.
We set the distances so the first two merges will always be $p_a$ to $q_a$, and $p_b$ to $q_b$.
These sets $\{p_a,q_a\}$ and $\{p_b,q_b\}$ will stay separated until the last few merge operations.
Throughout the analysis, at any point in the merging procedure, 
we denote the current superset containing $\{p_a,q_a\}$ by $A$, and we similarly denote the superset of
$\{p_b,q_b\}$ by $B$. 
Next, we construct the distances so that $p_i$ and $q_i$ will always merge before $p_j$ and $q_j$,
for $i<j$. Furthermore, for all $i$, $\{p_i\}$ will first merge to $A$ or $B$, and then $\{q_i\}$ will
merge to the other one. We call these merges `round $i$', for $1\leq i\leq N$.
Finally, there will be a set $C_A$ of size $N+2$ which merges together and then merges to $A$,
and similarly a set $C_B$ which merges to $B$. These sets will control the value of the resulting clusterings.
In our construction, the only freedom is whether $p_i$ merges to $A$ or to $B$, for all $i$, which is $2^{N}$ combinations total. 
The crux of the proof is to show there exists a unique $\alpha$ for each of these behaviors.

In round 1, the following equation specifies whether $p_1$ merges to $A$ or $B$:
\begin{equation*}
\frac{1}{2}(d(p_a,p_1)^\alpha+d(q_a,p_1)^\alpha)\lessgtr \frac{1}{2}(d(p_b,p_1)^\alpha+d(q_b,p_1)^\alpha)
\label{eq:round1}
\end{equation*}

If the LHS is smaller, then $p_1$ merges to $A$, otherwise it merges to $B$.
We set the distances to ensure there exists a value $\alpha'$ which is the only solution to the equation in the range $(1,3)$.
Then $p_1$ merges to $A$ for all $\alpha\in (1,\alpha')$, and $B$ for all $\alpha\in (\alpha',3)$.
We set $d(p_1,q_1)$ to be large, so that once $p_1$ merges to either $A$ or $B$, $q_1$ is forced to the other set, the one which does not contain $p_1$.

In round 2, there are two equations:
\begin{align*}
\frac{1}{3}(d(p_a,p_2)^\alpha+d(q_a,p_2)^\alpha+d(p_1,p_2)^\alpha)
\lessgtr \frac{1}{3}(d(p_b,p_2)^\alpha+d(q_b,p_2)^\alpha+d(q_1,p_2)^\alpha),\\
\frac{1}{3}(d(p_a,p_2)^\alpha+d(q_a,p_2)^\alpha+d(q_1,p_2)^\alpha)
\lessgtr \frac{1}{3}(d(p_b,p_2)^\alpha+d(q_b,p_2)^\alpha+d(p_1,p_2)^\alpha).
\end{align*}

The first equation specifies where $p_2$ merges in the case when $p_1\in A$, and the second equation is the
case when $p_1\in B$.
So we must ensure there exists a specific $\alpha_{[-1]}\in (1,\alpha')$ which solves equation 1, and
$\alpha_{[1]}\in (\alpha',3)$ which solves equation 2, and these are the only solutions in the corresponding
intervals.

In general, round $i$ has $2^{i-1}$ equations corresponding to the $2^{i-1}$ possible states for the partially
constructed tree. For each state, there is a specific $\alpha$ interval which will
cause the algorithm to reach that state.
We must ensure that the equation has exactly one solution in that interval.
By achieving this simultaneously for every equation, the next round will have $2\cdot 2^{i-1}$ states.
See Figure \ref{fig:general-lb} for a schematic of the clustering instance.
\begin{figure}\centering
\includegraphics[width=.7\textwidth]{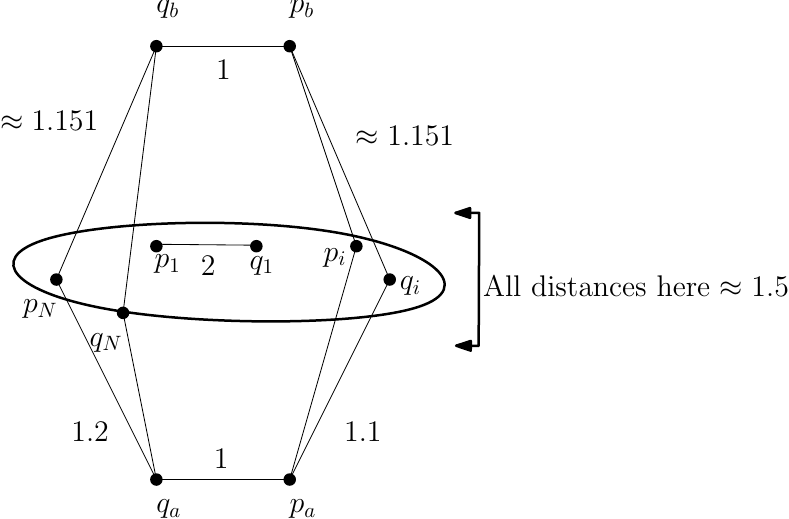}
\caption{The clustering instance used in Lemma \ref{lem:general_lb}}\label{fig:general-lb}
\end{figure}

For $1\leq i\leq N$, given ${\bf x}\in \{-1,1\}^{i-1}$, let $E_{\bf x}$ denote the equation 
in round $i$ which determines
where $p_i$ merges, in the case where for all $1\leq j<i$, 
$p_j$ merged to $A$ if $x_j=-1$, or $B$ if $x_j=1$
(and let $E'$ denote the single equation for round 1).
Let $\alpha_{\bf x}\in (1,3)$ denote the solution to $E_{\bf x}=0$.
Then we need to show the $\alpha$'s are well-defined and follow a specific ordering, shown in Figure \ref{fig:order}.
This ordering is completely specified by two conditions: \emph{(1)} $\alpha_{[{\bf x}~-1]}<\alpha_{[\bf x]}<\alpha_{[{\bf x}~1]}$
and \emph{(2)} $\alpha_{[{\bf x}~-1~{\bf y}]}<\alpha_{[{\bf x}~1~{\bf z}]}$
for all ${\bf x,y,z}\in\bigcup_{i< N}\{-1,1\}^i$ and $|{\bf y}|=|{\bf z}|$.
\begin{figure}
\centering
	{\includegraphics[width=.6\textwidth]{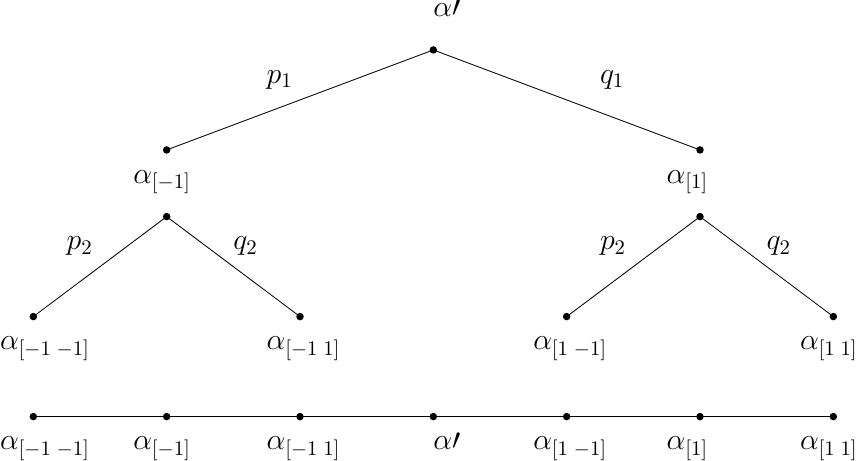}}
    \caption{A schematic for the $\alpha$ intervals. Each edge denotes whether to merge $p_i$ to $A$
		or $q_i$ to $A$.}
		\centering
    \label{fig:order}
\end{figure}

Now we show how to set up the distances to achieve all of these properties.
In the first round, we set the distances so that the merge equation is
$2\cdot 1.1^\alpha\lessgtr (1.1-q^*)^\alpha+(1.1+q^*)^\alpha$, for some offset value $q^*$ which solves the equation at $\alpha=2$.
Therefore, $\alpha\in (1,2)$ corresponds to $p_1\in A$, and $\alpha\in (2,3)$ corresponds to $p_1\in B$.
In the next round, there are three distances on each side of the merge equations, since $p_1$ and $q_1$ are added to sets $A$ and $B$.
In the first case, when $p_1\in A$, the merge equation for round 2 is
$2\cdot 1.1^\alpha+(1.5-o_1)^\alpha\lessgtr (1.1-q^*)^\alpha+(1.1+q^*)^\alpha+(1.5+o_1)^\alpha$,
and when $p_1\in B$ the equation is
$2\cdot 1.1^\alpha+(1.5+o_1)^\alpha\lessgtr (1.1-q^*)^\alpha+(1.1+q^*)^\alpha+(1.5-o_1)^\alpha$.
By setting the offset small enough, we ensure that both solutions to the equations fall in their respective ranges of $(1,2)$ and $(2,3)$.
This ensures that there are four distinct values of $\alpha$, such that we get four distinct merge trees after round 2.
The rest of the rounds repeat this pattern. For each new round $i$, the new distances added to the equation will be $1.5+o_i$ and $1.5-o_i$,
and we set these offsets $o_i$ smaller and smaller so that the solutions to the equations stay in the correct ranges.
To precisely show that such values of the offsets exist, we use an inductive proof.

Our inductive proof will need the following fact (true by elementary calculus).
\begin{fact} \label{fact:1.5}
For all $0\leq z\leq .01$ and $\alpha\in(1,3)$, the following are true about $g(z,\alpha)=(1.5-z)^\alpha-(1.5+z)^\alpha$
and $h(z,\alpha)=(1.1-z)^\alpha+(1.1+z)^\alpha-2\cdot (((1.1-z)^\alpha+(1.1+z)^\alpha)/2)^\frac{\alpha}{2}$.
\begin{enumerate}
\item For $z>0$, $g(z,\alpha)<0$,
\item for a fixed $z$, $g$ is nonincreasing in $\alpha$,
\item for a fixed $\alpha$, $g$ is nonincreasing in $z$,
\item $h(0,\alpha)=0$ and $h$ is nondecreasing in $z$.
\end{enumerate}
\end{fact}

Here are the details for the general construction.
All distances will be between 1 and 2 so that the triangle inequality is satisfied.
Given $N$, for all $i$,
\begin{align*}
&d(p_a,q_a)=d(p_b,q_b)=1,\\
&d(p_a,q_a)=d(p_a,q_b)=d(p_b,q_a)=d(p_b,q_b)=2,\\
\forall i\leq N,~&d(p_a,p_i)=d(p_a,q_i)=1.1-q,~d(q_a,p_i)=d(q_a,q_i)=1.1+q,\\
&d(p_b,p_i)=d(p_b,q_i)=d(q_b,p_i)=d(q_b,q_i)=
\sqrt{\frac{1}{2}((1.1-q)^2+(1.1+q)^2)},\\
&d(p_i,q_i)=2,\\
\forall 1\leq j<i\leq N,~&d(p_i,p_j)=d(p_i,q_j)=1.5+o_j\\
&d(q_i,p_j)=d(q_i,q_j)=1.5-o_j.\\
\end{align*}

where $q$ and $o_j$ are offset values in $(0,.01)$ which we will specify later.
Then for $\alpha\in (1,3)$, the following are true.
\begin{itemize}
\item The first two merges are $p_a$ to $q_a$ and $p_b$ to $q_b$,
\item $\{p_i\}$ and $\{q_i\}$ will always prefer merging to $A$ or $B$ instead of merging
to another singleton $\{p_j\}$ or $\{q_j\}$.
\end{itemize}

After the first two merges occur, all $p_i$ and $q_i$ are tied to first merge to $A$ or $B$.
For convenience, we specify the tiebreaking order as $\{p_1,q_1,\dots,p_{N},q_{N}\}$.
Alternatively, at the end we can make tiny perturbations to the distances so that tiebreaking does not occur.

Next, we choose the value for $q$, which must be small enough to ensure that $q_i$ always merges to the opposite cluster
as $p_i$.
Consider 
\begin{align*}
h(\alpha,q,o_1,\dots,o_{N},{\bf x})=&\frac{N+2}{N+3}\left((1.1+q)^\alpha+(1.1-q)^\alpha+\sum_{i<N}{\bf x}_i(1.5+o_i)^\alpha+1.5^\alpha\right)\\
&-2\cdot(((1.1+q)^2+(1.1-q)^2)/2)^{\frac{\alpha}{2}}-\sum_{i<N}{\bf x}_i(1.5+o_i)^\alpha.
\end{align*}

If this equation is positive for all ${\bf x}\in\{-1,1\}^{N-1}$, then $q_{N}$ will always merge to the opposite cluster as $p_{N}$
(and $q_i$ will always merge to the opposite cluster as $p_i$,
which we can similarly show by setting $o_j=0$ in $h$ for all $j>i$).

Note
\begin{equation*}
h(\alpha,0,0,\dots,0,{\bf x})=\frac{N+2}{N+3}\left(2\cdot 1.1^\alpha+(N+1)\cdot 1.5^\alpha\right)-2\cdot 1.1^\alpha-N\cdot 1.5^\alpha>0
\end{equation*}
for all ${\bf x}$ and all $\alpha\in(1,3)$.
Fact \ref{fact:1.5} implies there exists a $0<q^*<.01$ such that $h(\alpha,q,0,\dots,0,{\bf x})$ stays positive. Similarly,
there exists a cutoff value $\delta>0$ such that for all $0<o_1,\dots,o_{N}<\delta$, $\alpha\in (1,3)$, and ${\bf x}\in\{-1,1\}^{N-1}$,
$h(\alpha,q^*,o_1,\dots,o_k,{\bf x})>0$.
Therefore, as long as we set all the offsets $o_i$ less than $\delta$, 
the merges will be as follows: 
\begin{enumerate}
\item $p_a$ merges to $q_a$ and $p_b$ merges to $q_b$. 
\item For $1\dots,N$, $p_i$ merges to $A$ or $B$, and $q_i$ merges to the opposite cluster.
Then $q_{N}$ will always merge to the opposite cluster as $p_{N}$.
\end{enumerate}

Now we show that there are $2^{N}$ intervals for $\alpha\in (1,3)$ which give unique behavior.
Recall for ${\bf x}\in\bigcup_{i< N}\{-1,1\}^i$, $E_{\bf x}$ is defined as
\begin{equation*}
(1.1-q^*)^\alpha+(1.1+q^*)^\alpha-2\cdot(\frac{1}{2}((1.1-q^*)^2+(1.1+q^*)^2))^\frac{\alpha}{2}+
\sum_{i<N}{\bf x}_i((1.5-o_i)^\alpha-(1.5+o_i)^\alpha).
\end{equation*}

For brevity, we denote
\begin{equation*}
d=(\frac{1}{2}((1.1-q^*)^2+(1.1+q^*)^2))^\frac{1}{2}.
\end{equation*}

We show the $\alpha$s are correctly ordered by proving the following three statements with induction.
The first statement is sufficient to order the $\alpha$s, and the second two will help to prove the first.

\begin{enumerate}
\item
There exist $0<o_1,\dots,o_{N}<\delta$ such that if we solve $E_{\bf x}=0$ for $\alpha_{\bf x}$ for all ${\bf x}\in\bigcup_{i< N}\{-1,1\}^i$, 
then the $\alpha$'s satisfy $\alpha_{[{\bf x}~-1]}<\alpha_{[\bf x]}<\alpha_{[{\bf x}~1]}$
and for all $i<N$, $\alpha_{[{\bf x}~1]}<\alpha_{[{\bf y}~-1]}$
for ${\bf x,y}\in\{-1,1\}^i$ and ${\bf x}_1\dots{\bf x}_i<{\bf y}_1\dots{\bf y}_i$.
\item For all $k'\leq N$ and $\alpha\in(1,3)$,
\begin{equation*}
(1.5+o_{k'})^\alpha-(1.5-o_{k'})^\alpha+\sum_{k'<i<N}((1.5-o_i)^\alpha-(1.5+o_i)^\alpha)>0.
\end{equation*}
\item
\begin{align*}
&(1.1-q^*)^3+(1.1+q^*)^3-2\cdot d^3+\sum_{i<N}((1.5-o_i)^3-(1.5+o_i)^3)>0,\text{ and}\\
&(1.1-q^*)+(1.1+q^*)-2\cdot d+\sum_{i<N}((1.5+o_i)-(1.5-o_i))<0.
\end{align*}
\end{enumerate}

We proved the base case in our earlier example for $n=10$. 
Assume for $k\leq N$, there exist $0<o_1,\dots,o_k<\delta$ which satisfy the three properties.
We first prove the inductive step for the second and third statements.

By inductive hypothesis, we know for all $k'\leq k$ and $\alpha\in (1,3)$,
\begin{equation*}
(1.5+o_{k'})^\alpha-(1.5-o_{k'})^\alpha+\sum_{k'<i\leq k}((1.5-o_i)^\alpha-(1.5+o_i)^\alpha)>0,
\end{equation*}

Since there are finite integral values of $k'\leq k$, and the expression is $>0$ for all
values of $k'$, then there exists an $\epsilon>0$ such that the expression is $\geq\epsilon$ for
all values of $k'$.
Then we define $z_a$ such that $(1.5+z_a)^\alpha-(1.5-z_a)^\alpha<\frac{\epsilon}{2}$
for $\alpha\in (1,3)$. Then for all $0<z<z_a$, $k'\leq k+1$, and $\alpha\in (1,3)$,
\begin{equation*}
(1.5+o_{k'})^\alpha-(1.5-o_{k'})^\alpha+\sum_{k'<i\leq k+1}((1.5-o_i)^\alpha-(1.5+o_i)^\alpha)>0.
\end{equation*}
So as long as we set $0<o_{k+1}<z_a$, the inductive step of the second property will be fulfilled.
Now we move to the third property.
We have the following from the inductive hypothesis:
\begin{align*}
(1.1-q^*)^3+(1.1+q^*)^3-2\cdot d^3+\sum_{i\leq k'}((1.5-o_i)^3-(1.5+o_i)^3)>0,\\
(1.1-q^*)+(1.1+q^*)-2\cdot d+\sum_{i\leq k'}((1.5+o_i)-(1.5-o_i))<0.
\end{align*}

We may similarly find $z_b$ such that for all $0<o_{k+1}<z_b$,

\begin{align*}
(1.1-q^*)^3+(1.1+q^*)^3-2\cdot d^3+\sum_{i\leq k+1}((1.5-o_i)^3-(1.5+o_i)^3)>0,\\
(1.1-q^*)+(1.1+q^*)-2\cdot d+\sum_{i\leq k+1}((1.5+o_i)-(1.5-o_i))<0.
\end{align*}

Now we move to proving the inductive step of the first property.
Given ${\bf x}\in \{-1,1\}^k$, let $p({\bf x}),n({\bf x})\in \{-1,1\}^k$ 
denote the vectors which sit on either side of $\alpha_{\bf x}$ in the ordering, 
i.e., $\alpha_{\bf x}$ is the only $\alpha_{\bf y}$ in the range 
$(\alpha_{p({\bf x})},\alpha_{n({\bf x})})$ such that $|{\bf y}|=k$.
If ${\bf x}=[1~\dots~1]$, then set $\alpha_{n({\bf x})}=3$, and if 
${\bf x}=[0~\dots~0]$, set $\alpha_{p({\bf x})}=1$.
Define
\begin{equation*}
f(\alpha,{\bf x},z)=E_{\bf x}+(1.5-z)^\alpha-(1.5+z)^\alpha.
\end{equation*}
By inductive hypothesis, we have that $f(\alpha_{\bf x},{\bf x},0)=0$.
We must show there exists $z_{\bf x}$ such that for all $0\leq z\leq z_{\bf x}$, 
$f(\alpha_{\bf x},{\bf x},z)<0$ and $f(\alpha_{n({\bf x})},{\bf x},z)>0$. This will imply that
if we choose $0<o_{k+1}<z_{\bf x}$, then
$\alpha_{\bf [x~1]}\in (\alpha_{\bf x},\alpha_{n({\bf x})})$.

Case 1: ${\bf x}\neq [1\dots 1]$.
Since $f(\alpha_{\bf x},{\bf x},0)=0$, and by Fact \ref{fact:1.5},
then for all $0<z<.01$, $f(\alpha_{\bf x},{\bf x},z)<0$.
Now denote $i^*$ as the greatest index such that ${\bf x}_{i^*}=-1$.
Then $n({\bf x})=[{\bf x}_1\dots {\bf x}_{i^*-1}~1~-1\dots -1]$.
By statement 1 of the inductive hypothesis ($\alpha_{n({\bf x})}$ is a root of $E_{n({\bf x})}=0$),
\begin{equation*} 
(1.1-q^*)^{\alpha_{n({\bf x})}}+(1.1+q^*)^{\alpha_{n({\bf x})}}-2\cdot d^{\alpha_{n({\bf x})}}+
\sum_{i\leq k} (n({\bf x})_i(1.5-o_i)^{\alpha_{n({\bf x})}}-n({\bf x})_i(1.5+o_i)^{\alpha_{n({\bf x})}})=0
\end{equation*}
From statement 2 of the inductive hypothesis, we know that
\begin{equation*}
(1.5-o_{i^*})^{\alpha_{n({\bf x})}}-(1.5+o_{i^*})^{\alpha_{n({\bf x})}}+\sum_{i^*<i\leq k}((1.5+o_i)^{\alpha_{n({\bf x})}}-(1.5-o_i)^{\alpha_{n({\bf x})}})<0.
\end{equation*}
It follows that 
\begin{equation*}
(1.1-q^*)^{\alpha_{n({\bf x})}}+(1.1+q^*)^{\alpha_{n({\bf x})}}-2\cdot d^{\alpha_{n({\bf x})}}+
\sum_{i<i^*} (n({\bf x})_i(1.5-o_i)^{\alpha_{n({\bf x})}}-n({\bf x})_i(1.5+o_i)^{\alpha_{n({\bf x})}})>0,
\end{equation*}
and furthermore,
\begin{equation*}
(1.1-q^*)^{\alpha_{n({\bf x})}}+(1.1+q^*)^{\alpha_{n({\bf x})}}-2\cdot d^{\alpha_{n({\bf x})}}+
\sum_{i<i^*} ({\bf x}_i(1.5-o_i)^{\alpha_{n({\bf x})}}-{\bf x}_i(1.5+o_i)^{\alpha_{n({\bf x})}})>0.
\end{equation*}
Therefore, $f(\alpha_{n({\bf x})},0)>0$, so denote $f(\alpha_{n({\bf x})},0)=\epsilon>0$.
Then because of Fact \ref{fact:1.5}, there exists $z_{\bf x}$ such that $\forall 0<z<z_{\bf x}$,
$f(\alpha_{n({\bf x})},z)>0$.

Case 2: ${\bf x}= [1\dots 1]$.
Since $f(\alpha_{\bf x},0)=0$, and by Fact \ref{fact:1.5},
then for all $0<z<.01$, $f(\alpha_{\bf x},z)<0$.
By property 3 of the inductive hypothesis, we have 
\begin{equation*}
(1.1-q^*)^3+(1.1+q^*)^3-2\cdot d^3+\sum_{i\leq k}((1.5-o_i)^3-(1.5+o_i)^3)>0,
\end{equation*}
so say this expression is equal to some $\epsilon>0$.
Then from Fact \ref{fact:1.5}, there exists $z_{\bf x}$ such that for all $0<z<z_{\bf x}$,
$0<(1.5+z)^3-(1.5-z)^3<\frac{\epsilon}{2}$.
Combining these, we have $f(3,z)>0$ for all $0<z<z_{\bf x}$.

To recap, in both cases we showed there exists $z_{\bf x}$ such that for all $0<z<\min(.01,z_{\bf x})$,
$f(\alpha_{\bf x},z)<0$ and $f(\alpha_{n({\bf x})},z)>0$.
We may perform a similar analysis on a related function $f'$, defined as
$f'(\alpha,{\bf x},z)=E_{\bf x}+(1.5+z)^\alpha-(1.5-z)^\alpha$
to show there exists $z'_{\bf x}$ such that for all
$0<z<z_{\bf x}'$, $f'(\alpha_{p({\bf x})},z)<0$ and $f'(\alpha_{\bf x},z)>0$.
We perform this analysis over all ${\bf x}\in\{-1,1\}^k$.

Finally, we set $o_{k+1}=\min_{\bf x}(z_{\bf x},z'_{\bf x},z_a,z_b,.01)$.
Given ${\bf x}\in\{-1,1\}^k$,
since $f(\alpha_{\bf x},o_{k+1})<0$ and $f(\alpha_{n({\bf x})},o_{k+1})>0$,
 there must exist a root $\alpha_{[{\bf x}~1]}\in (\alpha_{\bf x},\alpha_{n({\bf x})})$
(and by Fact \ref{fact:1.5}, the function is monotone in $\alpha$ in the short interval $(\alpha_{\bf x},\alpha_{n({\bf x})})$,
so there is exactly one root).
Similarly, there must exist a root 
$\alpha_{[{\bf x}~-1]}\in (\alpha_{p({\bf x})},\alpha_{\bf x})$.
Then we have shown $\alpha_{[{\bf x}~-1]}$ and $\alpha_{[{\bf x}~1]}$ are roots of $E_{[{\bf x}~-1]}$ and
$E_{[{\bf x}~1]}$, respectively.
By construction, $\alpha_{[{\bf x}~-1]}<\alpha_{\bf x}<\alpha_{[{\bf x}~1]}$, so condition 1 is satisfied.
Now we need to show condition 2 is satisfied. Given ${\bf x},{\bf y}\in\{-1,1\}^k$, let $k'$ be the largest number
for which ${\bf x}_i={\bf y}_i$, $\forall i\leq k'$. Let ${\bf z}={\bf x}_{[1\dots k']}={\bf y}_{[1\dots k']}$.
Then by the inductive hypothesis, 
\begin{equation*}
\alpha_{\bf x}<\alpha_{n({\bf x})}\leq\alpha_{\bf z}\leq\alpha_{p({\bf y})}<\alpha_{\bf y}.
\end{equation*}
It follows that
\begin{equation*}
\alpha_{[{\bf x}~-1]}<\alpha_{[{\bf x}~1]}<\alpha_{\bf z}<\alpha_{[{\bf y}~-1]}<\alpha_{[{\bf y}~1]},
\end{equation*}
proving condition 2.
This completes the induction.

\end{proof}

Now we are ready to prove Lemma~\ref{lem:general_lb}.

\begin{proof}[Proof of Lemma \ref{lem:general_lb}]
Given $n$, and setting $N=\lfloor(n-8)/4\rfloor$, we will show there exists
a clustering instance $(V,d)$ of size $|V|=n$, a witness $r$, and
a set of $2^N+2$ $\alpha$'s 
$1=\alpha_0<\alpha_1<\cdots<\alpha_{2^N}<\alpha_{2^N+1}=3$, such that
$\Phi^{(p)}_{\A_3(\alpha)}(\V)$ oscillates above and below
$r$ between each interval $(\alpha_i,\alpha_{i+1})$.

We start by using the construction from Lemma \ref{lem:general_lb},
which gives a clustering instance with $2N+8$ points and 
$2^N+2$ values of $\alpha$ for which
$\alpha$-linkage creates a unique merge tree.
The next part is to add $2N$ more points and define a witness $r$ so that the cost function
alternates above and below $r$ along each neighboring $\alpha$ interval, for a total
of $2^N$ oscillations.
Finally, we will finish off the proof in a manner similar to Lemma \ref{lem:a13lower}.

Starting with the clustering instance $(V,d)$ from Lemma \ref{lem:general_lb},
we add two sets of points, $C_A$ and $C_B$, which do not interfere with the previous merges,
and ensure the cost functions alternates.
Let $C_A=\{c_a,c_a',a_1,a_2,\dots,a_N\}$ and 
$C_B=\{c_b,c_b',b_1,b_2,\dots,b_N\}$.
All distances between two points in $C_A$ are 1, and similarly for $C_B$.
All distances between a point in $C_A$ and a point in $C_B$ are 2.
The distances between $C_A\cup C_B$ and $A\cup B$ are as follows (we defined the sets $A$ and $B$
in Lemma \ref{lem:general_lb}).
\begin{align*}
&d(p_a,c_a)=d(p_a,c_a')=d(q_a,c_a)=d(q_a,c_a')=1.51,\\
&d(p_b,c_b)=d(p_b,c_b')=d(q_b,c_b)=d(q_b,c_b')=1.51,\\
&d(p_a,c_b)=d(p_a,c_b')=d(q_a,c_b)=d(q_a,c_b')=2,\\
&d(p_b,c_a)=d(p_b,c_a')=d(q_b,c_a)=d(q_b,c_a')=2,\\
&d(p_a,c)=d(q_a,c)=d(p_b,c)=d(q_b,c)=2~\forall c\in C_A\cup C_B\setminus\{c_a,c_a',c_b,c_b'\},\\
&d(c,p_i)=d(c,q_i)=1.51~\forall 1\leq i\leq N-1\text{ and }c\in C_A\cup C_B. 
\end{align*}

We will specify the distances between $\{c_a,c_a',c_b,c_b'\}$ and $\{p_N,q_N\}$ 
soon, but they will be in $[1.6,2]$.
So at the start of the merge procedure, all points in $C_A$ merge together, and
all points in $C_B$ merge together. Then all merges from Lemma \ref{lem:general_lb} take place,
because all relevant distances are smaller than $1.51$.
We end up with four sets: $A$, $B$, $C_A$, and $C_B$.
The pairs $(A,B)$ and $(C_A,C_B)$ are dominated by distances of length 2, 
so the merges $(C_A,A)$ and $(C_B,B)$ will occur,
which dominate $(C_A,B)$ and $(C_B,A)$ because of the distances between
$\{p_a,q_a,p_b,q_b\}$ and $\{c_a,c_a',c_b,c_b'\}$.
The final merge to occur will be $(C_A\cup A,C_B\cup B)$,
however, the 2-median pruning step will clearly pick the 2-clustering
$C_A\cup A$, $C_B\cup B$, since no other clustering in the tree has almost all 
distances $\leq 1.51$.
Then by construction, $c_a$ or $c_a'$ will be the best center for $C_A\cup A$,
which beat $p_a$ and $q_a$ because $1.51\cdot(2N)<1.1\cdot N+2\cdot N=1.55\cdot (2N)$.
Similarly, $c_b$ or $c_b'$ will be the best center for $C_B\cup B$.
Note that centers $\{c_a,c_a'\}$ and $\{c_b,c_b'\}$ currently give
equivalent 2-median costs. 
Denote this cost by $r'$ (i.e., the cost before we set the distances to $p_N$ and $q_N$).

Now we set the final distances as follows.
\begin{align*}
d(c_a,p_N)=d(c_b,q_N)=1.6,\\
d(c_a',p_N)=d(c_b',q_N)=1.7,\\
d(c_a',q_N)=d(c_b',p_N)=1.8,\\
d(c_a,q_N)=d(c_b,p_N)=1.9.\\
\end{align*}

If $p_N\in A$ and $q_N\in B$, then $c_a$ and $c_b$ will be the best centers,
achieving cost $r'+3.2$ for $(C_A\cup A,C_B\cup B)$.
If $p_N\in B$ and $q_N\in A$, then $c_a'$ and $c_b'$ will be the best centers,
achieving cost $r'+3.6$ for $(C_A\cup A,C_B\cup B)$.

The distances are also constructed so that in the variant where the pruning outputs
the optimal centers, and then all points are allowed to move to their closest center,
the cost still oscillates.
First note that no points other than $p_N$ and $q_N$ are affected, since
$d(c_a,p_i)=d(c_a,q_i)$ for $i<N$, and similarly for $c_b$.
Then $p_N$ will move to the cluster with $c_a$ or $c_a'$,
and $q_N$ will move to the cluster with $c_b$ or $c_b'$.
If $p_N$ was originally in $A$, then the cost is $r'+3.2$,
otherwise the cost is $r'+3.4$.

In either scenario, we set $r=r'+3.3$. Then we have ensured for all ${\bf x}\in \{-1,1\}^{N-1}$,
the cost for $\alpha\in(\alpha_{p({\bf x})},\alpha_{\bf x})$ is $<r$,
and the cost for $\alpha\in(\alpha_{\bf x},\alpha_{n({\bf x})})$ is $>r$.
We have finished our construction of a clustering instance whose
cost function alternates $2^N$ times as $\alpha$ increases.

To finish the proof, we will show there exists a set $S=\{V_1,\dots,V_s\}$ of size 
$s=N=\lfloor(n-8)/4\rfloor\in \Omega(n)$ that is shattered by $\mathcal{A}$.
Such a set has 
$2^N$
orderings total.
For $V_1$, we use the construction which alternates $2^N$
times. For $V_2$, we use the same construction, but we eliminate $(p_N,q_N)$ so that there
are only $N-1$ rounds (the extra two points can be added to $C_A$ and $C_B$ to preserve
$|V_2|=n$). Then $V_2$'s cost will alternate $\frac{1}{2}\cdot 2^N$
times, between the intervals $(\alpha_{p({\bf x})},\alpha_{\bf x})$ and
$(\alpha_{\bf x},\alpha_{n({\bf x})})$, for ${\bf x}\in \{-1,1\}^{N-2}$.
So $V_2$ oscillates every other time $V_1$ oscillates, as $\alpha$ increases.
In general, $V_i$ will be the construction with only $N-i+1$ rounds, 
oscillating $2^{\frac{N}{2^{i-1}}}$ times, 
and each oscillation occurs every other time $V_{i-1}$ oscillates.
This ensures for every ${\bf x}\in \{-1,1\}^{N-1}$, 
$(\alpha_{p({\bf x})},\alpha_{\bf x})$ and
$(\alpha_{\bf x},\alpha_{n({\bf x})})$ will have unique labelings, for a total of 
$2^N$ labelings.
This completes the proof.
\end{proof}

\begin{note}
As in Lemma \ref{lem:a13lower}, this lower bound holds even if the cost function is the symmetric distance to the ground truth clustering. Merely let $p_N$ and $q_N$ belong to different ground truth clusters, but for all
$i<N$, $p_i$ and $q_i$ belong to the same ground truth cluster. Since in each adjacent $\alpha$ interval, $p_N$
and $q_N$ switch clusters, this shows the symmetric distance to the ground truth clustering oscillates between
every interval.

Furthermore, as was the case in Lemma \ref{lem:oscillate}, we can achieve a tradeoff between the number of oscillations, and the difference in cost between
the oscillations. Specifically, for all $0<\gamma<\leq 1$, we can show an instance which oscillates $2^N$ times
above $r$ and below $\gamma r$, where $N=\lfloor\gamma(n-8)/4\rfloor$.
We use $N$ points to create the gadgets above, and then we add $\frac{1-\gamma}{2}\cot n$ points to $a$, and $\frac{1-\gamma}{2}\cdot n$ points to $z_N$.
\end{note}

Now we give an ERM algorithm for $\A_2$, similar to Algorithm~\ref{alg:a13-erm}.

\begin{algorithm} 
\caption{An algorithm for finding an empirical cost minimizing algorithm in $\A_2$}
\label{alg:a2-erm}
\begin{algorithmic}[1]
\Require {Sample $\sample = \left\lbrace \V^{(1)}, \hdots, \V^{(m)} \right\rbrace.$}
\State Let $T=\emptyset$. For each sample $\V^{(i)} = \left(V^{(i)},d^{(i)}\right)\in \sample$, and for all $A,B,X,Y \subseteq V^{(i)}$, solve for $\alpha$ (if a solution exists) in the following equation and add the solutions to $T$:
\[\frac{1}{|A||B|}\sum_{p \in A, q \in B} (d(p,q))^\alpha = \frac{1}{|X||Y|}\sum_{x \in X, y \in Y} (d(x,y))^\alpha.\] \label{step:find_alphas_2}
\State Order the elements of set $T \cup \{-\infty, +\infty \}$ as $\alpha_1< \hdots < \alpha_{|T|}$.
For each $0\leq i\leq |T|$, pick an arbitrary $\alpha$ in the interval $(\alpha_i,\alpha_{i+1})$ 
and run $\A_2(\alpha)$ on all clustering instances in $\sample$ to compute $\sum_{\V \in \sample} \Phi_{\A_2(\alpha)}(\V)$. Let $\hat{\alpha}$ be the value which minimizes $\sum_{\V \in \sample} \Phi_{\A_2(\alpha)}(\V)$.
\Ensure{$\hat{\alpha}$}
\end{algorithmic} 
\end{algorithm}

\begin{theorem}\label{thm:a2algo}
Let $\Phi$ be a clustering objective and let $\prune$ be a pruning function. Given an input sample of size 
$m = O\left(\left(\frac{H}{\epsilon}\right)^2\left(n + \log \frac{1}{\delta} \right)\right)$,  Algorithm~\ref{alg:a2-erm} $(\epsilon,\delta)$-learns the class $\mathcal{A}_2\times\{\prune\}$ with respect to the cost function $\Phi$.
\end{theorem}

\begin{proof}
The sample complexity analysis follows the same logic as the proof of Theorem~\ref{thm:a13algo}. To prove that Algorithm~\ref{alg:a2-erm} indeed finds the empirically best $\alpha$, recall from the pseudo-dimension analysis that the cost as a function of $\alpha$ for any instance is a piecewise constant function with $O(n^23^{2n})$  discontinuities. In Step~\ref{step:find_alphas_2} of Algorithm~\ref{alg:a2-erm}, we solve for the values of $\alpha$ at which the discontinuities occur and add them to the set $T$. $T$ therefore partitions $\alpha$'s range into $O(mn^23^{2n})$ subintervals. Within each of these intervals, $\sum_{\V \in \sample} \Phi_{\A_2(\alpha)}(\V)$ is a constant function. Therefore, we pick any arbitrary $\alpha$ within each interval to evaluate the empirical cost over all samples, and find the empirically best $\alpha$. 
\end{proof}

\end{document}